# Recent Advancements in Battery State of Power Estimation Technology: A Comprehensive Overview and Error Source Analysis

Ruohan Guo[*], Weixiang Shen

School of Science, Computing and Engineering Technologies, Swinburne University of Technology, Hawthorn, Victoria 3122, Australia

**Graphic abstract:**

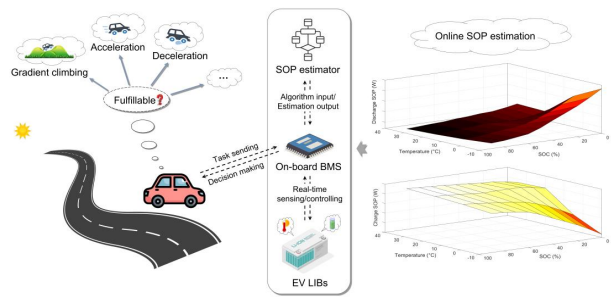

**Descriptive words:** The key functionalities of battery SOP estimation in providing decision-making references for electric vehicle power-intensive tasks like gradient climbing, acceleration, and deceleration.

**Abstract:** Accurate state of power (SOP) estimation is of great importance for lithium-ion batteries in safety-critical and power-intensive applications for electric vehicles. This review article delves deeply into the entire development flow of current SOP estimation technology, offering a systematic breakdown of all key aspects with their recent advancements. First, we review the design of battery safe operation area, summarizing diverse limitation factors and furnishing a profound comprehension of battery safety across a broad operational scale. Second, we illustrate the unique discharge and charge characteristics of various peak operation modes, such as constant current, constant voltage, constant current-constant voltage, and constant power, and explore their impacts on battery peak power performance. Third, we extensively survey the aspects of battery modelling and algorithm development in current SOP estimation technology, highlighting their technical contributions and specific considerations. Fourth, we present an in-depth dissection of all error sources to unveil their propagation pathways, providing insightful analysis into how each type of error impacts the SOP estimation performance. Finally, the technical challenges and complexities inherent in this field of research are addressed, suggesting potential directions for future development. Our goal is to inspire further efforts towards developing more accurate and intelligent SOP estimation technology for next-generation battery management systems.

---

[*] Corresponding author (Email: rguo@swin.edu.au)
Email address: rguo@swin.edu.au; wshen@swin.edu.au

| Abbreviations | |
|---|---|
| AI | Artificial intelligence |
| ANFIS | Adaptive neuro-fuzzy inference system |
| AR | Auto regression |
| ARMA | Auto regressive moving average |
| BMS | Battery management system |
| BV | Butler-Volmer |
| CC | Constant current |
| CC-CV | Constant current-constant voltage |
| CP | Constant power |
| CV | Constant voltage |
| CM | Characteristic mapping |
| DEKF | Dual extended Kalman filter |
| DP | Dual polarization |
| DT | Digital twin |
| ECM | Equivalent circuit model |
| EIS | Electrochemical impedance spectroscopy |
| EM | Electrochemical model |
| EV | Electric vehicle |
| FF-RLS | Forgetting-factor recursive least squares |
| FO | Fractional order |
| HPPC | Hybrid pulse power characterization |
| IA | Iterative approaching |
| ICT | Incremental OCV test |
| IO | Integer order |
| LCT | Low-current test |
| LIB | Lithium-ion battery |
| MPC | Model predictive control |
| NE | Negative electrode |
| NNM | Neural network model |
| OCV | Open circuit voltage |
| OLP | Open-loop prediction |
| PDE | Partial derivative equation |
| PE | Positive electrode |
| POM | Peak operation mode |
| RC | Resistor-capacitor |
| SOA | Safe operation area |
| SOC | State of charge |
| SOE | State of energy |
| SOH | State of health |
| SOP | State of power |
| SOT | State of temperature |
| SPM | Single particle model |
| UKF | Unscented Kalman filter |

1. **Introduction**

Global energy revolution is driving the rapid development of transportation electrification [1,2]. Lithium-ion batteries (LIBs), serving as critical energy storage components and power sources for electric vehicles (EVs), boast a multitude of desirable features. These include high volumetric energy and power density, extended cycle life, low self-discharge rate, lightweight construction, and relatively low manufacturing costs [3,4]. These attributes have garnered widespread favor in both industrial and academic circles. As the transition to zero-emission transportation gathers momentum, it is projected that the global EV fleet will exceed 300 million by 2030, necessitating an installed battery capacity of over 3000 GWh [5,6]. However, the high energy and power density of LIBs, combined with the use of flammable materials, present substantial challenges in ensuring their safe and reliable usage. In recent years, the increasing cases of safety failures and firing accidents of LIBs in EVs underline an urgent significance to mitigate the potential risks arising from over discharging and charging, internal short circuits, and thermal abuse or even thermal runaway [7–9]. Equipping LIBs with advanced battery management systems (BMSs) for real-time monitoring and regulation of their internal states offers a viable solution to this safety challenges [10].

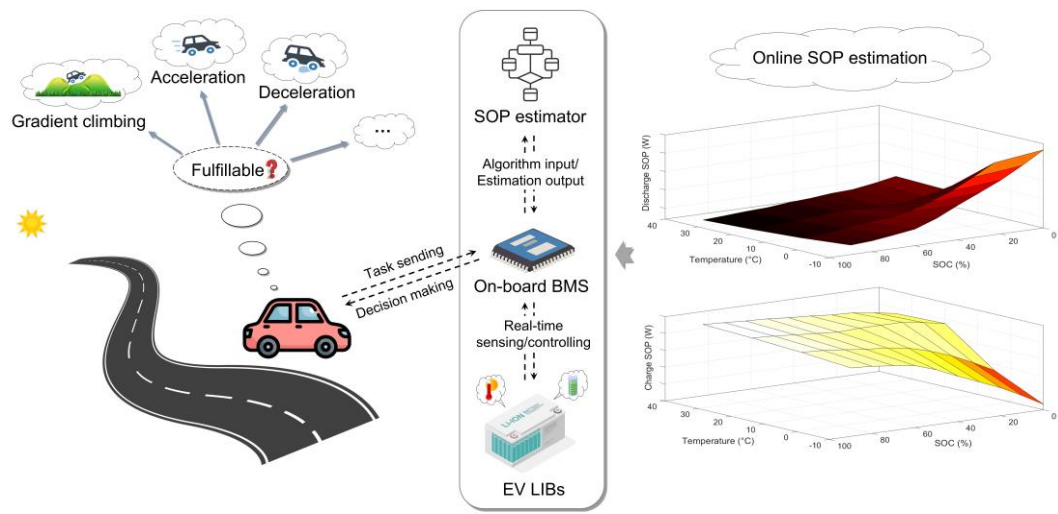

Fig. 1. The application scenarios of battery SOP estimation in EVs.

The field of battery state estimation, such as state of charge (SOC), state of energy (SOE), state of health (SOH), state of power (SOP), and state of temperature (SOT), has evolved rapidly over the past decades [11–16]. It has now become a vast area of research, rich with diverse methodologies and technical reviews. For instance, Hannan et al. [17] provided an extensive overview of available SOC estimation technology, focusing on battery modelling and algorithm development. Zheng et al. [18] conducted a thorough exploration of the error sources and their propagation pathways in SOC estimation, encompassing initial SOC deviation, measurement noise or bias, battery model error, open circuit voltage (OCV)-SOC curve error, closed-loop correction gain error, and SOH error. Quade et al. [19] surveyed SOE studies, categorizing different SOE definitions based on the concepts of inherently stored energy and usable energy, and they proposed a new, practically meaningful definition to better understand battery energy potential. Xiong et al. [20] summarized both experiment-based and model-based SOH estimation technology, highlighting their respective strengths and weaknesses. Wang et al. [21] and Hu et al. [10] comprehensively reviewed the challenges faced in this domain and discussed potential directions for future development of all key states.

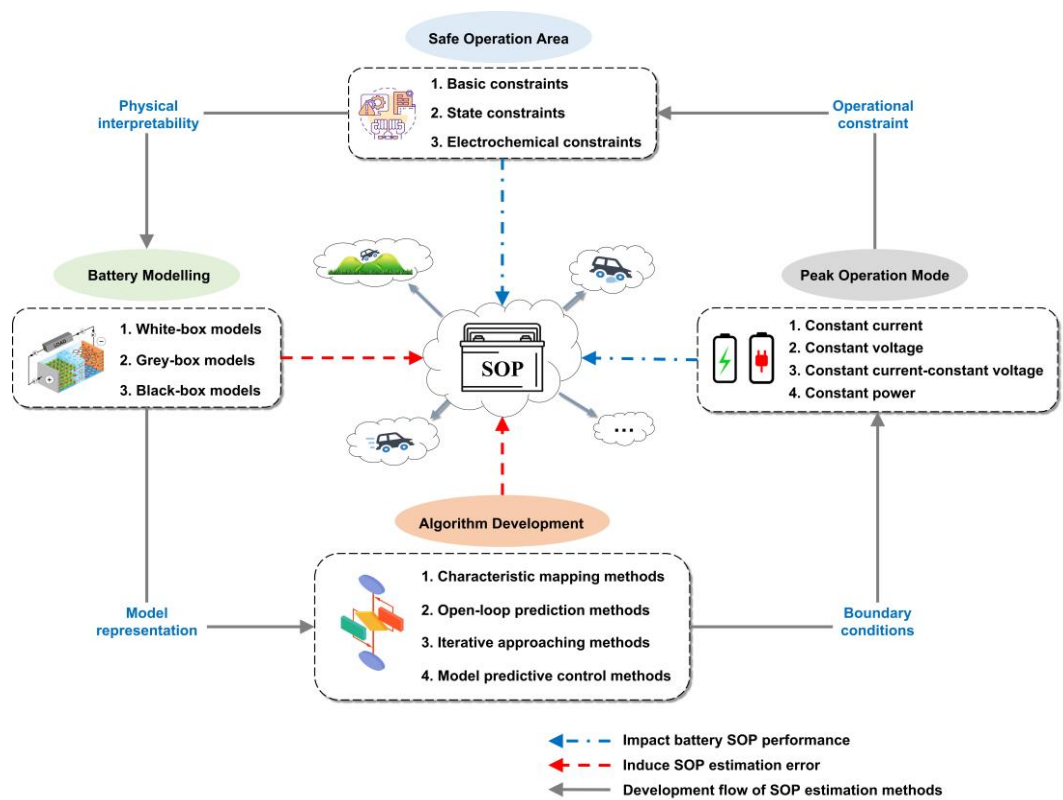

Fig. 2. An overview of key aspects in battery SOP estimation technology.

Of particular interest among these internal states is battery SOP, which measures the short-term peak power capability that batteries can deliver to or absorb from the EV powertrain. Given the peak power sequence over a prediction window, denoted as $P_{\max,k+i}^{dis/chg}$ ($i$=1,2,…,$K$), battery SOP is generally defined as the minimum of the achievable power within this sequence [2], as illustrated in Eq. (1).

$$SOP_k = \min\left\{\left|P_{\max,k+1}^{dis/chg}\right|, \left|P_{\max,k+2}^{dis/chg}\right|, \ldots, \left|P_{\max,k+K}^{dis/chg}\right|\right\} \tag{1}$$

The Idaho National Engineering & Environmental Laboratory first proposed a hybrid pulse power characterization (HPPC) method, as specified by the Partnership for New Generation Vehicles, to standardize the testing procedure for battery instantaneous peak power (also known as instantaneous SOP) [22]. Building on this, early studies investigated the characteristics of battery instantaneous SOP across the operation ranges of SOC, temperature, and aging level [23]. Despite ease of implementation, instantaneous SOP estimation enables limited contributions to optimize battery energy and power management, owing to its short prediction window of only one sampling interval. Acknowledging this limitation, the concept of continuous SOP was introduced in the literature, extending the prediction window to several tens of seconds or even a few minutes [24]. Battery SOP estimation over such an extended window plays a decisive role in determining if batteries can fulfill a relatively long-lasting task [25], such as to compete an EV travel under a specific route while avoiding preventive disconnections before arrival based on the remaining energy reserve or the stipulated power specification. In this context, the intuitive references provided by battery SOP are vital for on-board BMSs and EV drivers in making decisions and choosing driving strategies during acceleration, regenerative braking, and gradient climbing, as illustrated in Fig. 1. Beyond this, battery SOP (or peak current) is instrumental in planning pre-heating schedules [26,27] and fast charging strategies [28,29]. This is particularly important for rapid battery warm-up at extremely low temperatures and for minimizing battery downtime under energy depletion. Additionally, the studies in [30,31] have recognized the potentiality and viability of employing battery SOP (or peak current) in identifying battery degradation patterns and diagnosing short circuit faults. Given these diverse applications, battery SOP estimation has emerged as a significant research focus in recent years. While considerable efforts have been directed towards integrating battery SOP into joint or co-estimation frameworks, relevant research often treated it as an extension of studies on other internal states like SOC, SOE, and SOH. Previous works in [2,23] have reviewed those SOP-involved frameworks from perspectives of battery modelling, parameter identification and state estimation, however, these studies did not offer a systematic breakdown of all key aspects with their recent advancements.

Addressing this gap, this article presents a comprehensive overview of the entire development flow of current SOP estimation technology along with an in-depth analysis of their error sources. As depicted in Fig. 2, we delve into the recent advancements across a broad spectrum of aspects in battery SOP estimation and thoroughly examine how each type of error affects its performance, providing valuable insights for future research and applications. The structure of this paper is organized as follows: Section 2 discusses the design of battery safe operation area (SOA) with associated limitation factors. Section 3 illustrate the unique discharge and charge characteristics of various peak operation modes (POMs) while exploring their impacts on battery peak power performance. Sections 4 and 5 survey state-of-the-art battery modelling and algorithm development in current SOP estimation technology, respectively. Section 6 investigates the error sources and their propagation pathways in battery SOP estimation. Outlooks and future prospects are addressed in Section 7.

## 2. Safe operation area

High capacity and extensive serial-parallel connections of LIBs in EVs introduce potential challenges with regard to safety, durability, and uniformity. A SOA represents a pre-defined zone where batteries can function with desired stability and reliability. It is crucial to appropriately design a SOA for effective regulation and protection of battery electro-chemical-thermal behaviors under dynamic and variable operation conditions, contributing to enhanced safety and health management of all the cells in a battery pack. Extensively incorporating all the pertinent limitation factors and clearly defining their boundaries are, therefore, imperative to achieve a balance between battery discharge and charge capabilities and potential safety and health risks. In this section, available limitation factors and their operational constraints in battery SOP estimation are overviewed, spanning from macro scale to micro scale, as illustrated in Fig. 3.

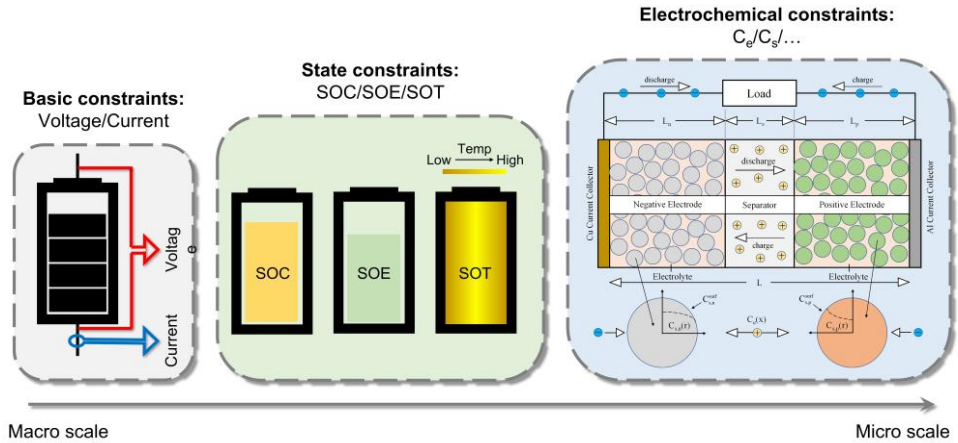

Fig. 3. Battery SOA design from macro scale to micro scale.

### 2.1. Basic constraint

Battery voltage and current are the most fundamental external characteristics that have been widely used as the basic constraints in SOA design for SOP estimation [32].

Battery voltage, or so-called terminal voltage, refers to a measurable voltage difference between the positive and negative poles of battery. Unlike OCV, which gauges the potential difference of two electrodes at an equilibrium state, the terminal voltage is influenced by the polarization dynamics originating from lithium surface concentration, electrolyte concentration, and side reaction overpotential during a discharge or charge process. Battery terminal voltage should be controlled tightly within an appropriate range to avoid over discharging or charging to an extremely low or high voltage. Either way may trigger unwanted side reactions with an irreversible structural damage that degrade battery power and energy performance and reduce cycle life. Specifically, it was reported in [33,34] that an extremely low voltage or over discharge can dissolve copper ions into electrolyte and form copper dendrites to penetrate the separator, causing an internal short circuit failure. On the flip side, an extremely high voltage or over charge can lead to metallic lithium deposition, consumption of electrolyte (common electrolyte starts to decompose at a voltage above 4.5 V), and significant heat generation, increasing the potential risk of internal short circuit and thermal abuse/runaway [35]. A rational voltage operation range differs with battery chemistries, which is around 2.5~4.2 V for electrode materials of graphite/LCO, graphite/NCA, graphite/NCM, and graphite/LMO, around 2.0~3.7 V for electrode materials of graphite/LFP, and around 1.5~2.7 V for electrode materials of LTO/LMO. Owing to the heightened internal resistance and the steep slope of OCV-SOC curve in the regions of extremely low and high SOCs, battery terminal voltage experiences rapid declines and increases with negligible charge throughput. For this reason, it is often recommended to adopt a slightly more conservative threshold of cut-off voltages for battery peak power operations in practical applications. Implementing such buffer zones can serve as a protective measure against an overaggressive SOP estimation, while their influence on battery available capacity and peak power performance is generally trivial [36].

Current is another critical external characteristic of batteries. To assess the discharge and charge capabilities of batteries in different capacities, C-rate is proposed as a metric of current magnitude at which the battery is discharged and charged relative to its nominal capacity. In real-world applications, batteries often experience pulse discharges and charges at high current rates. Some of these scenarios can pose risks to battery safety due to potential over-voltages and under-voltages, as seen in EV activities like acceleration, hill climbing, and regenerative braking. Meanwhile, other scenarios need a balance between reducing battery downtime and extending its lifespan, such as in the cases of preheating and fast charging. However, excessive discharge and charge current are detrimental to battery health [37,38]. Specifically, a high discharge rate can result in a large amount of heat generation and induce an irreversible mechanical damage to active materials while a high charge rate can significantly polarize the negative electrode (NE) and thus trigger lithium plating, both of which will accelerate battery degradation with an increased risk of internal short circuit [37]. Generally, battery discharge capability is greater than charge. The former can be designed up to 15~20 C-rate in some high-power LIBs while the latter can only be 1~2 C-rate suggested by most battery manufacturers, especially at low temperatures. These current constraints provided by battery manufacturers, sometimes, will be further separated into instantaneous pulse constraint (i.e., over 1 s) and continuous pulse constraint (i.e., over a few tens of seconds such as 30 s) that correspond to instantaneous SOP and continuous SOP, respectively.

## 2.2. State constraint

Although voltage and current showcase critical electrical behaviors of batteries, they are only a matter of outward manifestation of battery internal characteristics. For this reason, researchers have incorporated battery internal states as limitation factors while establishing associated state constraints to bolster battery safety.

SOC is a representative of non-measurable internal states, indicating the ratio of the remaining capacity to the nominal capacity that can be withdrawn from batteries. Intuitively, batteries should not be further discharged or charged when they are already out of capacity (i.e., 0% SOC) or full capacity (i.e., 100% SOC), and thus, it is reasonable to bring SOC constraint into battery SOP estimation to avert these extreme circumstances occurring within a prediction window. In practice, the cut-off thresholds of SOC will not be set ideally to 0% and 100%, and an appropriate SOC operation range recommended by [39] is from 20% to 80% for EVs and should be much narrower for stationary energy storage systems, considering the adverse impacts on battery lifetime by deep cycles (i.e., cyclic aging) [40] and energy storages (i.e., calendar aging) [41] in extreme SOC regions.

Since LIBs may incur significant energy losses at high current rates in comparison with relatively negligible capacity shrinks, an equal charge throughput at different SOCs can lead to discrepant energy consumptions or absorptions, especially when battery terminal voltage is approaching an end region close to the cut-off thresholds. However, the general index of SOC is incapable to account for energy dissipations on internal resistance, electrochemical reactions and OCV variations, bringing a risk in guaranteeing a sufficient energy reserve to sustain peak power operations of batteries over a prediction window. Unlike SOC, which depends solely on current, SOE represents the ratio of the remaining energy to the nominal energy of a battery, providing a more practically meaningful state constraint for preventing batteries from falling into energy depletion quickly. In light of this, Wang et al. [42] estimated SOE jointly with SOC through a model-based unscented Kalman filter (UKF), both of which were participated in the subsequent SOP estimation as two individual state constraints. Zhang et al. [43] completely replaced SOC with SOE in battery SOP estimation. In their method, SOE not only serves as a state variable in an adaptive UKF but also a limitation factor restricting the peak discharge/charge current over a prediction window. However, the above two works implemented the SOE-constraint SOP estimation by assuming that SOE varies linearly with current like SOC. This assumption is practically untrue as it overlooks energy dissipations inside batteries, thereby yielding over-optimistic SOP estimation results. To address this, Guo and Shen [44] analytically expressed the entire energy release or absorption over a window through a power integration method, and the SOE-constraint peak current estimation boiled down to a quadratic equation. Different from the SOC-constraint peak discharge or charge current estimation in the high or low SOC region, which can yield unrealistic results up to thousands of amperes, this quadratic equation offers no real solutions in such cases, indicating an inherent conflict of maintaining battery terminal voltage and SOE within a pre-defined SOA simultaneously. The simulation results confirmed that SOE is a more effective state constraint than SOC in both discharge and charge SOP estimation.

Temperature is also a key factor affecting battery discharge and charge performance, degradation rate, and operation safety. A suggested temperature operation range by most battery manufacturers for majority LIBs (e.g., graphite/LMO, graphite/NCM, graphite/LFP, graphite/NCA) is around -20~55 °C for discharge and 0~45 °C for charge [33]. Particularly, the minimum charge temperature can extend to -30 °C for LIBs with NE made of LTO. It has been further reported in [35,45,46] that the solid electrolyte interface (SEI) film starts to decompose exothermically when the battery temperature reaches 90 °C, combustible gas will be produced by side reactions between graphite and organic electrolyte when the battery temperature rises up to 120 °C, and the separator will start melting and shutting the cell down when the battery temperature exceeds 130 °C. On the flip side, the charge operations below 0 °C can accelerate battery degradation and shorten cycle life due to the metallic lithium deposition on particle surface at the NE. At extremely low temperatures, the positive electrode (PE) will break down, leading to internal short circuits [47]. Therefore, regardless of their envisioned end-use applications, LIBs need to be managed thermally at each level of assembly, from cell to pack [48]. The thermal state, also known as SOT, has emerged as another critical internal state, manifesting thermal dynamics of batteries. Due to the absence of a rigorous SOT definition, SOT usually stands for either battery surface temperature, volume-average temperature, or core temperature in existing literature [10]. However, for control purposes and safety concerns, the latter two are more indicative of actual internal thermal behaviors of batteries, instead of the readily measurable surface temperature. Since battery temperature changes as a result of electrochemical reactions, phase changes, mixing effects, and joule heating, the basic idea of SOT estimation/prediction is to simulate heat generations (i.e., reversible heat and irreversible heat) and dissipations (i.e., heat conduction, convection, and radiation) inside batteries. In [49,50], the battery thermal dynamics was characterized using the volume-average temperature. Assuming a uniform distribution of temperature and heat generation rate, the authors developed a one-state lumped thermal model to reproduce battery thermal behaviors over a prediction window, and they further estimated the SOT-constraint peak discharge/charge current by solving a quadratic equation. This estimation was then combined with other conventional operational constraints to determine battery SOP under extended multi-constraints. In [25], the core temperature was used to represent SOT, which also functioned as an extra system state that transformed the simplest one-state lumped thermal model into a two-state model. The coupling of thermal and electrical models facilitated the use of a common output equation, which was advantageous for the development of a joint estimation strategy encompassing both thermal and electrical states. The extended-multi-constraint SOP estimation was integrated into a control-oriented framework, aiming to maximize battery power potential while regulating associated electro-thermal behaviors. Niri et al. [51] proposed another type of two-state thermal model that simplified a set of partial derivative equations (PDEs) simulating a uniform and radial heat distribution inside batteries with the *r*-dimensional convective heat transfer boundary conditions. Unlike the aforementioned one, this two-state thermal model took the volume-average temperature and temperature gradient as state variables to implement battery SOP estimation and whole-cell thermal management based on the measurable surface temperature.

### 2.3. Electrochemical constraint

Battery peak power operations necessitate rapid lithium diffusion in and out of solid particles to sustain high current rates. With this concern, some electrochemical variables have been leveraged in recent studies to simulate the dynamics of lithium diffusion [52] and other side reactions [53]. In response, relevant electrochemical constraints have also been introduced into battery SOP estimation [54], enhancing safety protection at the micro scale and offering helpful insights into battery behaviors under the boundary condition.

Smith and Wang [55,56] first investigated the peak power characteristics of batteries based on an electro-chemical-thermal model. The simulation results demonstrated that the concentration gradients in the solid phase can lead to a substantial deviation between the lithium surface concentrations and those in the inner regions of solid particles, and the saturation (or depletion) of lithium surface concentration at PE (or NE) is the root cause for termination of high-rate pulse discharges longer than 10 s. Following this line of thought, the authors of [57–59] focused on simulating the lithium diffusion dynamics in the solid phase and took the lithium surface concentration as a limitation factor in battery instantaneous SOP estimation. Compared with the basic constraints, the simulation results indicated the advantages of this electrochemical constraint in battery safety protection and power exploitation. Nevertheless, these two studies were conducted under an ideal assumption that other electrochemical variables such as electrolyte concentration maintain constant within a short period of time. This assumption could compromise the SOP estimation accuracy over extended prediction windows or under conditions of high current rates or non-ambient temperatures. In addition, it might not be sufficient to guarantee battery safety for all circumstances relying solely on lithium surface concentration, and a more comprehensive approach would require incorporating additional factors to establish a SOA in high dimensions. For these reasons, the authors of [50,60] captured the dynamics of side-reaction overpotential and electrolyte concentration over a prediction window. They implemented associated electrochemical constraints to regulate these factors within safe limits, thereby preventing the formation of plated metallic lithium at NE while avoiding lithium depletion and over-saturation in the liquid phase. Li et al. [61] took this approach a step further by combining two basic constraints of current and voltage with several electrochemical constraints in their SOP estimation, involving lithium surface concentration, electrolyte concentration, and side-reaction overpotential. Both simulation and experimental results demonstrated the efficacy of this aggregation in enhancing battery safety management at both macro scale and micro scale. Taking into account the decreasing conductivity of electrode and electrolyte with battery degradation, Gu et al. [62] monitored the SEI growth and lithium plating using the Tafel equation. Besides the electrochemical variables, they also introduced energy conversion efficiency into battery SOP estimation as an operational constraint, allowing for an extensive supervision of battery health and safety throughout the entire lifespan.

## 2.4. Discussion

This section reviews the critical components and recent advancements in SOA design, categorizing existing operational constraints into three groups from macro scale to micro scale. Table 1 summarizes the pros and cons of these constraints in the context of battery SOP estimation. Among these, the basic constraints of voltage and current play the fundamental role in battery safety protection. While these electrical parameters are easily measurable through sensors and offer high accessibility, they fall short in revealing the intrinsic operation status of batteries. This limitation can lead to potential risks such as energy depletion and thermal abuse when batteries are functioning under the boundary condition during a prediction window. In contrast to voltage and current, the internal states of batteries allow for a deeper understanding of intrinsic operation status, the use of these states as the operational constraints contributes to improved safety protection. These non-measurable internal states have to be estimated from externally measurable signals, requiring specialized models and algorithms. Such tasks are generally achievable by on-board BMSs, making it an applicable solution for batteries of various types and chemistries. However, the internal states like SOC and SOE primarily manifest the static characteristics of batteries and do not characterize the electrochemical reactions that influence battery transient dynamics. Therefore, the state constraints may not pre-emptively address potential dangers at their root. This defects highlight the significance of introducing the operational constraints on electrochemical variables which offer high interpretability to kinetic processes inside batteries while signifying the occurrence of side reactions. This capability is crucial for enhancing battery safety protection and health prognosis before the manifestation of safety hazards. Nonetheless, the estimations of these electrochemical variables demand complex, specially designed models and algorithms, resulting in high computational burdens for on-board BMSs. In addition, the scalability of electrochemical constraints poses issues, as it is dependent on the unique design of batteries for different types or chemistries. Moreover, it is also challenging to prove the effectiveness of these electrochemical constraints in battery SOP estimation due to the difficulty in accessing their true values and thus the unavailability of reference SOP.

**Table 1**

A summary of pros and cons of different types of operational constraints in battery SOP estimation.

| Constraint types | Pros | Cons |
|---|---|---|
| Basic constraint | <ul><li>High accessibility</li><li>Ease of implementation</li></ul> | <ul><li>Low interpretability to battery dynamics</li><li>Limited capability in battery safety protection</li></ul> |
| State constraint | <ul><li>Relatively high accessibility</li><li>Ease of implementation</li><li>Enhanced capability in battery safety protection</li></ul> | <ul><li>Limited interpretability to battery kinetic contributions</li><li>Need specially designed models and algorithm for state estimation</li></ul> |
| Electrochemical constraint | <ul><li>High interpretability to battery kinetic contributions</li><li>Superior capability in battery safety protection</li><li>Prognosis for battery safety and health</li></ul> | <ul><li>Low accessibility</li><li>Need specially designed models and algorithm for electrochemical variable estimation</li><li>Challenge in validation for effectiveness</li></ul> |

## 3. Peak operation mode

While the definition of SOP is straightforward, it does not specify the way to discharge or charge batteries during a prediction window. Due to the inherent battery characteristic, it is possible to keep only one parameter—either current, voltage or power—constant to maximize the peak power capability [23,63]. Clearly, this nature can induce varied electrical behaviors of batteries functioning under different POMs, leading to diverse peak power performances and SOP estimation outcomes. Typically, constant current (CC), constant voltage (CV), CC-CV and constant power (CP) refer to four commonly used POMs [64]. Each mode, with its unique discharge and charge characteristics, is chosen to derive respective SOP estimation technique in the literature. In this section, we consider a SOP prediction window from time $k+1$ to $k+K$ (indicating a window length of $K$ time steps) and present a thorough exploration of these POMs, helping understand their advantages and disadvantages in regions governed by the basic constraints of current and voltage.

### 3.1. Constant current

CC-POM attains the peak power performance of batteries by maximizing a constant discharge/charge current over a prediction window [65]. It is usually favored for its ease of implementation and convenience in algorithm development, making it the most extensively researched POM for battery SOP estimation [21,66–68]. Fig. 4 (a) illustrates the typical evolutions of current, voltage and power when batteries are discharging under the CC-POM. In the high SOC region, current serves as the dominant factor limiting the peak power capability of batteries, where the peak discharge current is held at the maximum discharge current (i.e., current constraint for discharge), and the terminal voltage continues to decline throughout the window, yet it does not reach the lower cut-off threshold. With battery SOC decreases, the terminal voltage at the end of the window may touch the lower cut-off threshold at a moment, and thereafter the peak discharge current starts to drop, indicating a shift of the dominant limitation factor from current to voltage [44]. For the case that batteries are charging under the CC-POM, the peak charge current also keeps at the maximum charge constant (i.e., current constraint for charge), while the evolutions of voltage and power demonstrate the opposite trends to the discharge case in Fig. 4 (a), and their magnitudes continue to grow throughout the window [2].

Hence, the boundary conditions of CC-POM occur either throughout the prediction window from time $k+1$ to $k+K$ or at the end of the window, namely at the time of $k+K$. For the former case, the peak discharge/charge current equals to the associated current constraint from time $k+1$ to $k+K$. For the latter case, the terminal voltage reaches the cut-off threshold only at time $k+K$.

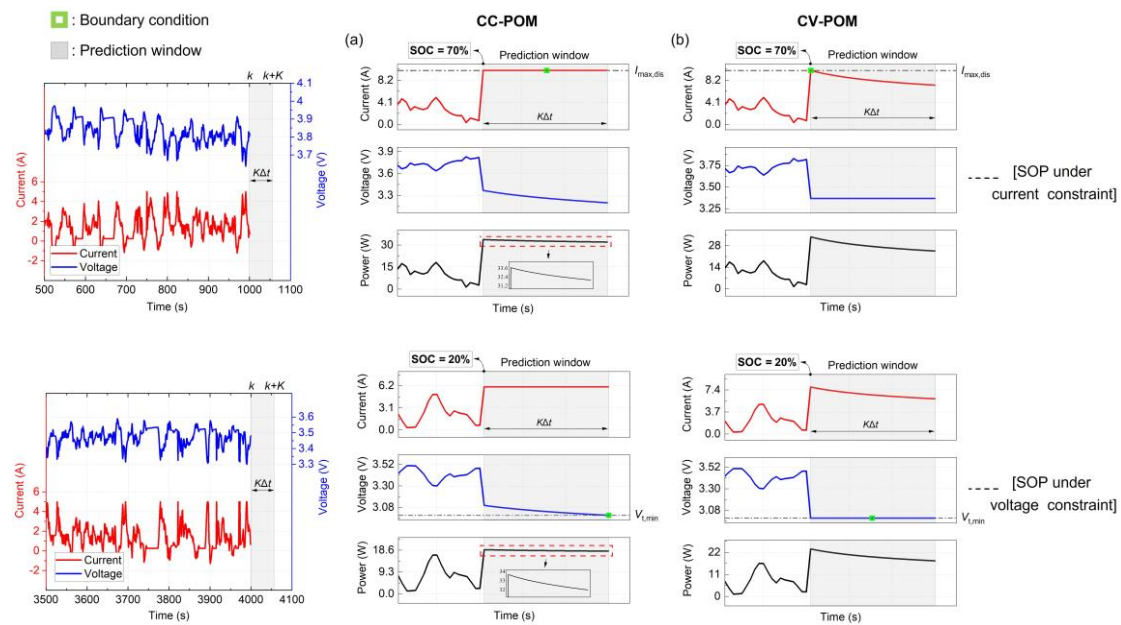

Fig. 4. Battery peak discharge behaviors under: (a) CC-POM; (b) CV-POM.

## 3.2. Constant voltage

In contrast to CC-POM, CV-POM has received far less attention in the literature [2]. Wang et al. [69] explored this area, examining its practical implementation. In their study, the peak power performance of batteries under the CV-POM was achieved by maximizing a variable discharge/charge current under a constant terminal voltage over the prediction window. Fig. 4 (b) illustrates the typical evolutions of current, voltage and power when batteries are discharging under the CV-POM. In the high SOC region, current serves as the dominant factor limiting the peak power capability of batteries, where the peak discharge current always reaches the maximum discharge current (i.e., current constraint for discharge) at the beginning of the prediction window, and thereafter it continues to decline in order to maintain constant terminal voltage concerning the growing polarization overpotential inside batteries. This constant terminal voltage will gradually decrease with SOC, and there will be a moment that it reaches the lower cut-off threshold. In the following, the peak discharge current at the beginning of the window will no longer reach the maximum discharge current, indicating a shift of the dominant limitation factor from current to voltage. For the case that batteries are charging under the CV-POM, the terminal voltage also keeps constant, while the evolutions of voltage and power demonstrate the same trends to the discharge case in Fig. 4 (b), and their magnitudes continue to decline throughout the window

Hence, the boundary conditions of CV-POM occur either at the beginning of the window, namely at the time of $k+1$, or throughout the window from time $k+1$ to $k+K$. For the former case, the peak discharge/charge current can reach the maximum current (i.e., current constraint) at time *k*+1. For the latter case, the terminal voltage is maintained at the cut-off threshold from time *k*+1 to *k*+*K*.

### 3.3. Constant current-constant voltage

Pei et al. [70] proposed a CC-CV-POM that combines the attributes of both CC-POM and CV-POM and implements these two POMs in the high and low SOC regions, with a transitional CC-CV phase in-between. As an advantage, this mode forces batteries to discharge or charge under either current constraint or voltage constraint at any instant throughout a prediction window, thereby exploiting the remaining capacity to the greatest extent [2]. Following this line of thought, the researchers in [71–73] have made efforts to convert the original SOP estimation problem into an optimization problem, solved by model predictive control (MPC) theory. The contributions of these works will be detailed in the later part of this paper.

Fig. 5 illustrates the typical evolutions of current, voltage and power when batteries are discharging under the CC-CV-POM. As can be seen, besides a transitional CC-CV phase in Fig. 5 (b), the electrical behaviors of batteries in Fig. 5 (a) and (c) are completely the same as those under the CC-POM and CV-POM, respectively. Assuming the mode shift from CC-POM to CV-POM takes place at time $k+K_c$, we have to identify this moment precisely which is crucial for battery SOP estimation under the CC-CV-POM. In this context, three possible cases may arise:

1) Case 1: $K_c > K$ (see Fig. 5 (a)).

It implies that the mode shift will occur beyond the prediction window of concern. In such a case, current serves as the sole factor limiting the peak power capability of batteries, which can be referred to battery SOP estimation under the CC-POM.

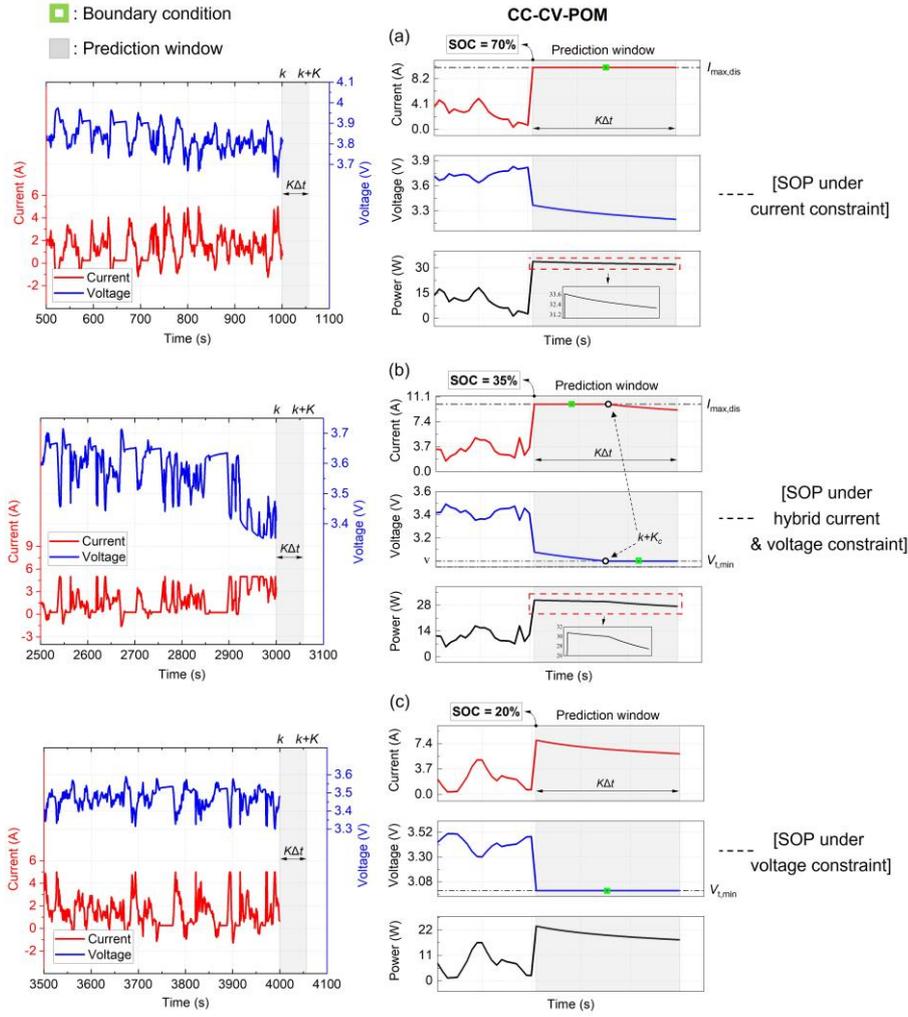

Fig. 5. Battery peak discharge behaviors under the CC-CV-POM.

2) Case 2: $0 \leq K_c \leq K$ (see Fig. 5 (b)).

It implies the occurrence of a transitional CC-CV phase and a mode shift within the prediction window. The moment of the mode shift marks the instance when the voltage reaches the cut-off threshold for batteries operating under the current constraint from the beginning of the prediction window. In such a case, SOP estimation can be referred to the one under the CC-POM from time $k$ to $k+K_c$ and the one under the CV-POM from time $k+K_c$ to $k+K$.

3) Case 3: $K_c < 0$ (see Fig. 5 (c)).

It implies that the mode shift has already completed at an earlier time prior to the prediction window. In such a case, voltage serves as the sole factor limiting the peak power capability of batteries, which can be referred to battery SOP estimation under the CV-POM.

### 3.4. Constant power

While the aforementioned three POMs are intuitively simple with batteries operating under either CC or CV, they result in a time-varying peak power delivery to or absorption from the EV powertrain throughout a prediction window [64]. This variability becomes even more pronounced in low or high SOC regions due to rapid changes in battery terminal voltage or peak current. On the contrary, CP-POM is more representative of actual battery loadings in EV applications (e.g., EV acceleration, gradient climbing and regenerative braking) than CC or CV-POM. Rahman et al. [74] highlighted that CP-POM could either maximize vehicle acceleration for a given power rating or minimize the power rating required for a specific level of vehicle acceleration. They further demonstrated that EV powertrains operating at a CP-POM could achieve optimal initial acceleration and gradeability with the minimum necessary power rating [75]. On this basis, Anun et al. [76] and Cao et al. [77] applied different control methods to address CP load instability in EV power systems. Besides, battery power is usually viewed as a direct variable in EVs rather than current or voltage in velocity/cruise control for the pursuit of co-optimization of vehicle speed and powertrain energy management [78,79]. Considering this, battery SOP estimation under the CP-POM is practically important for on-board BMSs to assess the peak power performance of batteries and determine the feasibility of specific driving scenarios, such as acceleration, gradient climbing, and regenerative braking.

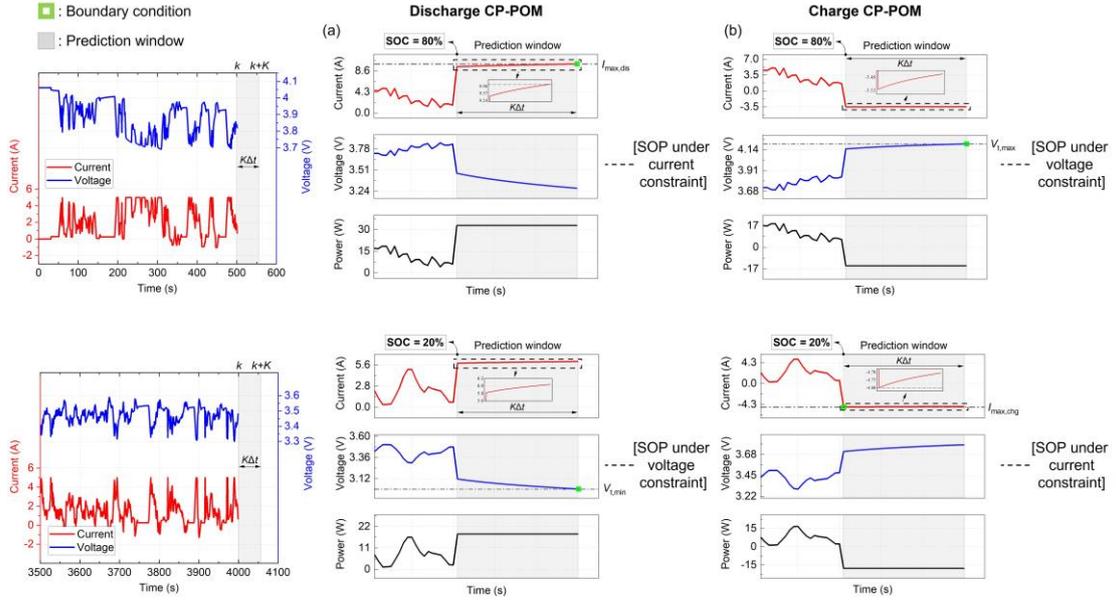

Fig. 6. Battery peak behaviors under the CP-POM: (a) discharge case; and (b) charge case.

In a pioneering effort, Guo et al. [64] introduced a model-switching approach for battery SOP estimation under the CP-POM. This innovative method straightforwardly maximizes the discharge/charge power of batteries as the product of peak current and terminal voltage and maintains it throughout the prediction window. It is obvious that a given CP corresponds to two unique sequences of peak current and terminal voltage within the window. However, the boundary condition occurs differently for battery discharging and charging under the CP-POM, as illustrated in Fig. 6 (a) and (b), respectively. As can be seen, the peak discharge current and terminal voltage grow and decline monotonically for the discharge case, but their magnitudes show the opposite trends for the charge case. In view of this, the boundary conditions for the CP-POM appear at the end of the prediction window for both discharge and charge SOP estimation when voltage is the dominant limitation factor. Conversely, if current is the dominant limitation factor, the boundary condition appears at the end of the window for discharge SOP estimation but shifts to the beginning of the window for charge SOP estimation. In response, they proposed a UKF based correction strategy to pinpoint the dominant factor limiting the peak power performance of batteries, thus successfully addressing the challenge of formulating battery SOP in an analytic form under the CP-POM.

**Table 2**

A summary of different boundary conditions for battery SOP estimation under each POM.

| CC-POM | Boundary condition | |
|---|---|---|
| | Peak current | Terminal voltage |
| Current constraint | $k+1 \sim k+K$ | - |
| Voltage constraint | - | $k+K$ |
| **CV-POM** | **Boundary condition** | |
| | Peak current | Terminal voltage |
| Current constraint | $k+1$ | - |
| Voltage constraint | - | $k+1 \sim k+K$ |
| **CC-CV-POM** | **Boundary condition** | |
| | Peak current | Terminal voltage |
| Current constraint | $k+1 \sim k+K$ | - |
| Dual constraint | $k+1 \sim k+K_c$ | $k+K_c \sim k+K$ |
| Voltage constraint | - | $k+1 \sim k+K$ |
| **CP-POM** | **Boundary condition** | |
| | Peak current | Terminal voltage |
| Current constraint | $k+K$ (dis)/$k+1$ (chg) | - |
| Voltage constraint | - | $k+K$ |

## 3.5. Discussion

Based on the above analysis, we summarize the boundary conditions for each POM in Table 2. Although battery SOP can be achieved by these four POMs, their impact on the peak power performance of batteries and associated interrelationship remain unclear. To address this, Guo et al. [80] conducted a comparative analysis and proposed four key indices (i.e., maximum and minimum instant magnitudes, time-averaged magnitude and falling/rising rate) for evaluating current, voltage, and power performance of batteries under each POM. In their work, a strong correlation was recognized between time-averaged power and current. This suggested that higher Ah-throughput over a prediction window leads to increased energy delivery or absorption. However, such a correlation was not evident between time-averaged power and voltage, indicating a weak dependency of energy throughput on terminal voltage. Additionally, the rankings of some key indices among the four POMs offered a clear understanding of each POM with their distinct characteristics.

- The CC-POM excels in current and power performance in the region governed by current, topping the rankings in time-averaged current and power as well as maximum instant current and power. However, when the dominant operational constraint turns to voltage, there is a swift decline in peak discharge/charge current, which impairs the peak power capability over a prediction window. Notably, the CC-POM maintains the highest one among the minimum instant currents across all the regions.

- The CV-POM yields subpar current and power performances in the region governed by current, but it ascends to the highest rankings in time-averaged current and power when the dominant operational constraint turns to voltage. Despite the CV-POM holds the highest one among the minimum instant voltages across all the regions, it tops the rankings in falling/rising rate of current and power, indicating the worst discharge/charge stability among the four POMs.

- The CC-CV-POM holds the advantages of both CC-POM and CV-POM and proves to have the optimal discharge/charge capability across all the regions as it achieves the highest rankings in time-averaged current and power as well as maximum instant current and power. This superior performance is realized by consistently discharging or charging the battery under the operational constraint of either current or voltage throughout the prediction window, thereby maximizing the utilization of the remaining capacity. On the other hand, such an aggressive strategy causes a high ranking in falling/rising rate of power, reflecting a poor discharge/charge stability of the CC-CV-POM.

- Among the four POMs, the CP-POM exhibits relatively compromised performances in time-averaged current and power across all regions. However, it stands out with the lowest falling/rising rate of power and highest minimum instant power, indicating the capability to supply persistent and stable power. In addition, unlike the other three POMs, the CP-POM only reaches the boundary condition at either beginning or end of a prediction window. Although this attribute limits the current and voltage performances within a prediction window, it contributes to the reduced risk of exceeding the design limit when batteries are discharging/charging at the peak power, thereby enhancing battery operational safety.

## 4. Battery modelling

Accurate battery modelling plays a pivotal role in battery SOP estimation as it crucially characterizes battery behaviors under the boundary condition. From a systematic point of view, current battery models for SOP estimation fall into three groups, in terms of their degrees of physical interpretation of battery dynamics. These range from the most to the least interpretable: white-box models, grey-box models, and black-box models.

### 4.1. White-box model

In the field of battery management technology, white-box models encompass a range of electrochemical models (EMs). These models are meticulously crafted to explicitly describe a set of electrochemical reactions occurring inside batteries, allowing for the mathematical expressions of spatial and temporal distributions of electrochemical variables, such as lithium concentrations and electrode and electrolyte potentials [81].

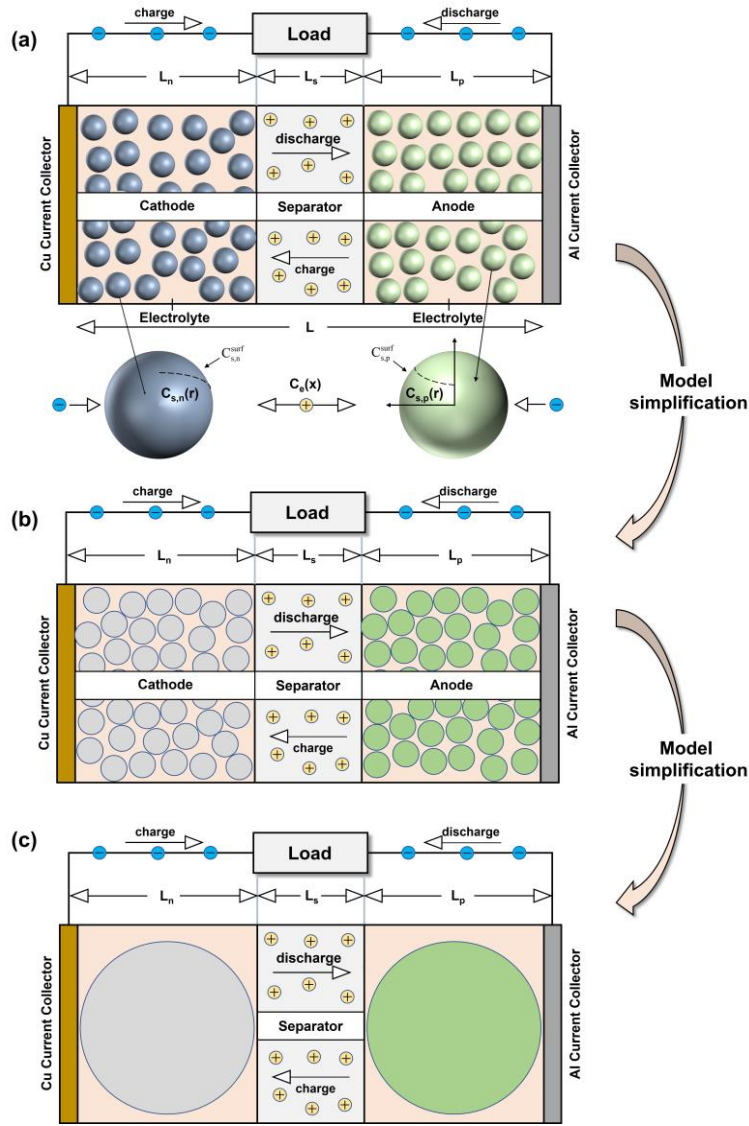

Fig. 7. Simplification from a rigorous two-dimensional model (a) to P2D model (b) and SPM (c).

Doyle–Fuller–Newman model, commonly known as the pseudo-two-dimensional (P2D) model, stands as a typical EM, which has served as a foundation for numerous applications. Initially proposed in [82,83], the P2D model is grounded on the theories of porous electrodes and concentrated solutions, where the microstructure of electrode particles is conceptualized as having a one-dimensional spherical geometry, while the electrodes and separator are modelled with a one-dimensional planar geometry. These dual assumptions facilitate the free movement of lithium ions across three domains in the liquid phase and allow for diffusion along the radial dimension in the solid phase. Although P2D model is a homogenized simplification derived from a rigorous two-dimensional model (see Fig. 7), it still comprises five inter-coupled nonlinear PDEs [84]. These equations capture the dynamics of concentration and over-potential in both solid and liquid phases, as well as simulate the intercalation kinetics at the solid-liquid interface. The intricate mathematical formulations of P2D model require substantial computational resources for accurate numerical solutions. To facilitate engineering practices, single particle model (SPM) family emerges from a model-order reduction applied to P2D model, which achieved simplified model structures with less parameters, making them possible to meet the requirements of on-board BMSs in EV applications.

In [57], a P2D model was simplified into a control-oriented SPM with a set of ordinary differential equations. This simplification was achieved through the finite difference method, which divided the sphere radius into several nodes with uneven discretization. The linear state-space representation of this SPM well suited for use with a filter or observer to estimate lithium ion concentration and lithium intercalation dynamics at an electrode level. The information obtained from these electrochemical states was then utilized in battery instantaneous SOP estimation. Another SPM was developed in [58] with each solid particle being meshed into 40 nodes for trading-off model accuracy and complexity. A theoretical analysis of Gibbs power and dissipation power on battery internal resistance was presented to access a quantitative relationship between lithium surface concentration and instantaneous SOP. Smith et al. [67] modified a P2D model by transforming the infinite-dimensional impedance into a finite-dimensional linear state-space representation. This approach, building on their earlier work in [85], significantly expedited the computations of peak discharge/charge current and thus battery SOP. However, this research did not account for the electrolyte dynamics across various domains, which affected the performance of this SPM in battery SOP estimation at high current rates. To address this, Perez et al. [50] and Li et al. [61] developed an extended SPM (eSPM), considering the inhomogeneous distribution of electrolyte concentration in the liquid phase. In their research, the Padé approximation and central difference method were leveraged to transform the coupled PDEs of P2D model into finite-dimensional differential-algebraic equations, enabling the descriptions of electrolyte dynamics with reduced computational complexity. In spite of this, the state-space representation of this eSPM was not entirely linear. Consequently, numerical methods were required to solve the peak discharge/charge current and SOP of a battery. In [62], an improved health-conscious eSPM was introduced for battery SOP estimation, which took into account the non-uniformities of reaction rates, current density, potentials, and lithium ion concentrations in both solid and liquid phases. Alongside this, a lumped thermal model was developed to capture the temperature dependency on model parameters, and the Tafel equation was incorporated to monitor the growth of SEI, thereby realizing enhanced model fidelity and health supervision. This approach was proven effective in accurately estimating the peak power characteristics of batteries, which in turn aided in reducing energy conversion losses, prolonging battery lifespan and increasing economic returns within an integrated energy-transportation system.

### 4.2. Grey box model

Compared to white-box models, grey-box models offer useful but ambiguous physical insights with a variable extent of details. Widely studied equivalent circuit models (ECMs) belong to this kind of models. They reproduce battery external dynamics through electrical components such as resistor, capacitor and voltage source. Various combinations of electrical components result in different ECMs with diverse structures and electrical characteristics. This underscores the importance of choosing suitable components to accurately depict specific aspects of battery electrochemical dynamics like charge transfer process and double-layer effect. Relevant reviews on ECMs for battery voltage prediction and state estimation can be referred to [86,87].

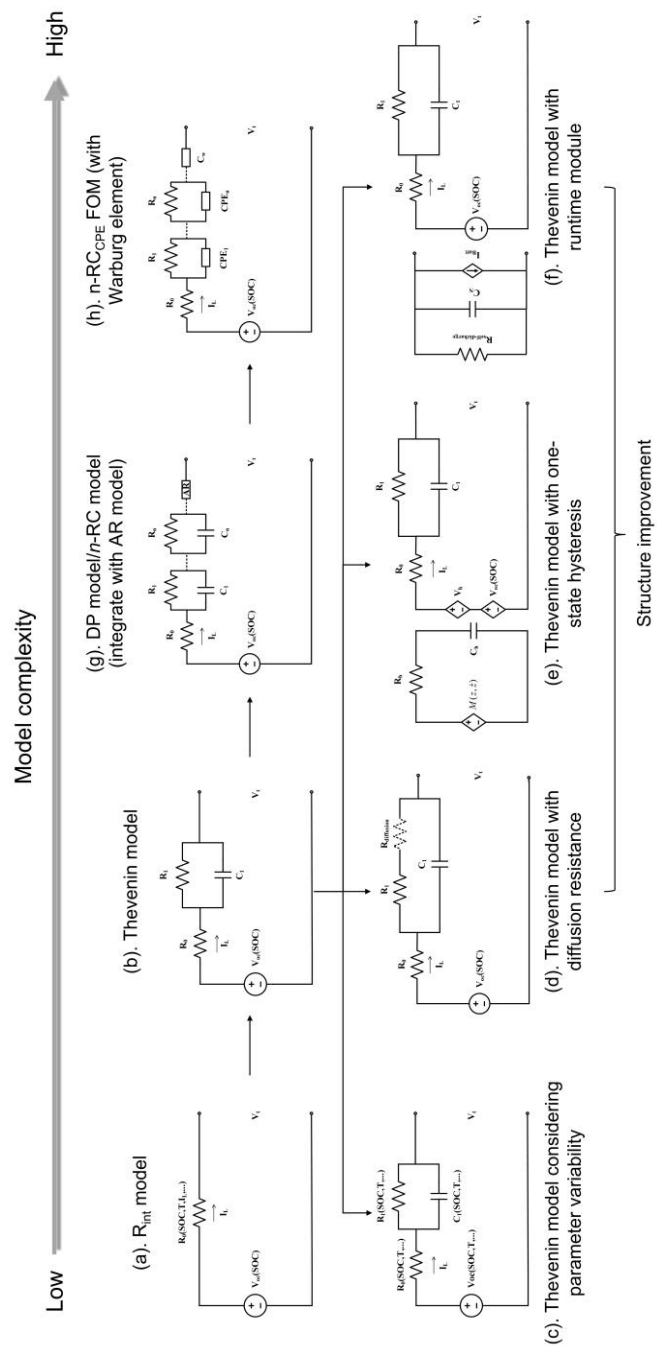

Fig. 8. Developments of available ECMs in battery SOP estimation.

Among the myriad of ECMs documented in the literature, $R_{int}$ model, as illustrated in Fig. 8 (a), stands out as the simplest. It attributes the entire voltage response of batteries to a single lumped resistance. This straightforward topology facilitates the easy calculation of instantaneous SOP. Plett [88] conducted a study focusing on the estimation of instantaneous SOP under the multi-constraints of current, voltage and SOC, based on an experimentally parameterized $R_{int}$ model. In [89,90], the authors characterized the lumped resistance of a $R_{int}$ model over wide ranges of SOC and temperature to effectively capture the multi-faceted dependencies of battery peak power. In [91], a fuzzy $R_{int}$ model was developed on a set of well-designed fuzzy rules to realize accurate nonlinear mappings of current and SOC-dependent lumped resistance. This approach demonstrated improved adaptability, which was proven effective for estimating instantaneous SOP under relatively complex load conditions.

He et al. [92] highlighted a notable limitation of $R_{int}$ model with its inability to capture the transient response of batteries. In their comparative study, which assessed the performance of both $R_{int}$ and Thevenin models in dynamic voltage simulations, a maximum percentage error of 2.82% was reported for $R_{int}$ model, as opposed to only 0.52% for Thevenin model. As a mainstream type of ECM, the Thevenin model, as depicted in Fig. 8 (b), integrates an additional parallel-connected resistor-capacitor (RC) pair into a $R_{int}$ model. This enhancement significantly improves the model capability to mimicg the polarization dynamics of batteries, enabling accurate tracking of nonlinear voltage responses under dynamic load conditions. Therefore, the Thevenin model has found extensive applications in various domains, including state estimation, cell balancing, and fast charging [93]. Early research, as seen in [69,94], utilized a Thevenin model for battery SOP estimation and yielded satisfactory results, particularly for fresh cells at room temperature. In [95], the parameter sensitivity of the Thevenin model was investigated based on a Cramer-Rao bound analysis, leading to the design of an optimally suited current excitation sequence for precise parameter identification and contributing to the improved performance of battery SOP estimation.

However, generic Thevenin model usually faces two notable limitations that affect its practical efficacy: (1) Structural imperfections: it falls shorts in reproducing some specific reactions and nonlinearities of batteries; and (2). Parameters variability: its parameters are influenced by various intrinsic and extrinsic factors. Overlooking these limitations can compromise model accuracy under variable internal states and changeable operation conditions of batteries. Several enhanced Thevenin models have been introduced to improve the SOP estimation accuracy. These adaptations are achieved either by incorporating specially designed components, such as diffusion resistance [67], one-state hysteresis term [42,96–98], self-discharge and runtime module [99] (see Fig. 8 (d-f)), or by characterizing the multi-faceted dependencies of model parameters, such as SOC [44,100], SOH [101], temperature [102], and current [65,103,104] (see Fig. 8 (c)).

Many researchers [105,106] have suggested that incorporating an additional RC pair in series with the existing one in a generic Thevenin model can more effectively account for charge transfer process and diffusion phenomena of batteries across both fast and relatively slow dynamics. This leads to an establishment of what is so-called dual polarization (DP) model, as depicted in Fig. 8 (g) with $n$=2. Benefiting from two RC pairs with distinct time constants, DP model outperforms Thevenin model in capturing the significant diffusion overpotential inside batteries under persistent and intense current excitations, particularly in long-term SOP estimation [25,107–110]. Based on a DP model, a joint SOC-SOP estimation method was presented in [111], which demonstrated a SOC error of less than 2% over the entire battery operation range. This indicated an improvement compared to 3% error when taking a Thevenin model as the baseline. However, the superiority of the DP model over the Thevenin model for battery SOP estimation was not experimentally validated in this study. Shen et al. [108] developed a hybrid-parameterized DP model for multi-state co-estimation of SOC, SOH and instantaneous SOP. This model facilitated online adaptations of OCV and ohmic resistance using a forgetting-factor recursive least squares (FF-RLS) algorithm, and additionally, it characterized the other parameters offline into three-dimensional surfaces with respect to SOC and temperature. The consideration of both time-varying characteristics and multi-faceted dependencies of model parameters was instrumental in enhancing the co-estimation accuracy of all three states. Another hybrid parameterization strategy for DP model was proposed in [109]. This approach utilized a dual extended Kalman filter (DEKF) to simultaneously conduct online parameter identification and SOC estimation when batteries were under loads, otherwise, a pseudo random binary sequence would be injected to excite the unloaded batteries for parameter recalibration. As an advantage, it guaranteed the availability of reliable prior knowledge for DEKF, thereby enhancing the robustness of battery SOP estimation, particularly in scenarios with unknown initial conditions. In [25,110], a electrothermal model was proposed for joint SOC-SOP estimation by coupling a DP model with a thermal model to reproduce battery dynamics from both electrical and thermal perspectives. Accurate estimations of electrical states such as SOC and polarization voltages captured heat generation inside batteries and thus contributed to the estimations of thermal states, involving volume-average temperature and core temperature. This, in turn, helped the parameter recalibration of DP model at varying temperatures.

Due to a lack of comprehensive interpretations to battery dynamics, researchers have complemented ECMs with an auto-regression (AR) model to offset their model residuals. In [112], a hybrid structure, combining a $R_{int}$ model with an AR model, was employed for battery SOP estimation, reducing associated reliance on detailed battery knowledge and extensive parameter identification. This approach simulated the polarization state of batteries as a linear combination of its historical states within a past rolling window, with the regression coefficients being adapted online using a FF-RLS algorithm in a data-driven manner. However, limitations exist as neither $R_{int}$ model nor AR model can capture the battery dynamics under an unseen load condition, leading to poor performance in battery SOP estimation over a lengthy prediction window. As an enhancement, Feng et al. [113] combined a Thevenin model with an auto-regressive moving average (ARMA) model for joint SOC-SOP estimation. This integration not only ensured a certain physical interpretation of battery dynamics but also improved voltage fitting performance in a noise-driven manner. Further study in [114] developed a hybrid battery model, comprising an *n*-RC model in series with an ARMA model. The optimal order of the *n*-RC model was identified based on the Akaike information criterion, effectively balancing model accuracy and complexity. A differentiated strategy was also derived for timely updating the regression coefficients in terms of their profile-dependent characteristics. The experimental results unveiled that a hybrid structure of 3-RC model/4-RC model and ARMA model achieved the best performance in joint SOC-SOP estimation in a noise-free/noise-corrupted working environment.

Owing to the limitation of the previously mentioned integer-order ECMs (IO-ECMs) with finite RC pairs in characterizing the frequency domain impedances of batteries, various fractional-order ECMs (FO-ECMs), incorporating constant phase elements (CPEs) and Warburg elements, have been introduced for battery SOP estimation, as depicted in Fig. 8 (h). The superiority of FO-ECMs over IO-ECMs has been recognized in [115–117]. This superiority lies not only in their ability to interpret the nonlinear characteristics of batteries over a broad frequency domain but also in providing explicit physical insights into the mechanisms of battery degradation. For instance, De Sutter et al. [118] highlighted the prominence of FO-ECMs in accurately capturing the intrinsic fractional differentiation properties of batteries in the middle and low frequency regions, primarily including diffusion dynamics, charge transfer and memory hysteresis. Lu et al. [119] disclosed that the fractional orders of FO-ECMs can be expressed as a function of recursive factors, defining the fractality of charge distribution on porous electrodes. Consequently, there is an intrinsic link between these fractional orders and electrode degradation, making them a valuable indicator of battery health.

As of now, FO-ECMs have found wide applications in the estimations for SOC [120–122], SOE [123] and SOH [115,124], while relevant studies on SOP estimation are still few [125]. In [126], a simplified FO-ECM was taken for rapid SOP estimation, where the ohmic loss and charge transfer process inside batteries were represented by a linear resistance coupled with a Warburg element, considering that the voltage responses of batteries under most EV driving profiles are concentrated in the low frequency region (below 1 Hz). To better interpret both fast and relatively slow dynamics of batteries, an enhanced FO-ECM, featuring two serial-connected resistor-CPE ($RC_{CPE}$) pairs, was introduced in [127]. Referring to the Grünwald-Letnikov definition and short-memory principle [128], the nonlinear dynamics of FO-ECMs was in relation to all of their historical states within a memory window. These FO-ECMs with complex mathematical formulations, however, presented challenges in expressing the peak discharge/charge current and SOP in an analytic form. A compromised solution was, therefore, leveraged by the authors, which linearized the model representation with a reduced memory window of only one time step, thereby significantly sacrificing model nonlinearity and affecting SOP estimation performance. Focusing on this issue, Guo and Shen [63] separately characterized the zero-state and zero-input transient dynamics of batteries through a 2-RC$_{CPE}$ FO-ECM and a Thevenin model, which were combined together in subsequent SOP estimation. The advantage of this model-fusion approach lies in its high computational efficiency while unlocking the need for model linearization. In their subsequent study [129], a fractional-order multi-model system, constituting three sub FO-ECMs, was developed to accommodate battery dynamics under complex operation conditions of load, SOC and temperature. A UKF-based correction strategy that compensated the change in battery peak power between two adjacent time steps was also developed for efficient SOP estimation without model linearization. In [130,131], comparative studies were carried out to assess the performance of IO-ECMs and FO-ECMs in battery SOP estimation. Farmann and Sauer [130] compared seven IO-ECMs and FO-ECMs with different numbers of RC and RC$_{CPE}$ pairs (up to three) in battery SOP estimation under various load profiles. Electrochemical impedance spectroscopy (EIS) method, ranging from 5 kHz to 1 mHz, was performed for model parameter identification, while HPPC test method was employed to evaluate model accuracy and SOP estimation performance. As reported, the FO-ECM with three RC$_{CPE}$ pairs achieved the highest accuracy, with a model error of around 20 m$\Omega$ in EIS fittings, however, it encountered convergence issues at non-room temperatures and high current rates, probably because of the complexity of identifying its numerous parameters. As a trade-off between model accuracy and complexity, the authors leaned towards favoring the Thevenin model for battery SOP estimation in EV applications, although it underperformed FO-ECMs in most comparison results. In a similar research presented in [131], the optimal ECM for joint

SOC-SOP estimation was examined experimentally. The findings from this research revealed several key insights: (1) An increase in the number of serial-connected RC or $RC_{CPE}$ pairs would be conducive to model accuracy; (2) FO-ECMs outperformed IO-ECMs with the same structure in SOC and SOP estimation over the whole battery operation range; and (3) The FO-ECM, incorporating two $RC_{CPE}$ pairs and one Warburg element, struck an effective balance between model accuracy and complexity. This configuration was identified as the optimal structure among all the models tested for joint SOC-SOP estimation, demonstrating its suitability and effectiveness in this application area.

### 4.3. Black-box model

Empirical models and neural network models (NNMs) are often categorized as black-box models in battery SOP estimation. From a physical standpoint, the parameters of these models do not provide insights into battery dynamics. However, on the positive side, these models can be fully parameterized in an automated manner, without requiring prior knowledge of battery mechanics.

#### 4.3.1. Empirical model

Empirical models are essentially obtained by using experiment-and-fit techniques and leveraging a variety of nonlinear functions (e.g., polynomial function and exponential function) to depict the peak power characteristics of batteries [132]. Several studies, such as [133,134], have endeavored to characterize the multi-faceted dependencies of SOP on a range of internal states and external parameters. For instance, Kim and Yang [135] considered factors like SOC, temperature, and aging level in an empirical model, capturing their relationships with battery instantaneous SOP through two fifth-order and one second-order polynomial models. In [136,137], a second-order polynomial model was employed to establish connections between battery SOP and various factors like voltage, SOC, temperature, and length of a prediction window. Additionally, a self-updating strategy was introduced to recalibrate the polynomial coefficients, thereby enhancing the estimation robustness.

**4.3.2. Neural network model**

Given the exceptional capability of machine learning in modelling battery nonlinearities, several NNMs have been explored for battery SOP estimation to achieve enhanced accuracy while maintaining a reasonable computational cost. In [138], an adaptive neuro-fuzzy inference system (ANFIS) was built to simulate battery dynamics for battery SOP estimation. This model incorporated current, accumulated charge, SOC, temperature, and time-averaged voltage during a pulse as input variables, with the output being the predicted terminal voltage at the end of the pulse. Using the ANFIS, the peak discharge/charge current and SOP of a battery can be estimated through iteratively running the model until they gradually converge towards the true values. In [139], an extreme learning machine was constructed for battery SOP estimation, innovatively substituting traditional activation functions in each neuron with a set of sub-models. Each of these sub-models comprised a Thevenin model and a thermal model with the initial SOC and model parameters being randomly selected within a reasonable range. By doing so, this approach was capable to reproduce the electrical and electrothermal dynamics of batteries with minimal requirements for prior knowledge, thereby elevating the estimation robustness and adaptability. In [140], extensive experimental work was undertaken to investigate the distorted polarization dynamics of batteries under peak operation conditions. It involved conducting CC pulses with durations ranging from 30 s to 120 s at different high current rates (exceeding 5 C-rate). The results revealed that such pronounced battery nonlinearity exceeded the explanatory capacity of conventional ECMs. For this reason, a deep neural network (DNN) was developed to characterize the time-varying polarization dynamics, considering the dependencies on current and SOC. On the basis of this DNN, a data-model fusion approach was proposed to realize battery SOP estimation in high fidelity.

## 4.4. Discussion

This section reviews available battery models and their recent advancements for battery SOP estimation. Table 3 summarizes the pros and cons of different model types in the context of battery SOP estimation.

White-box models are recognized for their explicit physical interpretations to battery electrochemical reactions, such as charge transfer process and diffusion phenomenon, offering preferable model accuracy over other model types. Due to this nature, white-box models are highly compatible with all types of operational constraints discussed in Section 2, enabling advanced battery safety and health prognosis that pre-emptively addresses potential risks at their root. On the flip side, constructing white-box models for battery SOP estimation requires extensive knowledge of battery structure and properties, including particle dimension, chemical composition, material conductivity, etc. However, this level of details is often impractical for many system engineers. Besides, the unique design of batteries for different types or chemistries further complicates the acquisition of these physical parameters, leading to low identifiability. Often, specialized physical-electrochemical techniques are needed for parameter identification, which elevates the effort and cost in model construction and limits their scalability to other battery types or chemistries. Moreover, deploying white-box models in on-board BMSs can be challenging due to their complex model representations. This complexity also poses difficulties in algorithm development for battery SOP estimation, bringing another challenge to the practical applications of white-box models.

Grey-box models employ various combinations of electrical components to replicate the external characteristics of batteries. Their simple structures and limited parameters are highly compatible with the implementations of state-of-the-art filtering or observer techniques, enhancing model adaptability to varied battery dynamics under changeable operation conditions while offering scalable solutions suitable for batteries of diverse types and chemistries. Despite the provision of moderate physical interpretability and less explicit depictions of battery electrochemical reactions, grey-box models maintain promising model accuracy. This allows for the effective integrations of both basic and state constraints in battery SOP estimation. These attributes make grey-box models currently the optimal choice for EV applications. Nevertheless, it should be recognized that grey-box models are less effective than other model types in capturing highly nonlinear dynamics of batteries, especially under extreme current rates, SOCs, and temperatures. This limitation can be a significant consideration in certain application contexts where these extreme conditions are prevalent.

Different from the above-discussed two models, black-box models have less dependency on domain expertise while exhibiting promising performance in simulating strong nonlinearities of batteries. Nonetheless, several notable drawbacks limit their widespread in on-board BMSs. Firstly, these models are essentially data-driven, requiring a large amount of high-quality experimental data. In the field of battery SOP estimation, it is of needs to test batteries under the boundary conditions of each operational constraint so that the most pertinent data are collected to build the models in high fidelity, which is considerably time-consuming and labor-intensive. Secondly, transferring these models from one battery chemistry to another remains a challenging endeavor due to the need for substantial investment of time and resources again. Furthermore, a rigid one (or many)-to-one mapping relationship restricts the flexibility of black-box models, making them less adaptable for battery SOP estimation in the prediction windows of varying lengths.

**Table 3**

A summary of pros and cons of different types of battery models in SOP estimation.

| Model types | Pros | Cons |
|---|---|---|
| White-box models | • High physical interpretability<br>• High model accuracy<br>• High compatibility with all types of operational constraints<br>• Battery safety and health prognosis | • High complexity/low identifiability<br>• Need for high computational resources<br>• Poor model adaptability<br>• High requirement on domain expertise<br>• Difficulty in algorithm development |
| Grey-box models | • Moderate physical interpretability<br>• Simple structure/few parameters<br>• Promising model accuracy<br>• Ease of implementation<br>• High adaptability and scalability | • Inexplicit description of battery electrochemical reactions<br>• Limited capability in simulating strong battery nonlinearities<br>• Inability to incorporate electrochemical constraints |
| Black-box models | • High capability in simulating strong battery nonlinearities<br>• Ease of implementation<br>• No need for domain expertise | • Strong reliance on high-quality data<br>• Need substantial training effort<br>• Low adaptability and scalability<br>• Low compatibility with only basic constraints |

## 5. Algorithm development

This section provides a comprehensive overview, detailing state-of-the-art algorithms for battery SOP estimation. Generally, SOP estimation algorithms can be methodically categorized into four groups based on their distinct basic principles, offering a clear exposition of their technical contributions.

### 5.1. Characteristic mapping method

Characteristic Mapping (CM) methods establish a direct relationship between battery peak power and several interdependent factors such as SOC, voltage, and temperature. These algorithms utilize relevant factors as input variables, feeding them into a multi-dimensional map to generate battery SOP estimation through a look-up table interpolation, as illustrated in Fig. *9* (a) [23]. Although these algorithms are relatively straightforward, they demand extensive experimental works for comprehensive data collection to build these input-output relationships. This requirement poses a significant challenge: battery peak power with inherently multi-faceted dependencies must be formulated in high dimension, necessitating a substantial amount of memory capacity. Such extensive memory demands are often infeasible for on-board BMSs in EV applications, making the practical implementation of conventional CM methods in these contexts a challenging endeavor. As an alternative strategy, some CM methods approximate battery peak power characteristics by fitting extensive experimental data into nonlinear function-based empirical models, considerably reducing storage needs [136,137].

### 5.2. Open-loop prediction method

Open-loop prediction (OLP) methods stand as a group of the most widely used algorithm in battery SOP estimation [141]. These algorithms are combined with battery models of linear representations, such as Thevenin and DP models, to reproduce the dynamic behaviors of batteries over a given prediction window, enabling the analytical formulations of peak discharge/charge current and battery SOP, as illustrated in Fig. *9* (b) [142]. The evolution of OLP methods has diversified into two directions, each influenced by specific technical considerations or requirements of their envisioned applications.

The first direction of OLP methods is to incorporate battery SOP into a co-estimation scheme together with SOC and SOH. Hu et al. [143,144] implemented the co-estimation scheme of these three inter-coupled states in a multi-timescale framework, giving a boost to OLP methods. They recognized the importance of SOH as a dynamic factor that evolves over time, affecting the overall accuracy of both SOC and SOP estimation. Following this line of thought, Shu et al. [145] developed an enhanced accumulated ampere-hour method to update battery SOH in a co-estimation scheme, aiming for robust estimations of SOC and SOP against battery degradation. The authors observed that the OCV-SOC curve remains relatively stable above 62.5% SOC over the whole battery lifetime, with a voltage deviation of less than 5 mV. In light of this finding, they recommended using the accumulated ampere-hour method in a range from 62.5% to 100% SOC for SOH estimation. The results demonstrated that this approach achieved commendable accuracy in SOH estimation with a maximum error of less than 4.13%. This accuracy, in turn, reduced the SOC estimation error below 2% for batteries at various aging levels. However, the specific impact of this improvement on SOP estimation performance was not quantitatively analyzed in their study. In a different co-estimation scheme [146], the authors proposed updating battery SOH either quarterly or semi-annually by minimizing an objective function related to the accumulated charge between any of two separate SOCs within a specified range. Compared to the conventional accumulated ampere-hour method, this approach offers greater tolerance against errors and noises, arising from improper selection of SOC points. The precision of SOH estimation subsequently improved the accuracies of UKF for SOC estimation and OLP method for SOP estimation, showcasing the effectiveness of the co-estimation scheme. Rahimifard et al. [101] developed a interacting-multi-model method to assess the likelihood functions of each sub-model at associated aging levels, which enabled the adaptations of SOC, SOH and SOP in a model-fusion manner, enhancing the overall performance of the co-estimation scheme. In [147], a four-layer network architecture was developed for next-generation BMS, integrating cloud-edge-end interactions and digital twin (DT) technology. It addressed key functionalities, covering the co-estimation of SOC, SOH and SOP, as well as cell equalization. The simulation results highlighted the effectiveness of this architecture in leveraging big data for enhanced battery protection and management throughout the entire battery lifespan, demonstrating the potential of advanced data-driven and interconnected BMS architectures in optimizing battery performance and safety over time.

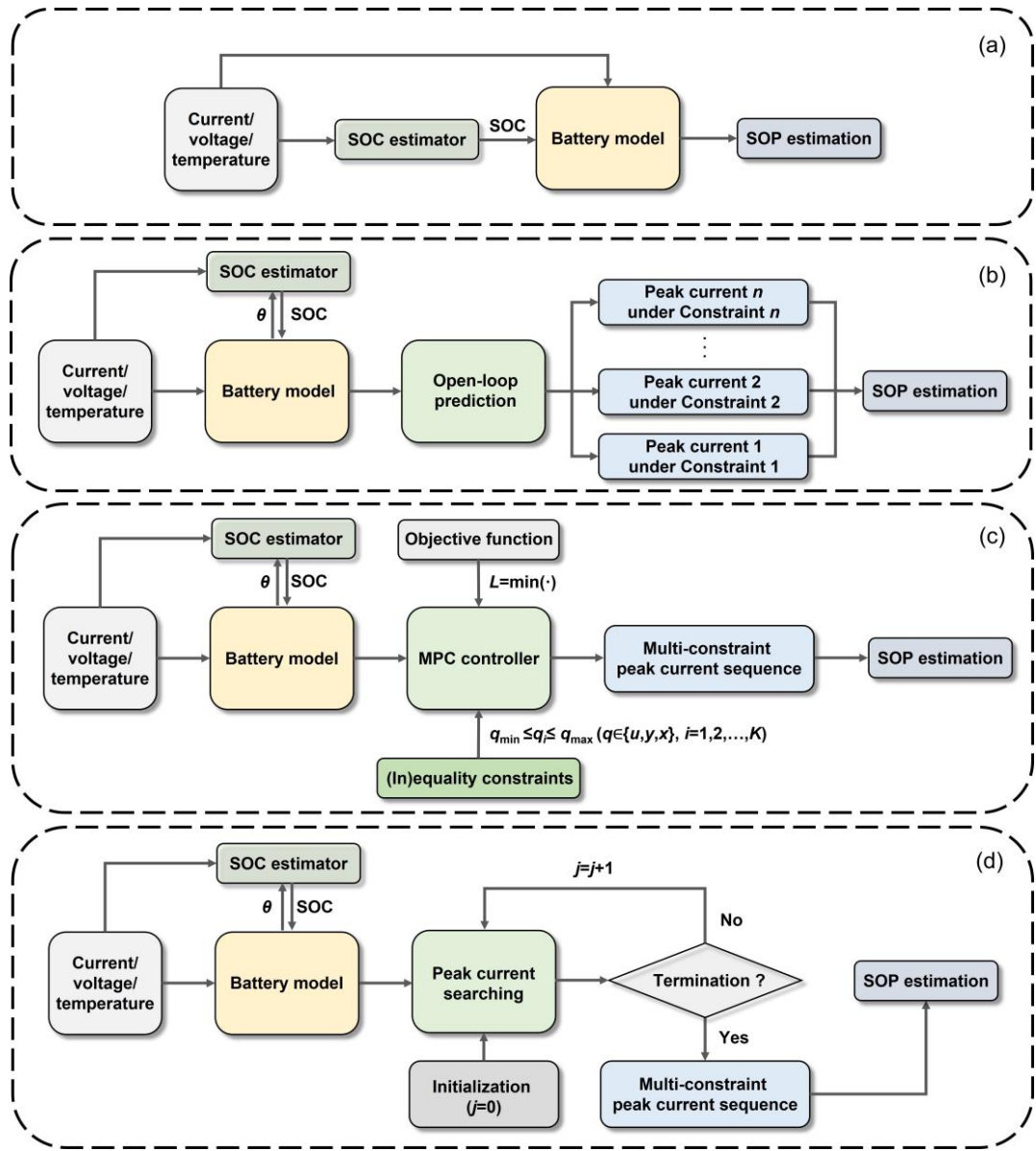

Fig. 9. Systematic diagram of state-of-the-art SOP estimation algorithms: (a) CM method; (b) OLP method; (c) MPC method; and (d) IA method.

Another direction of OLP methods is centered on battery SOP estimation at a pack level, addressing the challenge posed by inconsistencies among individual cells [148]. This focus indicates the critical need to precisely account for cell variations and disparities to enhance the SOP estimation reliability within battery packs. In EVs, battery packs often consist of hundreds or even thousands of individual cells to meet the required energy and power demands. The inhomogeneity in cell characteristics, arising from manufacturing variations and daily usage, significantly impacts the peak power performance at a pack level because the overall pack performance depends on the 'weakest cell' (or so-called representative cell) that first reaches either of the pre-set operational constraints [22]. In this regard, conventional idea [88], [66] that scales up battery SOP estimation from an individual cell to the entire battery pack can result in significant errors. This underscores the need for pack-level SOP estimation to take into account the operational constraints of all in-pack cells, instead of a simplistic scaling-up.

Generally, a battery pack is made up of two basic topologies—serial connection and parallel connection [149]. For serial-connected topology, cell characteristic comparison and "Mean+Difference model" are two effective strategy in the representative cell selection for pack-level SOP estimation [150]. In [96,151], the representative cell for pack-level SOP estimation was selected by comparing the estimated peak discharge/charge currents of all the individual cells. While this approach is straightforward, it necessitates a substantial memory capacity in on-board BMSs to establish cell-level models for storing their parameters. Additionally, the OLP method for peak current estimation at a cell level involves significant computational resources, making this approach less practical for real-world applications. In [152], a pack-level SOP estimation method was developed. This approach identified inherent cell characteristics, such as OCV and ohmic resistance, to select the representative cell. This preliminary step is crucial to enhance the computational efficiency of the OLP method for SOP estimation of a battery pack. To circumvent the intricate requirements of cell-level modelling and parameterization, a "Mean+Difference model" was introduced in [65,153]. This innovative model captured the overall electrical behaviors of the entire battery pack via a 'Mean model', while accounting for the distinct characteristics of each individual cell through a 'Difference model'. The selection of the representative cell was informed by various inhomogeneous factors identified from the 'Difference model'. This facilitated the pack-level SOP estimation by integrating the peak current estimation for the representative cell with the average cell voltage ascertained from the "Mean model".

In contrast to the serial-connected topology, the parallel-connected topology has received significantly less attention in area of battery SOP estimation. The primary challenge here stems from the heterogeneous distribution of branch currents and the resulted non-monotonic evolutions of cell voltages within a prediction window. Addressing this research gap, Han et al. [154] proposed a SOP estimation method for parallel-connected battery packs based on a generalized state-space system. This approach monitored the internal states and external behaviors of a representative cell, not only at the beginning and end of a prediction window but also throughout the intermediate period, to characterize the boundary condition of battery peak power operations.

### 5.3. Model predictive control method

Model predictive control (MPC), a widely utilized tool in the field of control engineering, excels in managing complex systems within specified boundaries. This capability enables MPC methods to simultaneously handle multi-variables, such as system states, inputs and outputs. From a MPC point of view, battery SOP estimation can be conceptualized as a finite-horizon optimization problem, which is solved by forecasting and adjusting battery behaviors dynamically to achieve the peak power performance within a prediction window, as illustrated in Fig. 9 (c).

Based on MPC theory, a dynamic matrix control method [155] and a receding-horizon control method [72] have been developed for battery SOP estimation. The key principle of these methods involved manipulating battery current change as a control action, which in turn regulated the system output (i.e., battery terminal voltage) as a linear function of future input changes (i.e., peak current). This approach facilitated the maximization of battery peak power over a prediction window by solving a nonlinear objective function. In [156], the SOP estimation for a serial-connected battery pack was achieved using a hybrid control paradigm. This approach combined a MPC method for estimation of battery peak discharge/charge current, along with a fuzzy control method for compensation of cell heterogeneity. In [25], an economic MPC method was applied for battery SOP estimation under extended multi-constraints, including current, voltage, SOC and SOT. Unlike conventional MPC method primarily designed for tracking purposes, this economic MPC method eliminated the need for laborious weight-tuning work and achieved enhanced closed-loop performance, particularly for nonlinear systems. In [157], an order-reduction MPC method was introduced. This approach simplified the original two-state control model into a more manageable one-state equivalent model, streamlining the SOP estimation process while maintaining effective control and prediction capabilities. In [71], the fractional-order calculus was integrated into the MPC theory, leading to the development of a factional-order MPC (FO-MPC) method that effectively adapted a FO-ECM in battery SOP estimation. The validation results indicated that this FO-MPC method optimized the peak discharge/charge current of batteries without impairing model nonlinearities. This approach demonstrated a significant advantage over ordinary MPC methods based on linear battery models, showcasing its effectiveness in handling the complexities of FO-ECMs.

### 5.4. Iterative approaching method

A multitude of factors, such as SOC, SOH and temperature, contribute to the complex electro-chemical-thermal characteristics of batteries. This makes battery models with time-invariant parameters fail to characterize such a complex system, which leads to an increased demand on battery modelling through either enhanced structures or time-variant parameters for guaranteeing SOP estimation performance over a prediction window. While advanced modelling techniques offer effectiveness in interpreting the significant nonlinearities inherent in battery behaviors, they also introduce challenges for battery SOP estimation. Due to the high complexity of these models, it becomes unavailable for conventional OLP methods to formulate an analytic form of peak discharge/charge current and SOP straightforwardly. Therefore, the iterative approaching (IA) method is required to iteratively run a battery model to gradually converge the estimations of peak discharge/charge current and SOP towards their true values (see Fig. 9 (d)).

Burgos-Mellado et al. [66] developed a fuzzy $R_{int}$ model to characterize the nonlinear dependencies of battery internal resistance on current and SOC. In their approach, the original problem of battery SOP estimation was transformed into a multi-inequality-constrained optimization problem. This problem was then addressed under the Karush–Kuhn–Tucker conditions and solved using a genetic algorithm-based IA method. In [103,158], the studies were conducted on the nonlinear charge transfer dynamics of batteries with regard to load currents. The results showed that the higher the load current, the lower the charge transfer resistance is. This phenomenon can be effectively simulated using the Butler-Volmer (BV) equation. Building on this understanding,

Waag et al. [65] introduced the BV equation into a Thevenin model and resorted a Newton-Raphson algorithm-based IA method to yielding the estimations of peak discharge/charge current and battery SOP. Although their approach demonstrated an advancement in battery modelling, particularly in response to varying load currents, for more precise SOP estimation, this type of IA methods is generally not recommended as calculating the gradient of a highly nonlinear function, especially the one incorporating the BV equation, can be quite computational complex. Tang et al. [102] proposed an adaptive method for joint SOC-SOP estimation on the basis of a so-called migrated Thevenin model. In this model, parameters were coupled with SOC and temperature through two particle filter-based linear transformations. The online adaptation of a total of ten migration coefficients markedly improved the model ability in interpreting battery nonlinearities under varying temperature and aging conditions. To provide the estimations of peak discharge/charge current and battery SOP, an IA method was derived to converge their initial guesses until a specific limitation factor—either current, voltage or SOC—reaches the associated operational constraint. In [36,129], a state-space system with an implicit representation was established to capture the intrinsic correlation between battery peak discharge/charge current and SOP. In this system, the pre-set operational constraints were treated as output references, and a UKF was employed as an adaptive algorithm for self-correction, ensuring the convergence of SOP estimation. In [61], the adoption of an eSPM presented a strong nonlinear representation. Hence, the authors proposed a hybrid approach for battery SOP estimation, combining a Gaussian process regression for prior knowledge providing along with a bisection algorithm-based IA method for peak discharge/charge current searching. This approach achieved desired performance in accurate and efficient SOP estimation, offering great opportunity in scenarios requiring rapid decision-making.

On the other hand, Chen et al. [159] facilitated the performance of SOP estimation by accounting for the SOC dependency of ohmic resistance in a Thevenin model, particularly focusing on its potential variation with notably changing SOC. However, this approach introduced complexity in predicting this ohmic resistance through the Taylor-series expansion. To address this challenge, they developed a case-dependent IA method to identify the peak discharge/charge current for battery SOP estimation, adapting to the specific circumstances of each case. In [160], the same authors engaged all model parameters in a forward prediction and employed a genetic algorithm-based IA method to work out the peak discharge/charge current from a highly nonlinear equation. This advancement realized a more precise characterization of battery behavior over a lengthy prediction window. Likewise, Hu and Xiong [161] built a linear-parameter-varying Thevenin model to track parameter variabilities over a prediction window. Levenberg–Marquardt algorithm and bisection algorithm were served as foundation to derive two IA methods for seeking numerical solutions of peak discharge/charge current. This approach was validated on batteries with different chemistries, and the results showed a preferable efficiency of the Levenberg–Marquardt algorithm-based IA method in battery SOP estimation than that of the bisection algorithm-based one. Malysz et al. [100] highlighted that the linear forward prediction strategy, as discussed in previous studies, could result in negative model parameters within a prediction window. In light of this, they introduced a root searching-based IA method integrated with a new forward prediction strategy for battery SOP estimation. In their approach, the model parameters were predicted forward along a slope that connects their current values to minimum values, thereby circumventing the generation of unrealistic parameter values. However, there is a possibility that this slope might deviate significantly from their original fitting curves. To address this concern, a regression-based IA method was proposed in [31], which not only guaranteed a reliable forward prediction of model parameters along their original fitting curves but also compensated for the OCV linearization errors over a prediction window.

### 5.5. Discussion

To help readers have a quick and thorough understanding of the above-mentioned SOP estimation algorithms, we summarize the model-algorithm compatibleness in Table 4 while analyzing the pros and cons of different algorithms in Table 5.

CM methods stand for a look-up table-based methods that map a range of input variables to battery SOP. Although these algorithms are straightforward and ease of implementation, they are only compatible with empirical models with the difficulties in extending to batteries of different types or chemistries, necessitating substantial offline efforts to conduct experiments and collect the data. With these concerns, CM methods are more prevalent in experimental investigations of battery peak power characteristics.

OLP methods are a representative of mainstream SOP estimation algorithms, enabling a case-by-case analysis for analytically expressing the peak discharge/charge current and SOP under different boundary conditions. As an advantage, these algorithms are ease of implementation, offer intuitive outcomes for battery SOP estimation under each operational constraint, and achieve promising accuracy with low computational cost, all of which make them highly applicable to on-board BMSs. Nevertheless, they are only compatible with EMs and ECMs with linear representations, which indicates their limitation of being coupled with those battery models of nonlinear representations, thereby affecting their precision.

Compared to OLP methods, MPC methods are more versatile in battery SOP estimation suitable for EMs and ECMs with both linear and nonlinear representations.

These approaches introduce a set of inequality constraints and adopt an aggregate consideration for battery SOP estimation under the multi-constraints, thereby avoiding a case-by-base analysis while achieving promising performance. However, the MPC methods incur a considerably increased complexity and time cost in solving an objective function when processing battery models with highly nonlinear representations, highlighting the significance of remaining model simplicity for MPC methods in online SOP estimation.

Unlocking the limitation of OLP methods, IA methods offer an opportunity for battery SOP estimation that integrates those models with highly nonlinear representations and those operational constraints across a broad operational scale. Consequently, these algorithms support a comprehensive consideration of battery peak power performance from all aspects and have potential to achieve superior performance over the other three algorithms. Nonetheless, it inevitably sacrifices computational efficiency to search battery SOP as precise as possible and may have difficulty in algorithm development for trading-off between accuracy and complexity. As of now, these algorithms are not ready to be deployed in on-board BMSs due to their limited computational power, but they may find applications in DT on cloud platforms to transmit high-fidelity estimations of battery SOP through wireless communication in real time.

**Table 4**

A summary of model-algorithm compatibleness in battery SOP estimation.

| Algorithm types | Battery model types | | |
| --- | --- | --- | --- |
| | White-box models | Grey-box models | Black-box models |
| CM methods | - | - | Empirical model |
| OLP methods | EM (L) | ECM (L) | - |
| MPC methods | EM (L/NL) | ECM (L/NL) | - |
| IA methods | EM (NL) | ECM (NL) | NNM |

Note: The notation L/NL denotes the linear/nonlinear representation of a battery model.

**Table 5**

A summary of pros and cons of state-of-the-art SOP estimation algorithms.

| Algorithm types | Pros | Cons |
| --- | --- | --- |
| CM methods | • Ease of algorithm development<br>• Low computational cost | • Limited performance<br>• Suitability only to empirical models |
| OLP methods | • Ease of algorithm development<br>• Low computational cost<br>• Case-by-case analysis of battery SOP under each constraint<br>• Promising accuracy<br>• High applicability in on-board BMS | • Incompatibleness with models of nonlinear representations |
| MPC methods | • Ease of algorithm development<br>• Relatively low computational cost<br>• Promising accuracy<br>• Aggregate analysis of battery SOP under multi-constraints<br>• High compatibility with models of linear and nonlinear representations<br>• High applicability in on-board BMS | • Significantly increased complexity for models of nonlinear representations |
| IA methods | • High accuracy<br>• High compatibility with all types of models<br>• High compatibility with all types of operational constraints | • High computational cost<br>• Difficulty in algorithm development |

## 6. Error source analysis

In this section, we conduct an in-depth analysis of the primary error sources in battery SOP estimation, along with their propagation pathways. For illustrative purposes, we focus on battery SOP estimation under the most commonly used CC-POM, utilizing the Thevenin model as our foundation and considering a SOA defined by two basic constraints (current and voltage) and one state constraint (SOC). The mathematical derivations of battery SOP under each operational constraint and the analytical expressions of SOP errors induced by various contributors are provided in Appendix I and Appendix II, respectively. It is worth noting that this approach can be easily adapted to other POMs (CC, CC-CV and CP) in a similar fashion. It is also important to point out that when we investigate the impact of a specific contributor all other potential error sources will be isolated to ensure clarity of the analysis.

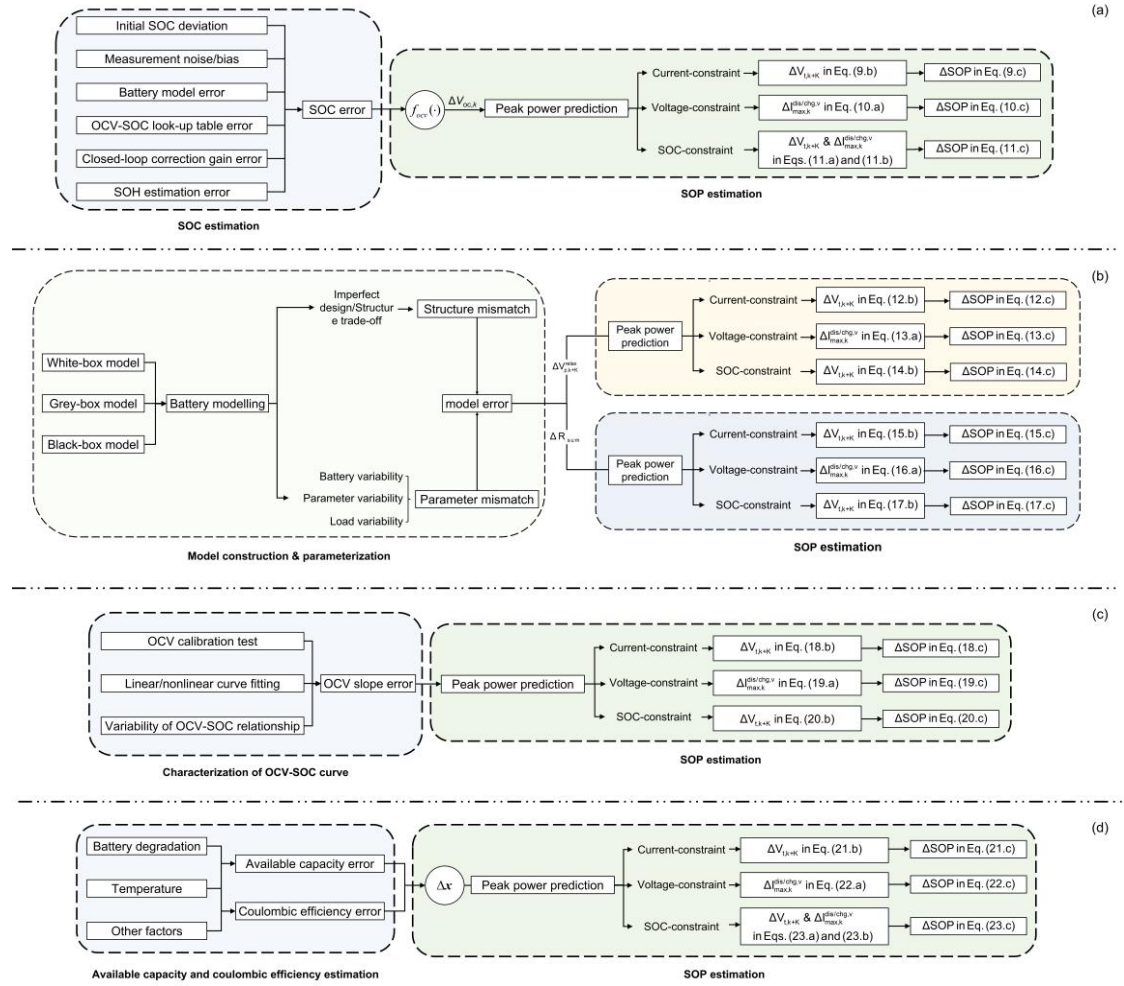

Fig. 10. Propagation pathway of each error source in battery SOP estimation: (a) SOC error; (b) Model error; (c) OCV slope error; and (d) available capacity/coulombic efficiency error.

### 6.1. SOC error

SOC is a pivotal state variable in Thevenin model and a critical limitation factor in battery SOP estimation. However, available sensing technologies do not allow for a direct measurement of SOC in real-world applications. This limitation necessitates reliance on high-quality SOC estimation derived from externally measurable signals, such as current, voltage and temperature. Despite the claims of excellent performance by some researchers for their developed SOC estimation approaches, a SOC error remains an intractable issue. In [18], the primary factors influencing SOC estimation have been summarized, involving initial SOC deviation, measurement noise or bias, battery model error, OCV-SOC curve error, closed-loop correction gain error and SOH error. These sources contribute to SOC inaccuracies, which then follow the propagation pathway, as illustrated in Fig. 10 (a), and impact battery SOP estimation to varying extents, depending on the specific operational constraints engaged.

Without loss of generality, we express the true SOC at time $k$ as the sum of a biased estimation in association with an error, denoted as $SOC_k = S\hat{O}C_k + \Delta SOC_k$. Fig. 11 depicts the current-voltage trajectories of batteries in discharge SOP estimation under both SOC-error-free and SOC-error-corrupted conditions, and Table 7 provides the analytical expressions of SOP errors that occur in the presence of a SOC error. As can be seen in Fig. 11 (a) and (b), a SOC error will deviate either the terminal voltage prediction or the peak current estimation in battery SOP estimation under the operational constraints of current and voltage. Assuming that the OCV-SOC curve maintains a relatively constant slope at both the true and estimated SOCs while this slope is linear over a window, we express the voltage prediction error at time $k+K$ and the peak current estimation error from time $k$ to $k+K$ in Eq. (9.b) and Eq. (10.a), respectively. This yields the associated SOP errors in Eq. (9.c) and Eq. (10.c). Evidently, the SOP error under the current constraint is influenced not only by the magnitude of SOC error itself but also by the slope of OCV-SOC curve and the manufacturer-defined maximum allowable current. It is reasonable to infer that this impact might be somewhat mitigated in the regions with a flat OCV-SOC curve or for batteries with relatively mild discharge and charge capabilities. On the other hand, the impact of a SOC error on battery SOP estimation under the voltage constraint is subject to multiple factors, such as the OCV-SOC curve, available capacity, the length of a prediction window and coulombic efficiency. Particularly, for those batteries with small internal resistance, the SOP estimation under the voltage constraint will be more sensitive to SOC error, highlighting the increased significance of accurate battery modelling. Unlike the SOP estimation under the operational constraint of current or voltage, it is more complex for the impact of a SOC error on SOP estimation under the SOC constraint, as illustrated in Fig. 11 (c), where the deviations arise not only in the voltage prediction at time $k+K$ but also in the peak current estimation from time $k$ to $k+K$. As indicated in Eqs. (11.a)-(11.c), the SOP error can be modelled as a parabolic function of SOC error, indicating a nonlinear sensitivity in their relationship.

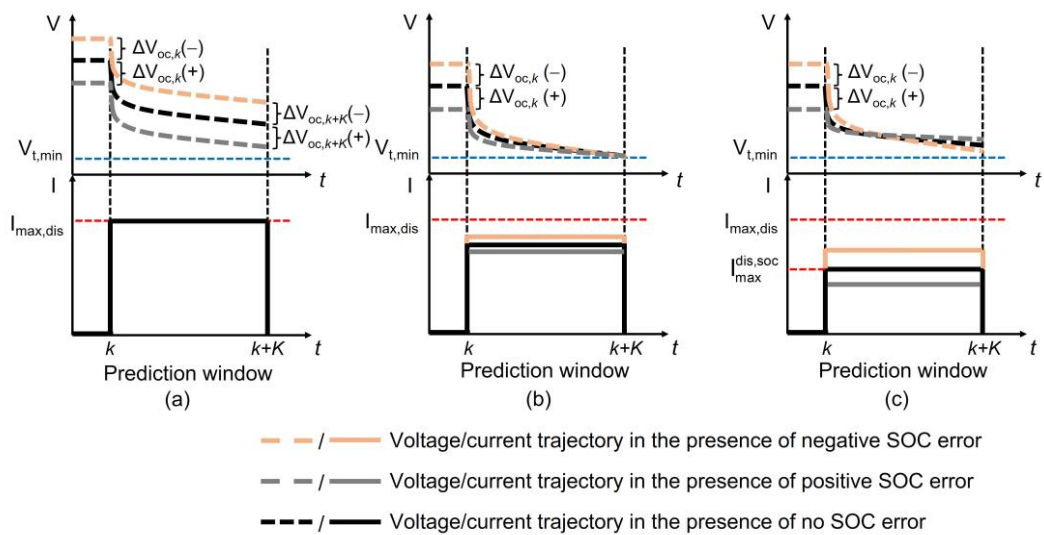

Fig. 11. Typical current-voltage trajectories of batteries in discharge SOP estimation for both SOC-error-free and SOC-error-corrupted conditions under: (a) current constraint; (b) voltage constraint; and (c) SOC constraint.

### 6.2. Model error

Batteries are undeniably a complex system, characterized by a range of electrochemical and thermal reactions occurring across a broad timescale and frequency domain, as depicted in Fig. 12. This complexity poses significant challenges in developing a comprehensive and explicit battery model that accurately encapsulates all the internal and external dynamics of batteries. In the meantime, it is practically important to strike a balance between model accuracy and complexity, especially for models to be compatible with on-board BMSs. In this context, a compromise in model structure can lead to a mismatch with the actual behaviors of real batteries. This issue often arises when the modelling focus is placed on some dominant reactions while overlooking the other less dominant ones.

In addition, model parameters play a crucial role in unlocking the full potential of a battery model. Despite the extensive research and a wealth of studies dedicated to techniques for accurate parameter identification, practical applications still grapple with an issue of parameter mismatch. This primarily arises from the following three aspects:

*Battery variability*: Battery variability pertains to the minor inconsistencies during cell manufacturing, such as the production of electrode slurry in a pre-designed proportion with active materials, conductive additives, solvents and binders. Such variability can cause a parameter mismatch between batteries that have been tested for offline parameter identification in a laboratory environment and those actually used in real-world applications, even if they originate from the same production batch.

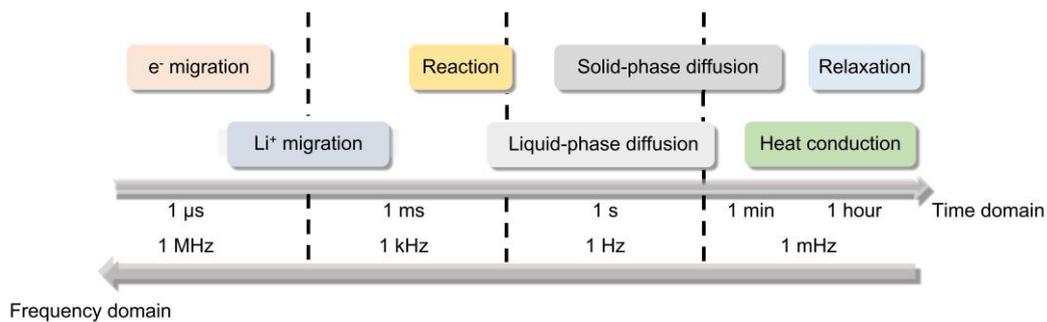

Fig. 12. Cell inherent mechanisms related to their typical time constants.

*Parameter variability*: Parameter variability refers to the time-varying characteristics of model parameters, influenced by various factors, such as SOC, SOH and temperature. Given the multitude of possible scenarios, it is exceedingly challenging and impractical to characterize model parameters for every conceivable condition. Consequently, there is often a mismatch between the model parameters identified offline in a laboratory environment and those after being used in real-world applications.

*Load variability*: Load variability relates to the differences in load characteristics encountered during model parameter identification and in battery state estimation. Specifically, offline parameter identification for test batteries is typically conducted using carefully designed pulse profiles, such as HPPC profile and dynamic stress test profile. In contrast, online parameter identification for in-service batteries occurs under dynamic load profiles, often with noise corruptions. However, neither approach can extensively capture the distinct characteristics of batteries in SOC estimation under dynamic load conditions and in SOP estimation under peak operation conditions, thereby giving rise to a parameter mismatch in their joint estimation.

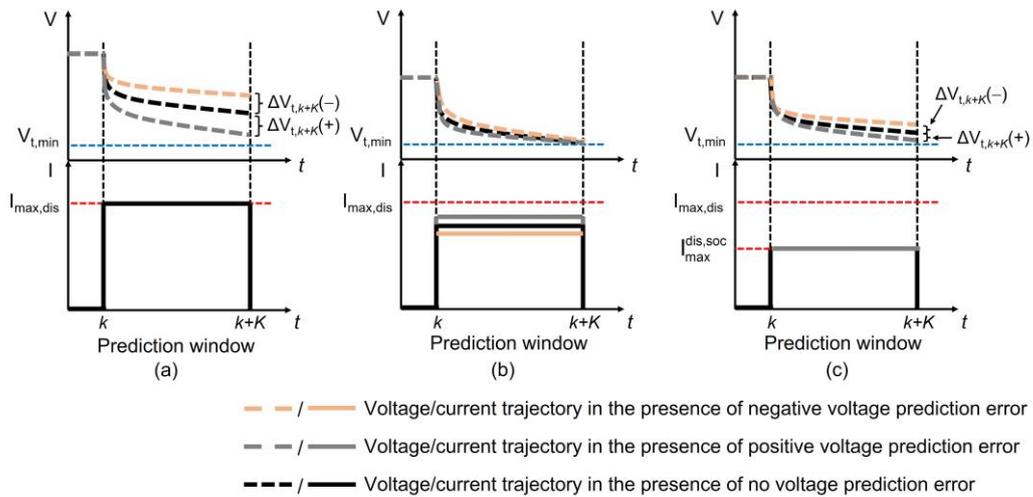

Fig. 13. Typical current-voltage trajectories of batteries in discharge SOP estimation for both model-error-free and model-error-corrupted conditions under: (a) current constraint; (b) voltage constraint; and (c) SOC constraint.

The mismatch in model structures or parameters will differ the simulated battery dynamics from their true behaviors over a prediction window, thereby producing a model error and affecting SOP estimation performance. Fig. 13 depicts the current-voltage trajectories of batteries in discharge SOP estimation under both model-error-free and model-error-corrupted conditions. In the context of Thevenin model, we categorize the model error into two distinct types: one arising from the description of battery polarization dynamics during relaxation (denoted as $\Delta V_p^{relax}$), and the other from the characterization of the sum of internal resistance (denoted as $\Delta R_{sum}$). The impact of these two types of model errors on battery SOP estimation under the operational constraint of current, voltage and SOC are summarized in Table 8 and Table 9, respectively, and the associated error propagation pathways are illustrated in Fig. 10 (b). Essentially, both $\Delta V_p^{relax}$ and $\Delta R_{sum}$ contribute to a voltage prediction error at the end of a prediction window, thereby leading to a SOP error under the current and SOC constraints. However, their impact on SOP estimation under the voltage constraint is slightly different in the way they affect the peak current estimation. A closer examination of Eq. (13.a) and Eq. (16.a) reveals that the peak current estimation error in the presence of $\Delta R_{sum}$ has certain dependency on the true peak current. Contrarily, the peak current estimation error does not exhibit such a dependency on the true peak current when the contributor becomes $\Delta V_p^{relax}$. This discrepancy, despite an unbiased voltage prediction at the end of the window, yields different forms of SOP errors, as illustrated in Eq. (13.c) and Eq. (16.c). Clearly, the SOP error has a linear relationship with $\Delta V_p^{relax}$ under the voltage constraint but it is nonlinear with $\Delta R_{sum}$. This distinction highlights the nuanced ways in which different types of model errors can influence SOP estimation.

**6.3. OCV slope error**

Battery OCV-SOC curve is critically important in battery management, not only enabling a look-up table approach for SOC estimation but also delineating an OCV evolution trend (either downward or upward) for SOP estimation. An inaccurate characterization of this curve can lead to an OCV prediction error at the end of a prediction window, thereby contributing to a source of SOP error. Generally, the OCV-SOC relationship is calibrated offline in a laboratory environment and then characterized through linear or nonlinear curve fittings for online applications. Therefore, the slope error of an OCV-SOC curve can stem from the following three aspects:

*OCV calibration test*: Low-current OCV test (LOT) and incremental OCV test (IOT) are two prevalent testing methodologies employed for OCV calibration. The LOT involves discharging or charging the batteries under a CC protocol at a minimal C-rate, typically around 1/20. Under such a minimal current, the voltage readings obtained are generally considered to directly represent battery OCVs. On the other hand, The IOT utilizes a pulse-relaxation protocol at specific SOC points. Each pulse-relaxation routine comprises a pulse discharge or charge corresponding to a certain amount of capacity (e.g., 10% SOC) as well as a sufficiently long rest period for mitigating the kinetic contributions inside batteries.

However, previous studies have indicated some discrepancies in the OCV calibration outcomes derived from these two tests. Zheng et al. [162] observed that while the results from the IOT and LOT for NCM batteries were relatively consistent in the middle SOC region, significant deviations occurred in the other regions. Notably, the OCVs at 0% and 100% SOCs from the LOT appeared strong rebounds after reaching the lower and upper cut-off thresholds, indicating a non-negligible polarization overpotential inside batteries. Petzl and Danzer [163] pointed out that even very small C-rates can cause an observable deviation between the OCV calibration results of LFP battery from the IOT and LOT in the middle SOC region. For instance, at 50% SOC, there could be a difference of around 10 mV between the two methods. Roscher and Sauer [164] have elaborated that the IOT, particularly with extended rest periods, more accurately reflects the OCV characteristics of batteries. They noted that the calibration error could be as low as approximately 3 mV after a half-hour rest, and further reduced to below 1 mV with a three-hour rest period. For this reason, IOT is generally recommended over LOT for OCV calibrations. The LOT, in contrast, may not effectively characterize the OCV-SOC relationship, potentially leading to more pronounced slope errors in an OCV-SOC curve.

*Curve fitting*: After gathering the experimental data from an OCV test, the subsequent step involves curve fitting to interpolate the OCV data across the entire SOC operation range. This step is crucial for avoiding the need to store extensive OCV data directly in on-board BMSs. In this context, Yu et al. [145] conducted a comprehensive comparison of 18 nonlinear functions for curve fitting, specifically analyzing their applicability to capture the OCV-SOC relationship for LFP and NCM batteries. Their findings revealed that while a twelfth-order polynomial function provided satisfactory results for both types of battery chemistries, there was still a noticeable accuracy loss — approximately 5 to 10 mV — between the raw data and the fitted data.

*Variability of OCV-SOC relationship*: While some studies have indicated a relatively stable relationship between OCV and SOC, there is still a slight degree of variability influenced by temperature and aging level. Liu et al. [165] explored how the OCVs of LMO batteries depend on temperature across various SOC levels, and their results demonstrated that temperature significantly impacts OCV in the low SOC region, while its effect diminishes above 40% SOC. For example, at 0% SOC, OCV changes by 3 mV when the temperature drops from 25°C to 10°C, but the change reduces to 1 mV at 50% SOC. Another study [18] found that the OCVs of LFP batteries and LTO batteries exhibit greater variations at lower temperatures. In the case of LFP batteries, the OCV difference within a temperature range of 15 to 45°C is minimal (<5 mV), but it increases to 15 mV between 0°C and 25°C. For LTO batteries, there is little OCV difference observed over a wide range of temperature from 0 to 45°C, which then escalates to 10 mV when the temperature drops to -15°C.

Regarding battery degradation, its impact on the OCV-SOC relationship is even more significant, though the affected SOC regions vary with different battery chemistries. For instance, it was shown that the OCV difference between fresh and aged LFP batteries will exceed 20 mV in a region between 70% and 80% SOC, and this difference can be up to 100 mV for NCM batteries below 10% SOC [18]. Supported by these studies, temperature and degradation are recognized with a variable degree of impacts on the OCV-SOC relationship, which may enlarge the OCV slope error and thus SOP error.

In this work, we characterize the true OCV slope over a prediction window as the sum of a biased slope in association with a slope error, denoted as $\kappa = \hat{\kappa} + \Delta\kappa$. Fig. 10 (c) depicts the propagation pathway of $\Delta\kappa$ in battery SOP estimation, and Table 10 details the associated contributions to SOP errors under the operational constraints of current, voltage and SOC. As observed, $\Delta\kappa$ deviates the OCV prediction over a prediction window, thereby contributing to a voltage prediction error at the end of the window in SOP estimation under the current and SOC constraints (see Eq. (18.b) and Eq. (20.b)). In these two cases, the SOP error varies linearly with $\Delta\kappa$, as illustrated in Eq. (18.c) and Eq. (20.c). For SOP estimation under the voltage constraint, $\Delta\kappa$ leads to a peak current estimation error, which is recognized with certain dependency on the true peak current (see Eq. (19.a)). As a result, the SOP error under the voltage constraint varies nonlinearly when $\Delta\kappa$ changes at different temperatures or aging levels (see Eq. (19.c)). Furthermore, the influence of $\Delta\kappa$ on battery SOP estimation is intricately linked to the discharge and charge capabilities of batteries as well as the length of a prediction window. These factors are crucial because they essentially dictate the extent to which the SOC can increase or decrease over the window, thereby determining the significance of the impact of $\Delta\kappa$ on SOP estimation.

### 6.4. Available capacity/coulombic efficiency error

Available capacity and coulombic efficiency are fundamental indices that characterize the discharge and charge performance of batteries. Previous research has thoroughly explored the interrelationship between these two key parameters and the range of factors that could potentially impact their values. For instance, Yang et al. [166,167] carried out experimental investigations into specific aging mechanisms that predominantly affect coulombic efficiency. Their findings revealed that, apart from the loss of lithium inventory, a loss of active material also accelerates battery degradation and brings down coulombic efficiency. In addition, the long-term evolution of coulombic efficiency is confirmed with a strong and intimate correlation with battery capacity decay. Specifically, a coulombic efficiency curve that decreases steadily over the battery lifetime typically aligns with a smooth decay of available capacity, while a sharp drop of the curve often signifies an accelerated decay rate. Smith et al. [168] examined the temperature dependencies of battery available capacity and coulombic efficiency. The results demonstrated that temperature significantly influences these parameters for all the batteries tested. Notably, as the temperature increases, both available capacity and coulombic efficiency tend to deviate more substantially from their original values of a fresh battery at room temperature. The authors attributed this phenomenon primarily to the intensification of side reactions, such as electrolyte oxidation, that occur at high temperatures.

In light of the interrelationship between battery available capacity and coulombic efficiency, we define a new parameter $x = \eta/3600C_a$ by collectively considering their impact on battery SOP estimation. Also, we characterize $x$ as the sum of a biased one in association with an error, denoted as $x = \hat{x} + \Delta x$. Fig. 10 (d) depicts the propagation pathway of $\Delta x$ in battery SOP estimation, and Table 11 details the associated contributions to SOP errors under the operational constraints of current, voltage and SOC. Referring to Eq. (21.b) and Eq. (22.a), $\Delta x$ influences either the terminal voltage prediction or the peak current estimation in battery SOP estimation under the operational constraints of current and voltage. Particularly, the nonlinear relationship of $\Delta x$ with the peak current estimation error under the voltage constraint is similar to those of $\Delta R_{sum}$ and $\Delta \kappa$, as discussed in previous subsections. However, when considering SOP estimation under the SOC constraint, $\Delta x$ impacts both the terminal voltage prediction and peak current estimation, posing a more intricate influence on SOP error (see Eq. (23.c)). The reason for this complexity lies in that the SOC prediction over a prediction window tends to diverge from the true SOC in the presence of $\Delta x$. This divergence affects the peak current estimation under the SOC constraint and causes a deviation in the terminal voltage prediction when there is a peak current estimation error.

## 7. Challenges and future outlooks

The review of current SOP estimation technology with associated key aspects in the previous sections indicates great strides in this area. However, challenges still persist and require future attention. In this section, we synthesize these challenges and address potential research directions, as illustrated in Fig. 14.

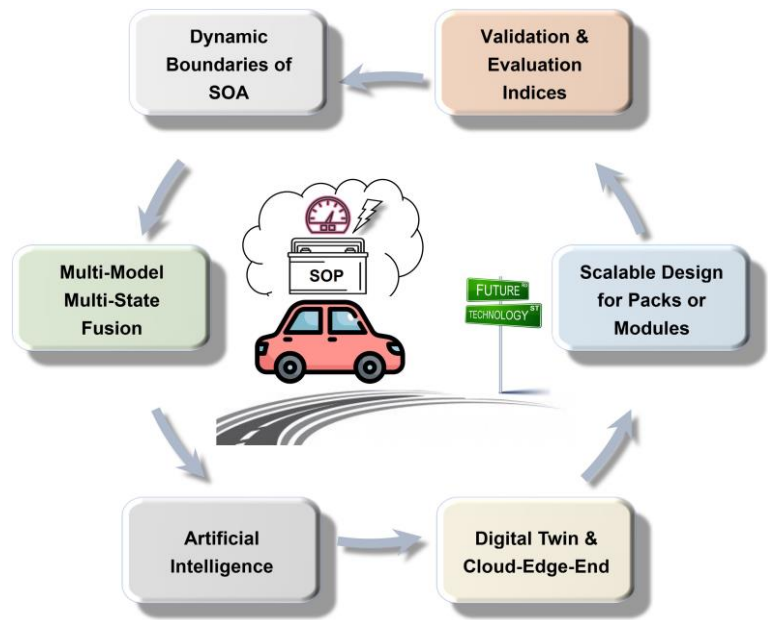

Fig. 14. Challenges and future outlooks of battery SOP estimation technology.

## 7.1. Safe operation area with dynamic boundaries

Battery safety and health is always a critical concern for SOA design in pursuit of an acceptable compromise between the peak power performance and the short term operational safety/long-term cycle life. However, due to the complex side reaction mechanisms, it is challenging for a conventional SOA with fixed boundaries to achieve such an optimal balance. An ideal solution is to fully consider these side reactions, capture the thresholds of their occurrence in the form of probability functions, and design a flexible SOA with dynamic boundaries. By constructing a set of probability functions with regard to factors such as temperature and current rate, it is expected that the side reactions can be effectively mitigated and the overall performance of battery safety and health management can be greatly enhanced when batteries are functioning under the peak power condition. Nevertheless, it may require substantial experimental efforts supported by post-mortem analyses from field failures as well as safety and abuse tolerance tests to establish the probability functions that unravel the underlying correlation between the thresholds for the occurrence of various side reactions and their multi-dependencies. This difficulty is likely to be resolved by continually advancing DT technology [169] with multi-physics health-conscious virtual models [170,171]. The battery DT can assist intricate studies that correlate microscopic changes (e.g., electrode, electrolyte, and separator) to macroscopic consequences (e.g., voltage, internal resistance and capacity), creating more opportunities for characterizing the probability functions of the thresholds for the occurrence of various side reactions and indicating a great potential to be extended in the flexible SOA design with dynamic boundaries for battery SOP estimation.

### 7.2. Multi-model multi-state fusion

The literature abounds with methods for battery single state estimation, yet there remains a scarcity of research on jointly estimating more than two internal states, also referring to as multi-state co-estimation [21]. Given the intricate interplay and mutual influence of different internal states of batteries, independently estimating one state without considering others often yields satisfactory results only under specific conditions. To date, only a handful of studies have focused on multi-state co-estimation, with pioneering works in [143–146] contributing to the development of co-estimation schemes that include SOC, SOH, and SOP. Compared to traditional joint SOC-SOP estimation methods, periodic updates of SOH (capacity or resistance), typically achieved through approaches like accumulated ampere-hour method or filtering/observer techniques, enhance the estimation accuracy of SOC and SOP to certain extents [124,172,173]. Although these approaches improved the overall performance, they have not delved deeply into the inherent coupling mechanisms among SOC, SOH, and SOP, and notably, SOT, a crucial factor affecting the peak power performance of batteries, is rarely involved into co-estimation schemes. In light of these limitations, a multi-model multi-state fusion framework (as illustrated in Fig. 15) that accounts for the multi-physics coupling of internal electro-chemical-thermal-aging dynamics of batteries presents a promising yet challenging direction for future research. Furthermore, such a comprehensive framework will inevitably complicate the entire estimation process, presenting another challenge for efficient integration of this framework into on-board BMSs.

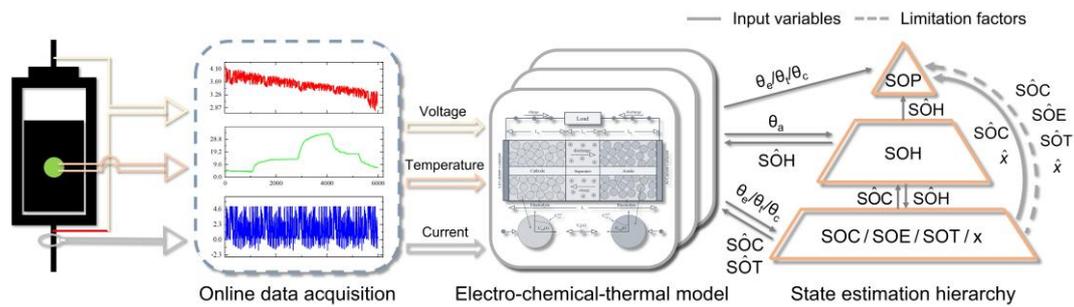

Fig. 15. Battery multi-model multi-state co-estimation scheme and hierarchical relationship of SOX. ($\theta_e/\theta_t/\theta_c/\theta_a$ stands for the model parameters of electrical/thermal/chemical/aging model; $x$ denotes the state/electrochemical variables, such as polarization voltage, lithium surface concentration, and side reaction potential).

### 7.3. Artificial intelligence

Artificial intelligence (AI), as an emerging technology, complements traditional mathematical algorithms with its promising capabilities in classification and regression. With the ongoing development and expansion of big data, the fusion of BMSs with AI is anticipated to become increasingly prevalent in EVs. Recent studies have demonstrated significant advancements in the applications of AI for state estimation [13,174], cell balancing [175], fault diagnosis [176], etc. Prominent among these technologies are support vector machines [177], relevance vector machines [178], fuzzy logic [179], convolutional neural networks [180], recurrent neural networks [181], transformers [182], etc. While AI presents significant potential in BMSs, there are some concerns associated with its applications in battery SOP estimation, primarily stemming from the following three aspects:

- **Lack of physical interpretability:** AI is essentially data-driven by characterizing the underlying input-output correlation, instead of interpreting electrochemical reactions inside batteries. In the context of battery SOP estimation, AI may struggle to correctly choose the dominant operational constraint and explicitly simulate the battery behaviors under associated boundary condition. This limitation could impact the reliability and accuracy of SOP estimation.

- **Limited data sources:** Battery SOP describes the peak power performance under the boundary condition, which is challenging to be reproduced due to the experimental randomness and noise disturbances. Collecting data under such extreme circumstances over the whole battery operation range can be considerably time-consuming and labor-intensive. In addition, repetitive SOP tests at high current rates can accelerate battery degradation, thereby compromising the quality of data. Moreover, a limited SOP sample size can lead to an overfitting with selected input variables, resulting in a poor generalization performance of the model under varying operation conditions.

- **Challenges in multi-state co-estimation:** Battery SOP is intricately coupled with other internal states such as SOC, SOH, and SOT. These states operate on different timescales, posing a challenge in identifying common input variables suitable for multi-state co-estimation. This complexity adds another challenge in developing robust AI methods capable of accurately estimating battery SOP while considering the interdependencies with other internal states.

Viable solutions to address these concerns may include physics-informed deep learning, data augmentation, and distributive topologies. Recently, the concept of physics-informed deep learning has garnered considerable interest [183]. With the aid of conventional battery models, physics-informed deep learning proves to be effective in interpreting battery dynamics, offering great potential to promote the reliability and accuracy of SOP estimation. Data augmentation presents a helpful pathway to the challenge of data scarcity in battery SOP estimation. NNMs such as generative adversarial networks and variational autoencoders are effective tools in generating synthetic data to augment limited sources of real SOP data. These NNMs have already shown success in areas like SOC estimation [184] and remaining useful life prediction [185]. Furthermore, the study in [186] developed a distributive topology comprising two parallel-connected DNNs for estimating SOC and SOH. In this framework, each DNN utilizes the estimation outcomes of the other as inputs, thereby enhancing estimation robustness and accounting for the interdependencies between SOC and SOH in their joint estimation. This strategy holds potential for extension in SOP-involved multi-state co-estimation, representing a forward-looking direction for research in this domain. With the increasing adoption of the above-mentioned technologies, it is believed that AI will find broad applications in battery SOP estimation, thereby shaping a promising direction for future research.

### 7.4. Battery digital twin and cloud-edge-end collaboration

Nowadays the wave of digital economy has swept the world [169]. The adoption of DT technology introduces a novel paradigm for the networked management and service of batteries. DT models represent a virtual representation created digitally to mirror a physical battery entity, providing valuable references for a range of BMS functionalities, such as data collection and integration, real-time analysis and simulation, energy management and optimization, as well as remote monitoring and control [187,188]. Especially, a high-fidelity DT model that encapsulates the multi-physics coupling of internal electro-chemical-thermal-aging dynamics of batteries is highly beneficial for battery SOP estimation under an unseen operation condition. However, deploying such a sophisticated DT model in on-board BMSs, which often have limited computational resources and memory storages, presents a significant challenge. Recent research on networked BMS architectures with a cloud-edge-end collaboration presents a promising solution [147]. These BMS architectures are usually composed of four layers, as illustrated in Fig. 16. Through online learning and model updating, they overcame the shortcomings of fixed model parameters in conventional BMSs and proved to be effective in maintaining high accuracy of multi-state co-estimation, including SOP. The detailed structure of these BMS architectures is outlined as follows:

- **Data-based application layer:** This layer contains multiple application modules for data visualization, report display, fault warning, etc. It can be event-driven with high concurrency, availability and rapid response to address cell balancing, monitoring, error recovery and other issues. Historic information of hardware, software and application modules can be retrospectively analysed to establish a comprehensive management mechanism, covering the entire lifecycle of battery systems.
- **Edge computing layer:** This layer contains a local BMS equipped for real-time sensing and data acquisition. It includes an edge computing server responsible for executing basic BMS algorithms and control strategies, utilizing lightweight battery models like ECMs and NNMs. Besides, it is furnished with communication infrastructures to facilitate the wireless transmission of collaborative control commands from the cloud.

- **Data access layer:** This layer establishes a bidirectional communication channel between the edge computing layer and the cloud. It parses and repackages the data sent from the lower layer for storage and analysis. Also, it distributes the outcomes of data analysis, such as the updated model parameters and the control strategies derived from a DT model, to the designated devices.

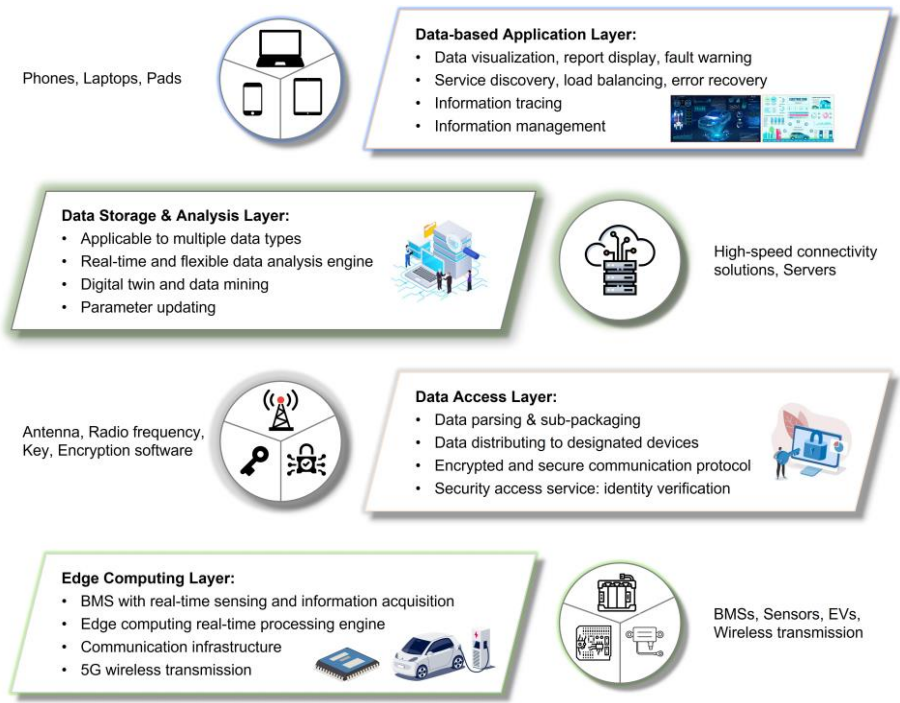

Fig. 16. Architecture of cloud-edge-end collaboration for next-generation BMSs in EVs. Reproduced with permission from Ref. *[147]*. Copyright 2022, Elsevier.

- **Data storage and analysis layer:** This layer serves as a centralized repository for all battery operation data, transforming raw data into formats suitable for processing, visualization, and AI-assisted analysis. In the meantime, it also provides essential data services to upper-layer applications. Leveraging the power of cloud computing with a high-fidelity DT model, this layer is adept at handling advanced algorithms with high complexity, allowing for effective data mining and incremental learning across various dimensions and scales.

Ongoing research on networked BMS architectures suggests that the advancements in battery DT and cloud-edge-end collaboration technology will offer an opportunity to enhance SOP estimation accuracy.

**7.5. Pack/module-level estimation with scalable design**

As SOP estimation technology continues to evolve, it is crucial to address their scalability from individual cells to modules and entire battery packs [10]. While most existing research concentrates on cell-level SOP estimation, adapting these schemes to battery modules or packs with various configurations, such as serial, parallel, or hybrid connection, remains a critical challenge [189]. This adaptation not only relates to re-identifying the boundary conditions of battery peak power operations but also involves striking a balance between estimation accuracy and computational efficiency. Different estimation strategies, including distributed and lumped approaches, must be thoroughly explored to meet system-level estimation requirements [190,191].

In a distributed approach, SOP estimation is developed for each cell or group of adjacent cells, resulting in an array of estimators for a module or pack. This approach leverages different sets of model parameters to respective cells or cell groups, aiming for precise cell-level modelling but at the expense of computational efficiency. Conversely, in lumped approach the entire battery system is treated as a modelling object, where just one state estimator is utilized to minimize the computational load while overlooking cell interactions and inconsistencies.

Extensive research is still needed to effectively extend available approaches to the module and pack levels, ensuring robust and efficient SOP estimation for an entire battery system.

## 7.6. Validation approach & evaluation index

A high-fidelity, laboratory-based validation approach is vital for quantitative analysis of SOP estimation results, facilitating a fair comparison among different methods to highlight any performance enhancement. Regrettably, the current literature does not offer a standardized approach for SOP validation. This absence has led to the use of diverse indices in both direct and indirect validation approaches.

*Direct validation:* Direct validation refers to a straightforward evaluation of SOP estimation results based on the experimentally calibrated reference SOPs (or peak discharge/charge currents). Well-known HPPC test and its derivatives belong to this category, which determine reference SOPs by characterizing the OCV-resistance (OCV-R) model at a few pre-selected test points. As they are easy to implement, they have been widely applied in SOP validation [90,107,109]. However, HPPC test-based validation approaches only consider voltage constraint and highly rely on the stationary OCV-R model without reflecting transient dynamics of batteries, which have made them tend to yield the over-optimistic reference SOPs and thus greatly affect the reliability of SOP validation. To obtain high-precision reference SOPs, some other studies [63,100,131,140] tried to reproduce the peak power operations of a battery under the boundary conditions in a trial-and-error manner by inserting a series of pulse tests with manually adjusted pulse amplitudes. This approach is labor-intensive and time-consuming, and a perfect reproduction can be sometimes unachievable due to the presence of external disturbances and experimental randomness.

*Indirect validation:* Indirect validation evaluates the accuracy of SOP estimation results in terms of some indirect indices. For instance, Zheng et al. [89] compared the measured and the estimated battery internal resistance to evaluate the SOP estimation performance. The studies in [65,100] discharged and charged a battery using the peak current estimation, and they evaluated the SOP estimation performance based on the relative voltage error between the measured end-of-pulse voltage and the cut-off voltage. While these approaches are intuitive for qualitative performance evaluation and comparison, a quantitative assessment of SOP estimation accuracy is still unavailable.

From the above discussions, it can be seen that there are challenges in obtaining high-precision reference SOPs. Firstly, a test battery has to be discharged and charged repeatedly under a close-to-overcurrent or a close-to-over/undervoltage condition. It is possible for the test battery to operate outside the SOA if the control and sensing of battery voltage, current, and temperature are not sufficiently precise and synchronized. In the meantime, the test battery operating under such harsh conditions incur an accelerated degradation rate, thereby bringing additional uncertainties and influencing the repeatability of the experimental results. Secondly, SOP calibration tests for high-power batteries (e.g., maximum discharge current ≥15 C-rate) demand a specialized experimental platform supporting hundreds or even thousands of amperes, which not only increases the experimental cost but also leaves potential safety issues during high-current tests. Moreover, SOP calibration tests will be only implemented at a few pre-selected test points of a dynamic load profile. The reference SOPs are usually scarce compared to the length of the load profile in the unit of hour (indicating thousands of time points at a commonly used sampling interval of 1 s). It is, therefore, questionable whether such limited reference SOPs are capable to reflect the overall SOP estimation accuracy.

To deal with these challenges, a safe, efficient, and reliable SOP validation approach is imperative for future research. Some potentially viable alternatives have been proposed in the literature. Tang et al. [192] designed an equivalent discharge test with a set of low-current pulses to obtain the approximated reference SOPs under high peak discharge currents, and a flexible softmax neural network was built for nonlinear inter/extrapolations of vacant current amplitudes. The results indicated the effectiveness of the proposed SOP calibration test with a 0.5% relative SOP error and a 33% pulse amplitude reduction in comparison with the actual peak discharge current. While this equivalent discharge test avoided those pulse tests under high currents, it still demanded labor-intensive experimental efforts to collect sufficient data for neural network training. Besides, the proposed method essentially obtains reference SOPs by extrapolating the experimental data from low-current tests, but batteries may embody strong nonlinearity under high currents, thereby deteriorating the extrapolation performance. Shen et al. [193] adopted a high-precision P2D model to reproduce the battery peak power operations under the boundary condition in order to calibrate reference SOPs. Similarly, Shu et al. [145] and Zou et al. [25] employed an ECM with well-characterized model parameters at various SOCs and temperatures for reference SOP calibrations. This simulation-based alternative allows to generate a continuous reference SOP curve over the whole load profiles of interest, and the reliability of SOP validation can be ensured as long as the battery model retains high accuracy.

**Appendix I. Preliminary of battery SOP estimation**

Battery SOP estimation is essentially a task of predicting the peak power behaviors of batteries under the boundary condition. Based on the definition provided in Eq. (1) and the detailed discussion presented in Section 3, battery SOP under the CC-POM occurs at the end of a prediction window, equalling to the multiplication of the peak discharge/charge current under the multi-constraints with the end-of-window voltage, namely $SOP_k = I_{\max,k}^{dis/chg,MC} \times V_{t,k+K}$. Intuitively, the SOP error can stem from inaccuracies in either peak current estimation, terminal voltage prediction, or a combination of both, where any of these deviations can impact the overall SOP estimation performance. For illustrative purpose, we assume a Thevenin model with well-characterized model parameters that perfectly simulate all battery dynamics over a prediction window. The derivation of reference SOP based on this ideal Thevenin model is presented in the following.

Given the governing equation of a Thevenin model in Eq. (2), the battery terminal voltage at time $k+K$ can be expressed by Eq. (3) based on step-wise model iterations.

$$\begin{cases} V_{p,k} = V_{p,k-1} \exp\left(\dfrac{-\Delta t}{\tau}\right) + I_{L,k-1} R_1 \left(1 - \exp\left(\dfrac{-\Delta t}{\tau}\right)\right) \\ SOC_k = SOC_{k-1} - \dfrac{\eta \Delta t I_{L,k-1}}{3600 C_a} \\ V_{t,k} = V_{oc,k} - V_{p,k} - V_{ohmic,k} \end{cases} \quad (2)$$

where $V_{oc,k}$, $V_{t,k}$, and $V_{p,k}$ denotes battery OCV, terminal voltage, and polarization voltage, respectively. $V_{ohmic,k} = I_{L,k} R_0$ gives the voltage loss on the sum of the ohmic resistances inside batteries, such as current collectors and electrolyte. $I_{L,k}$ is the load current. $R_1$ and $\tau$ represent the polarization resistance and time constant of RC pair. $C_a$ is the battery available capacity, $\eta$ is the coulombic efficiency, and $\Delta t$ is the sampling interval.

$$V_{t,k+K} = V_{oc,k+K} - V_{p,k+K} - V_{ohmic,k+K} \quad (3)$$

$$\begin{cases} V_{oc,k+K} = f_{ocv}(SOC_k) - \dfrac{\eta K \Delta t \kappa I_{k+1}}{3600 C_a} \\ V_{p,k+K} = \underbrace{V_{p,k} \exp\left(\dfrac{-K\Delta t}{\tau}\right)}_{V_{p,k+K}^{relax}} + \underbrace{I_{k+1} \overbrace{R_1 \left(1 - \exp\left(\dfrac{-K\Delta t}{\tau}\right)\right)}^{R_{1,k+K}}}_{V_{p,k+K}^{load}} \\ V_{ohmic,k+K} = I_{k+1} R_0 \end{cases} \quad (4)$$

where the OCV-SOC curve is characterized using a nonlinear function $f_{ocv}(\cdot)$, and $\kappa$ denotes the slope of this curve. Based on the analysis on battery transient dynamics, we separately consider the polarization voltage $V_{p,k+K}$ into the dynamics during relaxation, i.e., $V_{p,k+K}^{relax}$, and the dynamics under load, i.e., $V_{p,k+K}^{load}$, in terms of their current dependency. $R_{1,k+K}$ denotes the battery polarization resistance at time $k+K$. $I_{k+1} = I_{k+2} = \cdots = I_{k+K}$ represent the constant peak current within a prediction window. The battery peak current, terminal voltage and SOC should be limited within the pre-defined SOA throughout the prediction window, as described in Eq. (5).

$$\begin{cases} V_{t,\min} \leq V_{t,k+j} \leq V_{t,\max} \\ I_{\max,dis} \leq I_{k+j} \leq I_{\max,chg} \\ SOC_{\min} \leq SOC_{k+j} \leq SOC_{\max} \end{cases} \quad (5)$$
$$\forall j = 1,...,K$$

where $V_{t,\min/\max}$ denotes the lower or upper cut-off voltage, $I_{\max,dis/chg}$ denotes the maximum allowable current for discharging or charging, respectively, given by battery manufacturers (defined as negative for charging and positive for discharging), $SOC_{\min/\max}$ denotes the minimum or maximum battery SOC. Eqs. (3)-(5) allow for the mathematical derivations of reference SOP under each operational constraint, as listed in Table 6. Building on this, we carry out a set of case-by-case analysis to unveil the impact of each error on battery SOP estimation. It is worth noting that we define the SOP error as a difference between the estimation value and the reference value, i.e., $\Delta SOP = SOP - S\hat{O}P$.

**Table 6**
Reference SOP under the operational constraints of current, voltage and SOC.

| Current-constraint SOP estimation |
|---|

| | Current-constraint SOP estimation | |
|---|---|---|
| Peak current: | $I_{max,k}^{dis/chg,i} = I_{max,dis}$ or $I_{max,chg}$ | (6.a) |
| Terminal voltage: | $V_{t,k+K} = f_{ocv}(SOC_k) - V_{p,k+K}^{relax} - I_{max,k}^{dis/chg,i}\left(\dfrac{\eta K \Delta t \kappa}{3600C_a} + R_0 + R_{1,k+K}\right)$ | (6.b) |
| Reference SOP: | $SOP_k = I_{max,k}^{dis/chg,i}\left(f_{ocv}(SOC_k) - V_{p,k+K}^{relax}\right) - \left(I_{max,k}^{dis/chg,i}\right)^2 \left(\dfrac{\eta K \Delta t \kappa}{3600C_a} + R_0 + R_{1,k+K}\right)$ | (6.c) |
| | **Voltage-constraint SOP estimation** | |
| Peak current: | $I_{max,k}^{dis/chg,v} = \dfrac{f_{ocv}(SOC_k) - V_{p,k+K}^{relax} - V_{t,min/max}}{\dfrac{\eta K \Delta t \kappa}{3600C_a} + R_0 + R_{1,k+K}}$ | (7.a) |
| Terminal voltage: | $V_{t,k+K} = V_{t,min/max}$ | (7.b) |
| Reference SOP: | $SOP_k = \dfrac{V_{t,min/max}\left(f_{ocv}(SOC_k) - V_{p,k+K}^{relax} - V_{t,min/max}\right)}{\dfrac{\eta K \Delta t \kappa}{3600C_a} + R_0 + R_{1,k+K}}$ | (7.c) |
| | **SOC-constraint SOP estimation** | |
| Peak current: | $I_{max,k}^{dis/chg,soc} = \dfrac{SOC_k - SOC_{min/max}}{\eta K \Delta t / 3600C_a}$ | (8.a) |
| Terminal voltage: | $V_{t,k+K} = f_{ocv}(SOC_{min/max}) - V_{p,k+K}^{relax} - \dfrac{(SOC_k - SOC_{min/max})(R_0 + R_{1,k+K})}{\eta K \Delta t / 3600C_a}$ | (8.b) |
| Reference SOP: | $SOP_k = \dfrac{(SOC_k - SOC_{min/max})\left(f_{ocv}(SOC_{min/max}) - V_{p,k+K}^{relax}\right)}{\eta K \Delta t / 3600C_a} - \left(\dfrac{SOC_k - SOC_{min/max}}{\eta K \Delta t / 3600C_a}\right)^2 (R_0 + R_{1,k+K})$ | (8.c) |

# Appendix II. Error analysis of various contributors

## Table 7

Analytical expressions of SOP errors that occur in the presence of a SOC error.

| | Current-constraint SOP estimation | |
|---|---|---|
| Peak current error: | $\Delta I_{max,k}^{dis/chg,i} = 0$ | (9.a) |
| Terminal voltage error: | $\Delta V_{t,k+K} = \Delta V_{oc,k} = \kappa \Delta SOC_k$ | (9.b) |
| SOP error: | $\Delta SOP_k = \kappa I_{max,k}^{dis/chg,i} \Delta SOC_k$ | (9.c) |
| | **Voltage-constraint SOP estimation** | |
| Peak current error: | $\Delta I_{max,k}^{dis/chg,v} = \dfrac{\kappa \Delta SOC_k}{\dfrac{\eta K \Delta t \kappa}{3600C_a} + R_0 + R_{1,k+K}}$ | (10.a) |
| Terminal voltage error: | $\Delta V_{t,k+K} = 0$ | (10.b) |
| SOP error: | $\Delta SOP_k = \dfrac{\kappa V_{t,\min/\max}\Delta SOC_k}{\dfrac{\eta K \Delta t \kappa}{3600 C_a}+R_0+R_{1,k+K}}$ | (10.c) |
| **SOC-constraint SOP estimation** | | |
| Peak current error: | $\Delta I_{\max/\min,k}^{dis/chg,soc} = \dfrac{\Delta SOC_k}{\eta K \Delta t / 3600 C_a}$ | (11.a) |
| Terminal voltage error: | $\Delta V_{t,k+K} = -\Delta I_{\max/\min,k}^{dis/chg,soc}\left(R_0+R_{1,k+K}\right)$ | (11.b) |
| | $\Delta SOP_k = a\Delta SOC_k^2 + b\Delta SOC_k$ | (11.c) |
| | with | |
| SOP error: | $\begin{cases} a = \dfrac{R_0+R_{1,k+K}}{\left(\eta K \Delta t / 3600 C_a\right)^2} \\ b = \dfrac{f_{ocv}(SOC_{\min/\max})-V_{p,k+K}^{relax}}{\eta K \Delta t / 3600 C_a} - \dfrac{2(SOC_k-SOC_{\min/\max})(R_0+R_{1,k+K})}{\left(\eta K \Delta t / 3600 C_a\right)^2} \end{cases}$ | |

**Table 8**

Analytical expressions of SOP errors that occur in the presence of an error arising from the description of battery polarization dynamics during relaxation ($\Delta V_p^{relax}$).

| | **Current-constraint SOP estimation** | |
|---|---|---|
| Peak current error: | $\Delta I_{\max,k}^{dis/chg,i} = 0$ | (12.a) |
| Voltage error: | $\Delta V_{t,k+K} = -\Delta V_{p,k+K}^{relax}$ | (12.b) |
| SOP error: | $\Delta SOP_k = -I_{\max,k}^{dis/chg,i}\Delta V_{p,k+K}^{relax}$ | (12.c) |
| | **Voltage-constraint SOP estimation** | |
| Peak current error: | $\Delta I_{\max,k}^{dis/chg,v} = \dfrac{\Delta V_{p,k+K}^{relax}}{\dfrac{\eta K \Delta t \kappa}{3600 C_a}+R_0+R_{1,k+K}}$ | (13.a) |
| Voltage error: | $\Delta V_{t,k+K} = 0$ | (13.b) |
| SOP error: | $\Delta SOP_k = \dfrac{V_{t,\min/\max}\Delta V_{p,k+K}^{relax}}{\dfrac{\eta K \Delta t \kappa}{3600 C_a}+R_0+R_{1,k+K}}$ | (13.c) |
| | **SOC-constraint SOP estimation** | |
| Peak current error: | $\Delta I_{\max,k}^{dis/chg,soc} = 0$ | (14.a) |
| Voltage error: | $\Delta V_{t,k+K} = \Delta V_{p,k+K}^{relax}$ | (14.b) |
| SOP error: | $\Delta SOP_k = \dfrac{(SOC_k-SOC_{\min/\max})}{\eta K \Delta t / 3600 C_a}\Delta V_{p,k+K}^{relax}$ | (14.c) |

**Table 9**

Analytical expressions of SOP errors that occur in the presence of an error arising from the characterization of the sum of internal resistance ($\Delta R_{sum}$).

| | **Current-constraint SOP estimation** | |
|---|---|---|
| Peak current error: | $\Delta I_{max,k}^{dis/chg,i} = 0$ | (15.a) |
| Voltage error: | $\Delta V_{t,k+K} = -I_{max,k}^{dis/chg,i} \Delta R_{sum}$ | (15.b) |
| SOP error: | $\Delta SOP_k = -\left(I_{max,k}^{dis/chg,i}\right)^2 \Delta R_{sum}$ | (15.c) |
| | **Voltage-constraint SOP estimation** | |
| Peak current error: | $\dfrac{\Delta I_{max,k}^{dis/chg,v}}{I_{max,k}^{dis/chg,v}} = \dfrac{-\Delta R_{sum}}{\dfrac{\eta K \Delta t \kappa}{3600 C_a} + R_0 + R_{1,k+K} - \Delta R_{sum}}$ | (16.a) |
| Voltage error: | $\Delta V_{t,k+K} = 0$ | (16.b) |
| SOP error: | $\Delta SOP_k = \dfrac{f_{ocv}(SOC_k) - V_{p,k+K}^{relax} - V_{t,\min/\max}}{\dfrac{\eta K \Delta t \kappa}{3600 C_a} + R_0 + R_{1,k+K}} \times \dfrac{-V_{t,\min/\max} \Delta R_{sum}}{\dfrac{\eta K \Delta t \kappa}{3600 C_a} + R_0 + R_{1,k+K} - \Delta R_{sum}}$ | (16.c) |
| | **SOC-constraint SOP estimation** | |
| Peak current error: | $\Delta I_{max,k}^{dis/chg,soc} = 0$ | (17.a) |
| Voltage error: | $\Delta V_{t,k+K} = -\dfrac{(SOC_k - SOC_{\min/\max}) \Delta R_{sum}}{\eta K \Delta t / 3600 C_a}$ | (17.b) |
| SOP error: | $\Delta SOP_k = -\left(\dfrac{SOC_k - SOC_{\min/\max}}{\eta K \Delta t / 3600 C_a}\right)^2 \Delta R_{sum}$ | (17.c) |

**Table 10**

Analytical expressions of SOP errors that occur in the presence of an OCV slope error ($\Delta \kappa$).

| | **Current-constraint SOP estimation** | |
|---|---|---|
| Peak current error: | $\Delta I_{max,k}^{dis/chg,i} = 0$ | (18.a) |
| Voltage error: | $\Delta V_{t,k+K} = -I_{max,k}^{dis/chg,i} \dfrac{\eta K \Delta t \Delta \kappa}{3600 C_a}$ | (18.b) |
| SOP error: | $\Delta SOP_k = -\left(I_{max,k}^{dis/chg,i}\right)^2 \dfrac{\eta K \Delta t \Delta \kappa}{3600 C_a}$ | (18.c) |
| | **Voltage-constraint SOP estimation** | |
| Peak current error: | $\dfrac{\Delta I_{max,k}^{dis/chg,v}}{I_{max,k}^{dis/chg,v}} = \dfrac{-\eta K \Delta t \Delta \kappa / 3600 C_a}{\dfrac{\eta K \Delta t (\kappa - \Delta \kappa)}{3600 C_a} + R_0 + R_{1,k+K}}$ | (19.a) |
| Voltage error: | $\Delta V_{t,k+K} = 0$ | (19.b) |
| SOP error: | $\Delta SOP_k = \dfrac{f_{ocv}(SOC_k) - V_{p,k+K}^{relax} - V_{t,\min/\max}}{\dfrac{\eta K \Delta t \kappa}{3600 C_a} + R_0 + R_{1,k+K}}$ $\times \dfrac{-V_{t,\min/\max} \eta K \Delta t \Delta\kappa / 3600 C_a}{\dfrac{\eta K \Delta t (\kappa - \Delta\kappa)}{3600 C_a} + R_0 + R_{1,k+K}}$ | (19.c) |
| **SOC-constraint SOP estimation** | | |
| Peak current error: | $\Delta I_{\max,k}^{dis/chg,soc} = 0$ | (20.a) |
| Voltage error: | $\Delta V_{t,k+K} = \Delta\kappa (SOC_k - SOC_{\min/\max})$ | (20.b) |
| SOP error: | $\Delta SOP_k = \dfrac{(SOC_k - SOC_{\min/\max})^2}{\eta K \Delta t / 3600 C_a} \Delta\kappa$ | (20.c) |

**Table 11**

Analytical expressions of SOP errors that occur in the presence of an error arising from available capacity and coulombic efficiency ($\Delta x$).

| | | |
|---|---|---|
| | **Current-constraint SOP estimation** | |
| Peak current error: | $\Delta I_{\max,k}^{dis/chg,i} = 0$ | (21.a) |
| Voltage error: | $\Delta V_{t,k+K} = -I_{\max,k}^{dis/chg,i} K \Delta t \kappa \Delta x$ | (21.b) |
| SOP error: | $\Delta SOP_k = -\left(I_{\max,k}^{dis/chg,i}\right)^2 K \Delta t \kappa \Delta x$ | (21.c) |
| | **Voltage-constraint SOP estimation** | |
| Peak current error: | $\dfrac{\Delta I_{\max,k}^{dis/chg,v}}{I_{\max,k}^{dis/chg,v}} = \dfrac{-K \Delta t \kappa \Delta x}{K \Delta t \kappa (x - \Delta x) + R_0 + R_{1,k+K}}$ | (22.a) |
| Voltage error: | $\Delta V_{t,k+K} = 0$ | (22.b) |
| SOP error: | $\Delta SOP_k = \dfrac{f_{ocv}(SOC_k) - V_{p,k+K}^{relax} - V_{t,\min/\max}}{K \Delta t \kappa x + R_0 + R_{1,k+K}}$ $\times \dfrac{-V_{t,\min/\max} K \Delta t \kappa \Delta x}{K \Delta t \kappa (x - \Delta x) + R_0 + R_{1,k+K}}$ | (22.c) |
| | **SOC-constraint SOP estimation** | |
| Peak current error: | $\dfrac{\Delta I_{\max,k}^{dis/chg,soc}}{I_{\max,k}^{dis/chg,soc}} = \dfrac{-\Delta x}{x - \Delta x}$ | (23.a) |
| Voltage error: | $\Delta V_{t,k+K} = -\Delta I_{\max,k}^{dis/chg,soc} (R_0 + R_{1,k+K})$ | (23.b) |
| SOP error: | $\Delta SOP_k = \alpha \dfrac{\Delta x}{x(x - \Delta x)} + \beta \dfrac{2x\Delta x - \Delta x^2}{x^2 (x - \Delta x)^2}$ with | (23.c) |

$$\begin{cases} \alpha = -\dfrac{(SOC_k - SOC_{\min/\max})(f_{ocv}(SOC_{\min/\max}) - V_{p,k+K}^{relax})}{K\Delta t} \\ \beta = \left(\dfrac{SOC_k - SOC_{\min/\max}}{K\Delta t}\right)^2 (R_0 + R_{1,k+K}) \end{cases}$$

**Declaration of competing interest**

The authors declare that they have no known competing financial interests or personal relationships that could have appeared to influence the work reported in this paper.

**Acknowledgement**

This research is supported by an Australian government research training program scholarship offered to the first author of this study.

**References**

[1]   J. Tian, C. Chen, W. Shen, F. Sun, R. Xiong, Deep Learning Framework for Lithium-ion Battery State of Charge Estimation: Recent Advances and Future Perspectives, Energy Storage Mater. 61 (2023) 102883. https://doi.org/10.1016/j.ensm.2023.102883.

[2]   R. Guo, W. Shen, A Review of Equivalent Circuit Model Based Online State of Power Estimation for Lithium-Ion Batteries in Electric Vehicles, Vehicles 4 (2021) 1–31. https://doi.org/10.3390/vehicles4010001.

[3]   X. Hu, C. Zou, C. Zhang, Y. Li, Technological developments in batteries: a survey of principal roles, types, and management needs, IEEE Power Energy Mag. 15 (2017) 20–31.

[4]   R. Xiong, Y. Pan, W. Shen, H. Li, F. Sun, Lithium-ion battery aging mechanisms and diagnosis method for automotive applications: Recent advances and perspectives, Renew. Sustain. Energy Rev. 131 (2020) 110048.

[5]   N. Rietmann, B. Hügler, T. Lieven, Forecasting the trajectory of electric vehicle sales and the consequences for worldwide CO2 emissions, J. Clean. Prod. 261 (2020) 121038.

[6]   R. Guo, F. Wang, M. Akbar Rhamdhani, Y. Xu, W. Shen, Managing the surge: A comprehensive review of the entire disposal framework for retired lithium-ion batteries from electric vehicles, J. Energy Chem. 92 (2024) 648–680. https://doi.org/10.1016/j.jechem.2024.01.055.

[7]   X. Hu, K. Zhang, K. Liu, X. Lin, S. Dey, S. Onori, Advanced Fault Diagnosis for Lithium-Ion Battery Systems: A Review of Fault Mechanisms, Fault Features, and Diagnosis Procedures, IEEE Ind. Electron. Mag. 14 (2020) 65–91. https://doi.org/10.1109/MIE.2020.2964814.

[8]  Y. Xu, X. Ge, R. Guo, W. Shen, Online Soft Short Circuit Diagnosis of Electric Vehicle Li-ion Batteries Based on Constant Voltage Charging Current, IEEE Trans. Transp. Electrification (2022).

[9]  Y. Xu, X. Ge, R. Guo, W. Shen, Recent Advances in Model-Based Fault Diagnosis for Lithium-Ion Batteries: A Comprehensive Review, ArXiv Prepr. ArXiv240116682 (2024).

[10] X. Hu, F. Feng, K. Liu, L. Zhang, J. Xie, B. Liu, State estimation for advanced battery management: Key challenges and future trends, Renew. Sustain. Energy Rev. 114 (2019) 109334. https://doi.org/10.1016/j.rser.2019.109334.

[11] J. Tian, R. Xiong, W. Shen, State-of-Health Estimation Based on Differential Temperature for Lithium Ion Batteries, IEEE Trans. Power Electron. 35 (2020) 10363–10373. https://doi.org/10.1109/TPEL.2020.2978493.

[12] R. Guo, W. Shen, An Information Analysis Based Online Parameter Identification Method for Lithium-ion Batteries in Electric Vehicles, IEEE Trans. Ind. Electron. (2023).

[13] J. Tian, R. Xiong, W. Shen, J. Lu, F. Sun, Flexible battery state of health and state of charge estimation using partial charging data and deep learning, Energy Storage Mater. 51 (2022) 372–381. https://doi.org/10.1016/j.ensm.2022.06.053.

[14] R. Xiong, J. Tian, W. Shen, J. Lu, F. Sun, Semi-supervised estimation of capacity degradation for lithium ion batteries with electrochemical impedance spectroscopy, J. Energy Chem. 76 (2023) 404–413. https://doi.org/10.1016/j.jechem.2022.09.045.

[15] C. Chen, R. Xiong, R. Yang, W. Shen, F. Sun, State-of-charge estimation of lithium-ion battery using an improved neural network model and extended Kalman filter, J. Clean. Prod. 234 (2019) 1153–1164. https://doi.org/10.1016/j.jclepro.2019.06.273.

[16] R. Guo, Y. Xu, C. Hu, W. Shen, A Curve Relocation Approach for Robust Battery Open Circuit Voltage Reconstruction and Capacity Estimation Based on Partial Charging Data, IEEE Trans. Power Electron. (2023) 1–15. https://doi.org/10.1109/TPEL.2023.3347236.

[17] M.A. Hannan, M.S.H. Lipu, A. Hussain, A. Mohamed, A review of lithium-ion battery state of charge estimation and management system in electric vehicle applications: Challenges and recommendations, Renew. Sustain. Energy Rev. 78 (2017) 834–854. https://doi.org/10.1016/j.rser.2017.05.001.

[18] Y. Zheng, M. Ouyang, X. Han, L. Lu, J. Li, Investigating the error sources of the online state of charge estimation methods for lithium-ion batteries in electric vehicles, J. Power Sources 377 (2018) 161–188. https://doi.org/10.1016/j.jpowsour.2017.11.094.

[19] K.L. Quade, D. Jöst, D.U. Sauer, W. Li, Understanding the Energy Potential of Lithium-Ion Batteries: Definition and Estimation of the State of Energy, Batter. Supercaps n/a (n.d.) e202300152. https://doi.org/10.1002/batt.202300152.

[20] R. Xiong, L. Li, J. Tian, Towards a smarter battery management system: A critical review on battery state of health monitoring methods, J. Power Sources 405 (2018) 18–29. https://doi.org/10.1016/j.jpowsour.2018.10.019.

[21] Y. Wang, J. Tian, Z. Sun, L. Wang, R. Xu, M. Li, Z. Chen, A comprehensive review of battery modeling and state estimation approaches for advanced battery management systems, Renew. Sustain. Energy Rev. 131 (2020) 110015. https://doi.org/10.1016/j.rser.2020.110015.

[22] PNGV battery test manual, (1997). https://doi.org/10.2172/578702.

[23] A. Farmann, D.U. Sauer, A comprehensive review of on-board State-of-Available-Power prediction techniques for lithium-ion batteries in electric vehicles, J. Power Sources 329 (2016) 123–137. https://doi.org/10.1016/j.jpowsour.2016.08.031.

[24] L. Yang, Y. Cai, Y. Yang, Z. Deng, Supervisory long-term prediction of state of available power for lithium-ion batteries in electric vehicles, Appl. Energy 257 (2020) 114006. https://doi.org/10.1016/j.apenergy.2019.114006.

[25] C. Zou, A. Klintberg, Z. Wei, B. Fridholm, T. Wik, B. Egardt, Power capability prediction for lithium-ion batteries using economic nonlinear model predictive control, J. Power Sources 396 (2018) 580–589. https://doi.org/10.1016/j.jpowsour.2018.06.034.

[26] D. Huang, Z. Chen, S. Zhou, Model prediction-based battery-powered heating method for series-connected lithium-ion battery pack working at extremely cold temperatures, Energy 216 (2021) 119236. https://doi.org/10.1016/j.energy.2020.119236.

[27] H. Ruan, J. Jiang, B. Sun, X. Su, X. He, K. Zhao, An optimal internal-heating strategy for lithium-ion batteries at low temperature considering both heating time and lifetime reduction, Appl. Energy 256 (2019) 113797. https://doi.org/10.1016/j.apenergy.2019.113797.

[28] C. Zou, X. Hu, Z. Wei, X. Tang, Electrothermal dynamics-conscious lithium-ion battery cell-level charging management via state-monitored predictive control, Energy 141 (2017) 250–259. https://doi.org/10.1016/j.energy.2017.09.048.

[29] M. Xu, R. Wang, P. Zhao, X. Wang, Fast charging optimization for lithium-ion batteries based on dynamic programming algorithm and electrochemical-thermal-capacity fade coupled model, J. Power Sources 438 (2019) 227015. https://doi.org/10.1016/j.jpowsour.2019.227015.

[30] S. Bharathraj, S.P. Adiga, A. Kaushik, K.S. Mayya, M. Lee, Y. Sung, Towards in-situ detection of nascent short circuits and accurate estimation of state of short in Lithium-Ion Batteries, J. Power Sources 520 (2022) 230830. https://doi.org/10.1016/j.jpowsour.2021.230830.

[31] J. Shin, I. Joe, S. Hong, A State of Power Based Deep Learning Model for State of Health Estimation of Lithium-Ion Batteries, in: R. Silhavy, P. Silhavy, Z. Prokopova (Eds.), Data Sci. Intell. Syst., Springer International Publishing, Cham, 2021: pp. 922–931. https://doi.org/10.1007/978-3-030-90321-3_77.

[32] M. Wu, L. Qin, G. Wu, State of power estimation of power lithium-ion battery based on an equivalent circuit model, J. Energy Storage 51 (2022) 104538. https://doi.org/10.1016/j.est.2022.104538.

[33] L. Lu, X. Han, J. Li, J. Hua, M. Ouyang, A review on the key issues for lithium-ion battery management in electric vehicles, J. Power Sources 226 (2013) 272–288. https://doi.org/10.1016/j.jpowsour.2012.10.060.

[34] P. Arora, R.E. White, M. Doyle, Capacity Fade Mechanisms and Side Reactions in Lithium-Ion Batteries, J. Electrochem. Soc. 145 (1998) 3647–3667. https://doi.org/10.1149/1.1838857.

[35] T.M. Bandhauer, S. Garimella, T.F. Fuller, A Critical Review of Thermal Issues in Lithium-Ion Batteries, J. Electrochem. Soc. 158 (2011) R1. https://doi.org/10.1149/1.3515880.

[36] R. Guo, W. Shen, Lithium-Ion Battery State of Charge and State of Power Estimation Based on a Partial-Adaptive Fractional-Order Model in Electric Vehicles, IEEE Trans. Ind. Electron. 70 (2022) 10123–10133.

[37] J.G. Qu, Z.Y. Jiang, J.F. Zhang, Investigation on lithium-ion battery degradation induced by combined effect of current rate and operating temperature during fast charging, J. Energy Storage 52 (2022) 104811. https://doi.org/10.1016/j.est.2022.104811.

[38] G. Ning, B. Haran, B.N. Popov, Capacity fade study of lithium-ion batteries cycled at high discharge rates, J. Power Sources 117 (2003) 160–169. https://doi.org/10.1016/S0378-7753(03)00029-6.

[39] J. Jiang, W. Shi, J. Zheng, P. Zuo, J. Xiao, X. Chen, W. Xu, J.-G. Zhang, Optimized Operating Range for Large-Format LiFePO$_4$/Graphite Batteries, J. Electrochem. Soc. 161 (2014) A336–A341. https://doi.org/10.1149/2.052403jes.

[40] S. Watanabe, M. Kinoshita, T. Hosokawa, K. Morigaki, K. Nakura, Capacity fading of LiAlyNi1−x−yCoxO2 cathode for lithium-ion batteries during accelerated calendar and cycle life tests (effect of depth of discharge in charge–discharge cycling on the suppression of the micro-crack generation of LiAlyNi1−x−yCoxO2 particle), J. Power Sources 260 (2014) 50–56. https://doi.org/10.1016/j.jpowsour.2014.02.103.

[41] M. Ecker, N. Nieto, S. Käbitz, J. Schmalstieg, H. Blanke, A. Warnecke, D.U. Sauer, Calendar and cycle life study of Li(NiMnCo)O2-based 18650 lithium-ion batteries, J. Power Sources 248 (2014) 839–851. https://doi.org/10.1016/j.jpowsour.2013.09.143.

[42] Y. Wang, R. Pan, C. Liu, Z. Chen, Q. Ling, Power capability evaluation for lithium iron phosphate batteries based on multi-parameter constraints estimation, J. Power Sources 374 (2018) 12–23. https://doi.org/10.1016/j.jpowsour.2017.11.019.

[43] W. Zhang, W. Shi, Z. Ma, Adaptive unscented Kalman filter based state of energy and power capability estimation approach for lithium-ion battery, J. Power Sources 289 (2015) 50–62. https://doi.org/10.1016/j.jpowsour.2015.04.148.

[44] R. Guo, W. Shen, An enhanced multi-constraint state of power estimation algorithm for lithium-ion batteries in electric vehicles, J. Energy Storage 50 (2022) 104628. https://doi.org/10.1016/j.est.2022.104628.

[45] M. Richard, J. Dahn, Accelerating rate calorimetry study on the thermal stability of lithium intercalated graphite in electrolyte. II. Modeling the results and predicting differential scanning calorimeter curves, J. Electrochem. Soc. 146 (1999) 2078.

[46] D. Mishra, Simulation Modelling of Thermal Runaway Propagation in Li Ion Batteries, The University of Texas at Arlington, 2019.

[47] J. Fan, S. Tan, Studies on charging lithium-ion cells at low temperatures, J. Electrochem. Soc. 153 (2006) A1081.

[48] S. Abada, G. Marlair, A. Lecocq, M. Petit, V. Sauvant-Moynot, F. Huet, Safety focused modeling of lithium-ion batteries: A review, J. Power Sources 306 (2016) 178–192. https://doi.org/10.1016/j.jpowsour.2015.11.100.

[49] P. Qin, Y. Che, H. Li, Y. Cai, M. Jiang, Joint SOC–SOP estimation method for lithium-ion batteries based on electro-thermal model and multi-parameter constraints, J. Power Electron. 22 (2022) 490–502. https://doi.org/10.1007/s43236-021-00376-9.

[50] H. Perez, N. Shahmohammadhamedani, S. Moura, Enhanced Performance of Li-Ion Batteries via Modified Reference Governors and Electrochemical Models, IEEEASME Trans. Mechatron. 20 (2015) 1511–1520. https://doi.org/10.1109/TMECH.2014.2379695.

[51] M.F. Niri, T.Q. Dinh, T.F. Yu, J. Marco, T.M.N. Bui, State of Power Prediction for Lithium-Ion Batteries in Electric Vehicles via Wavelet-Markov Load Analysis, IEEE Trans. Intell. Transp. Syst. (2020) 1–16. https://doi.org/10.1109/TITS.2020.3028024.

[52] M.S. Islam, C.A. Fisher, Lithium and sodium battery cathode materials: computational insights into voltage, diffusion and nanostructural properties, Chem. Soc. Rev. 43 (2014) 185–204.

[53] S.S. Zhang, The effect of the charging protocol on the cycle life of a Li-ion battery, J. Power Sources 161 (2006) 1385–1391.

[54] N.A. Chaturvedi, R. Klein, J. Christensen, J. Ahmed, A. Kojic, Algorithms for Advanced Battery-Management Systems, IEEE Control Syst. Mag. 30 (2010) 49–68. https://doi.org/10.1109/MCS.2010.936293.

[55] K. Smith, C.-Y. Wang, Power and thermal characterization of a lithium-ion battery pack for hybrid-electric vehicles, J. Power Sources 160 (2006) 662–673. https://doi.org/10.1016/j.jpowsour.2006.01.038.

[56] K. Smith, C.-Y. Wang, Solid-state diffusion limitations on pulse operation of a lithium ion cell for hybrid electric vehicles, J. Power Sources 161 (2006) 628–639. https://doi.org/10.1016/j.jpowsour.2006.03.050.

[57] T.-K. Lee, Y. Kim, A. Stefanopoulou, Z.S. Filipi, Hybrid electric vehicle supervisory control design reflecting estimated lithium-ion battery electrochemical dynamics, in: Proc. 2011 Am. Control Conf., 2011: pp. 388–395. https://doi.org/10.1109/ACC.2011.5990985.

[58] L. Zheng, D.D.-C. Lu, Lithium-ion Battery Instantaneous Available Power Prediction Using Surface Lithium Concentration of Solid Particles in a Simplified Electrochemical Model, IEEE Trans. POWER Electron. 33 (2018) 10.

[59] X. Sun, N. Xu, Q. Chen, J. Yang, J. Zhu, J. Xu, L. Zheng, State of Power Capability Prediction of Lithium-Ion Battery from the Perspective of Electrochemical Mechanisms Considering Temperature Effect, IEEE Trans. Transp. Electrification (2022) 1–1. https://doi.org/10.1109/TTE.2022.3206452.

[60] K.A. Smith, C.D. Rahn, C.-Y. Wang, Model-Based Electrochemical Estimation and Constraint Management for Pulse Operation of Lithium Ion Batteries, IEEE Trans. Control Syst. Technol. 18 (2010) 654–663. https://doi.org/10.1109/TCST.2009.2027023.

[61] W. Li, Y. Fan, F. Ringbeck, D. Jöst, D.U. Sauer, Unlocking electrochemical model-based online power prediction for lithium-ion batteries via Gaussian process regression, Appl. Energy 306 (2022) 118114. https://doi.org/10.1016/j.apenergy.2021.118114.

[62] Y. Gu, Y. Chen, J. Wang, W. Xiao, Q. Chen, Enhancing Dispatchability of Lithium-ion Battery Sources in Integrated Energy-Transportation Systems with Feasible Power Characterization, IEEE Trans. Ind. Inform. (2022) 1–10. https://doi.org/10.1109/TII.2022.3195731.

[63] R. Guo, W. Shen, A model fusion method for online state of charge and state of power co-estimation of lithium-ion batteries in electric vehicles, IEEE Trans. Veh. Technol. 71 (2022) 11515–11525.

[64] R. Guo, C. Hu, W. Shen, An electric vehicle-oriented approach for battery multi-constraint state of power estimation under constant power operations, IEEE Trans. Veh. Technol. (2023).

[65] W. Waag, C. Fleischer, D.U. Sauer, Adaptive on-line prediction of the available power of lithium-ion batteries, J. Power Sources 242 (2013) 548–559. https://doi.org/10.1016/j.jpowsour.2013.05.111.

[66] F. Sun, R. Xiong, H. He, W. Li, J.E.E. Aussems, Model-based dynamic multi-parameter method for peak power estimation of lithium–ion batteries, Appl. Energy (2012) 9.

[67] S. Wang, M. Verbrugge, J.S. Wang, P. Liu, Power prediction from a battery state estimator that incorporates diffusion resistance, J. Power Sources 214 (2012) 399–406. https://doi.org/10.1016/j.jpowsour.2012.04.070.

[68] R. Xiong, H. He, F. Sun, K. Zhao, Online Estimation of Peak Power Capability of Li-Ion Batteries in Electric Vehicles by a Hardware-in-Loop Approach, Energies 5 (2012) 1455–1469. https://doi.org/10.3390/en5051455.

[69] S. Wang, M. Verbrugge, J.S. Wang, P. Liu, Multi-parameter battery state estimator based on the adaptive and direct solution of the governing differential equations, J. Power Sources 196 (2011) 8735–8741.

[70] L. Pei, C. Zhu, T. Wang, R. Lu, C.C. Chan, Online peak power prediction based on a parameter and state estimator for lithium-ion batteries in electric vehicles, Energy 66 (2014) 766–778. https://doi.org/10.1016/j.energy.2014.02.009.

[71] R. Guo, W. Shen, Online state of charge and state of power co-estimation of lithium-ion batteries based on fractional-order calculus and model predictive control theory, Appl. Energy 327 (2022) 120009. https://doi.org/10.1016/j.apenergy.2022.120009.

[72] B. Xiong, S. Dong, Y. Li, J. Tang, Y. Su, H.B. Gooi, Peak Power Estimation of Vanadium Redox Flow Batteries Based on Receding Horizon Control, IEEE J. Emerg. Sel. Top. Power Electron. (2022) 1–1. https://doi.org/10.1109/JESTPE.2022.3152588.

[73] M.J. Esfandyari, V. Esfahanian, M.R. Hairi Yazdi, H. Nehzati, O. Shekoofa, A new approach to consider the influence of aging state on Lithium-ion battery state of power estimation for hybrid electric vehicle, Energy 176 (2019) 505–520. https://doi.org/10.1016/j.energy.2019.03.176.

[74] Z. Rahman, K.L. Butler, M. Ehsani, Effect of extended-speed, constant-power operation of electric drives on the design and performance of EV-HEV propulsion system, SAE Technical Paper, 2000.

[75] K.M. Rahman, M. Ehsani, Performance analysis of electric motor drives for electric and hybrid electric vehicle applications, in: Power Electron. Transp., IEEE, 1996: pp. 49–56.

[76] M. Anun, M. Ordonez, I.G. Zurbriggen, G.G. Oggier, Circular switching surface technique: High-performance constant power load stabilization for electric vehicle systems, IEEE Trans. Power Electron. 30 (2014) 4560–4572.

[77] Z. Cao, H. Wen, Q. Bu, H. Shi, P. Xu, Y. Yang, Y. Du, Constant Power Load Stabilization with Fast Transient Boundary Control for DAB Converters-based Electric Drive Systems, IEEE Trans. Ind. Electron. (2023).

[78] B. Chen, S.A. Evangelou, R. Lot, Hybrid Electric Vehicle Two-Step Fuel Efficiency Optimization With Decoupled Energy Management and Speed Control, IEEE Trans. Veh. Technol. 68 (2019) 11492–11504. https://doi.org/10.1109/TVT.2019.2948192.

[79] L. Zhang, X. Ye, X. Xia, F. Barzegar, A real-time energy management and speed controller for an electric vehicle powered by a hybrid energy storage system, IEEE Trans. Ind. Inform. 16 (2020) 6272–6280.

[80] R. Guo, C. Hu, W. Shen, Battery Peak Power Assessment under Various Operational Scenarios: A Comparative Study, (2024).

[81] F. Brosa Planella, W. Ai, A.M. Boyce, A. Ghosh, I. Korotkin, S. Sahu, V. Sulzer, R. Timms, T.G. Tranter, M. Zyskin, S.J. Cooper, J.S. Edge, J.M. Foster, M.

Marinescu, B. Wu, G. Richardson, A continuum of physics-based lithium-ion battery models reviewed, Prog. Energy 4 (2022) 042003. https://doi.org/10.1088/2516-1083/ac7d31.

[82] M. Doyle, T.F. Fuller, J. Newman, Modeling of galvanostatic charge and discharge of the lithium/polymer/insertion cell, J. Electrochem. Soc. 140 (1993) 1526.

[83] T.F. Fuller, M. Doyle, J. Newman, Simulation and optimization of the dual lithium ion insertion cell, J. Electrochem. Soc. 141 (1994) 1.

[84] S. Santhanagopalan, Q. Guo, P. Ramadass, R.E. White, Review of models for predicting the cycling performance of lithium ion batteries, J. Power Sources 156 (2006) 620–628.

[85] K.A. Smith, C.D. Rahn, C.-Y. Wang, Model order reduction of 1D diffusion systems via residue grouping, (2008).

[86] X. Lai, Y. Zheng, T. Sun, A comparative study of different equivalent circuit models for estimating state-of-charge of lithium-ion batteries, Electrochimica Acta 259 (2018) 566–577. https://doi.org/10.1016/j.electacta.2017.10.153.

[87] L. Zhang, H. Peng, Z. Ning, Z. Mu, C. Sun, Comparative Research on RC Equivalent Circuit Models for Lithium-Ion Batteries of Electric Vehicles, Appl. Sci. 7 (2017) 1002. https://doi.org/10.3390/app7101002.

[88] G.L. Plett, High-Performance Battery-Pack Power Estimation Using a Dynamic Cell Model, IEEE Trans. Veh. Technol. 53 (2004) 1586–1593. https://doi.org/10.1109/TVT.2004.832408.

[89] L. Zheng, J. Zhu, G. Wang, D.D.-C. Lu, P. McLean, T. He, Experimental analysis and modeling of temperature dependence of lithium-ion battery direct current resistance for power capability prediction, in: 2017 20th Int. Conf. Electr. Mach. Syst. ICEMS, IEEE, Sydney, Australia, 2017: pp. 1–4. https://doi.org/10.1109/ICEMS.2017.8056426.

[90] F. Zheng, J. Jiang, B. Sun, W. Zhang, M. Pecht, Temperature dependent power capability estimation of lithium-ion batteries for hybrid electric vehicles, Energy 113 (2016) 64–75. https://doi.org/10.1016/j.energy.2016.06.010.

[91] C. Burgos-Mellado, M.E. Orchard, M. Kazerani, R. Cárdenas, D. Sáez, Particle-filtering-based estimation of maximum available power state in Lithium-Ion batteries, Appl. Energy 161 (2016) 349–363. https://doi.org/10.1016/j.apenergy.2015.09.092.

[92] H. He, R. Xiong, J. Fan, Evaluation of lithium-ion battery equivalent circuit models for state of charge estimation by an experimental approach, Energies 4 (2011) 582–598.

[93] A. Fotouhi, D.J. Auger, K. Propp, S. Longo, M. Wild, A review on electric vehicle battery modelling: From Lithium-ion toward Lithium–Sulphur, Renew. Sustain. Energy Rev. 56 (2016) 1008–1021.

[94] R. Xiong, H. He, F. Sun, X. Liu, Z. Liu, Model-based state of charge and peak power capability joint estimation of lithium-ion battery in plug-in hybrid electric vehicles, J. Power Sources 229 (2013) 159–169. https://doi.org/10.1016/j.jpowsour.2012.12.003.

[95] Z. Song, J. Hou, H.F. Hofmann, X. Lin, J. Sun, Parameter Identification and Maximum Power Estimation of Battery/Supercapacitor Hybrid Energy Storage System Based on Cramer–Rao Bound Analysis, IEEE Trans. Power Electron. 34 (2019) 4831–4843. https://doi.org/10.1109/TPEL.2018.2859317.

[96] X. Hu, R. Xiong, B. Egardt, Model-Based Dynamic Power Assessment of Lithium-Ion Batteries Considering Different Operating Conditions, IEEE Trans. Ind. Inform. 10 (2014) 1948–1959. https://doi.org/10.1109/TII.2013.2284713.

[97] C. Wei, M. Benosman, Extremum seeking-based parameter identification for state-of-power prediction of lithium-ion batteries, in: 2016 IEEE Int. Conf. Renew. Energy Res. Appl. ICRERA, IEEE, Birmingham, United Kingdom, 2016: pp. 67–72. https://doi.org/10.1109/ICRERA.2016.7884376.

[98] S. Xiang, G. Hu, R. Huang, F. Guo, P. Zhou, Lithium-Ion Battery Online Rapid State-of-Power Estimation under Multiple Constraints, Energies 11 (2018) 283. https://doi.org/10.3390/en11020283.

[99] G. Dong, J. Wei, Z. Chen, Kalman filter for onboard state of charge estimation and peak power capability analysis of lithium-ion batteries, J. Power Sources 328 (2016) 615–626. https://doi.org/10.1016/j.jpowsour.2016.08.065.

[100] P. Malysz, J. Ye, R. Gu, H. Yang, A. Emadi, Battery State-of-Power Peak Current Calculation and Verification Using an Asymmetric Parameter Equivalent Circuit Model, IEEE Trans. Veh. Technol. 65 (2016) 11.

[101] S. Rahimifard, R. Ahmed, S. Habibi, Interacting Multiple Model Strategy for Electric Vehicle Batteries State of Charge/Health/ Power Estimation, IEEE Access 9 (2021) 109875–109888. https://doi.org/10.1109/ACCESS.2021.3102607.

[102] X. Tang, Y. Wang, K. Yao, Z. He, F. Gao, Model migration based battery power capability evaluation considering uncertainties of temperature and aging, J. Power Sources 440 (2019) 227141. https://doi.org/10.1016/j.jpowsour.2019.227141.

[103] J. Jiang, S. Liu, Z. Ma, L.Y. Wang, K. Wu, Butler-Volmer equation-based model and its implementation on state of power prediction of high-power lithium titanate batteries considering temperature effects, Energy 117 (2016) 58–72. https://doi.org/10.1016/j.energy.2016.10.087.

[104] W. Zhang, L. Wang, L. Wang, C. Liao, Y. Zhang, Joint State-of-charge and State-of-available-power Estimation based on the Online Parameter Identification of Lithium-ion Battery Model, IEEE Trans. Ind. Electron. (2021).

[105] M. Dubarry, B.Y. Liaw, Development of a universal modeling tool for rechargeable lithium batteries, J. Power Sources 174 (2007) 856–860.

[106] H. He, X. Zhang, R. Xiong, Y. Xu, H. Guo, Online model-based estimation of state-of-charge and open-circuit voltage of lithium-ion batteries in electric vehicles, Energy 39 (2012) 310–318. https://doi.org/10.1016/j.energy.2012.01.009.

[107] Y. Tan, M. Luo, L. She, X. Cui, Joint Estimation of Ternary Lithium-ion Battery State of Charge and State of Power Based on Dual Polarization Model, Int J Electrochem Sci 15 (2020) 20.

[108] P. Shen, M. Ouyang, L. Lu, J. Li, X. Feng, The co-estimation of state of charge, state of health, and state of function for lithium-ion batteries in electric vehicles, IEEE Trans. Veh. Technol. 67 (2017) 92–103.

[109] S. Nejad, D.T. Gladwin, Online Battery State of Power Prediction Using PRBS and Extended Kalman Filter, IEEE Trans. Ind. Electron. 67 (2020) 3747–3755. https://doi.org/10.1109/TIE.2019.2921280.

[110] M.F. Niri, T.Q. Dinh, T.F. Yu, J. Marco, T.M.N. Bui, State of Power Prediction for Lithium-Ion Batteries in Electric Vehicles via Wavelet-Markov Load Analysis, IEEE Trans. Intell. Transp. Syst. (2020).

[111] Y. Tan, M. Luo, L. She, X. Cui, Joint Estimation of Ternary Lithium-ion Battery State of Charge and State of Power Based on Dual Polarization Model, Int J Electrochem Sci 15 (2020) 20.

[112] Y. Wang, Online estimation of battery power state based on improved equivalent circuit model, IOP Conf. Ser. Earth Environ. Sci. 651 (2021) 022080. https://doi.org/10.1088/1755-1315/651/2/022080.

[113] T. Feng, Online identification of lithium-ion battery parameters based on an improved equivalent-circuit model and its implementation on battery state-of-power prediction, J. Power Sources (2015) 12.

[114] F. Liu, D. Yu, W. Su, F. Bu, Multi-state joint estimation of series battery pack based on multi-model fusion, Electrochimica Acta 443 (2023) 141964. https://doi.org/10.1016/j.electacta.2023.141964.

[115] J. Tian, R. Xiong, Q. Yu, Fractional-order model-based incremental capacity analysis for degradation state recognition of lithium-ion batteries, IEEE Trans. Ind. Electron. 66 (2018) 1576–1584.

[116] H. Ruan, B. Sun, J. Jiang, W. Zhang, X. He, X. Su, J. Bian, W. Gao, A modified-electrochemical impedance spectroscopy-based multi-time-scale fractional-order model for lithium-ion batteries, Electrochimica Acta 394 (2021) 139066. https://doi.org/10.1016/j.electacta.2021.139066.

[117] D. Guo, G. Yang, X. Feng, X. Han, L. Lu, M. Ouyang, Physics-based fractional-order model with simplified solid phase diffusion of lithium-ion battery, J. Energy Storage 30 (2020) 101404. https://doi.org/10.1016/j.est.2020.101404.

[118] L. De Sutter, Y. Firouz, J. De Hoog, N. Omar, J. Van Mierlo, Battery aging assessment and parametric study of lithium-ion batteries by means of a fractional differential model, Electrochimica Acta 305 (2019) 24–36. https://doi.org/10.1016/j.electacta.2019.02.104.

[119] X. Lu, H. Li, N. Chen, An indicator for the electrode aging of lithium-ion batteries using a fractional variable order model, Electrochimica Acta 299 (2019) 378–387. https://doi.org/10.1016/j.electacta.2018.12.097.

[120] R. Xiong, J. Tian, W. Shen, F. Sun, A Novel Fractional Order Model for State of Charge Estimation in Lithium Ion Batteries, IEEE Trans. Veh. Technol. 68 (2019) 4130–4139. https://doi.org/10.1109/TVT.2018.2880085.

[121] Q. Zhu, M. Xu, W. Liu, M. Zheng, A state of charge estimation method for lithium-ion batteries based on fractional order adaptive extended kalman filter, Energy 187 (2019) 115880. https://doi.org/10.1016/j.energy.2019.115880.

[122] Y. Xu, M. Hu, A. Zhou, Y. Li, S. Li, C. Fu, C. Gong, State of charge estimation for lithium-ion batteries based on adaptive dual Kalman filter, Appl. Math. Model. 77 (2020) 1255–1272. https://doi.org/10.1016/j.apm.2019.09.011.

[123] X. Li, K. Pan, G. Fan, R. Lu, C. Zhu, G. Rizzoni, M. Canova, A physics-based fractional order model and state of energy estimation for lithium ion batteries. Part II: Parameter identification and state of energy estimation for LiFePO4 battery, J. Power Sources 367 (2017) 202–213.

[124] X. Hu, H. Yuan, C. Zou, Z. Li, L. Zhang, Co-Estimation of State of Charge and State of Health for Lithium-Ion Batteries Based on Fractional-Order Calculus, IEEE Trans. Veh. Technol. 67 (2018) 10319–10329. https://doi.org/10.1109/TVT.2018.2865664.

[125] R. Guo, W. Shen, Toward accurate online state of power estimation for lithium-ion batteries in electric vehicles, in: 2023 IEEE 6th Int. Electr. Energy Conf. CIEEC, IEEE, 2023: pp. 402–407.

[126] X. Li, J. Sun, Z. Hu, R. Lu, C. Zhu, G. Wu, A New Method of State of Peak Power Capability Prediction for Li-Ion Battery, in: 2015 IEEE Veh. Power Propuls. Conf. VPPC, IEEE, Montreal, QC, Canada, 2015: pp. 1–5. https://doi.org/10.1109/VPPC.2015.7352880.

[127] C. Liu, M. Hu, G. Jin, Y. Xu, J. Zhai, State of power estimation of lithium-ion battery based on fractional-order equivalent circuit model, J. Energy Storage 41 (2021) 102954. https://doi.org/10.1016/j.est.2021.102954.

[128] I. Podlubny, Fractional differential equations: an introduction to fractional derivatives, fractional differential equations, to methods of their solution and some of their applications, Elsevier, 1998.

[129] R. Guo, C. Hu, W. Shen, An Adaptive Approach for Battery State of Charge and State of Power Co-Estimation With a Fractional-Order Multi-Model System Considering Temperatures, IEEE Trans. Intell. Transp. Syst. 24 (2023) 15131–15145. https://doi.org/10.1109/TITS.2023.3299270.

[130] A. Farmann, D.U. Sauer, Comparative study of reduced order equivalent circuit models for on-board state-of-available-power prediction of lithium-ion batteries in electric vehicles, Appl. Energy 225 (2018) 1102–1122. https://doi.org/10.1016/j.apenergy.2018.05.066.

[131] X. Lai, L. He, S. Wang, L. Zhou, Y. Zhang, T. Sun, Y. Zheng, Co-estimation of state of charge and state of power for lithium-ion batteries based on fractional variable-order model, J. Clean. Prod. 255 (2020) 120203. https://doi.org/10.1016/j.jclepro.2020.120203.

[132] O. Bohlen, M. Roscher, Method for determining and/or predicting the maximum power capacity of a battery, (2014).

[133] A. Boehm, J. Weber, Adaptives verfahren zur bestimmung der maximal abgebaren oder aufnehmbaren leistung einer batterie, Pat. WO 2011095368 Al (2011).

[134] J. Belt, S. Jorgensen, Battery test manual for low-energy storage system for power-assist hybrid electric vehicles, Ida. Natl. Lab. US DoE (2013).

[135] J. Do Yang, others, Method of estimating maximum output of battery for hybrid electric vehicle, (2009).

[136] O. Bohlen, Impedance-based battery monitoring, Shaker, 2008.

[137] O. Bohlen, Robust algorithms for a reliable battery diagnosis-managing batteries in hybrid electric vehicles, in: Proc 22nd Int. Battery Hybrid Fuel Cell Electr. Veh. Symp. Expo. 2006, 2006.

[138] C. Fleischer, W. Waag, Ziou Bai, D.U. Sauer, Self-learning state-of-available-power prediction for lithium-ion batteries in electrical vehicles, in: 2012 IEEE Veh. Power Propuls. Conf., 2012: pp. 370–375. https://doi.org/10.1109/VPPC.2012.6422670.

[139] X. Tang, K. Yao, B. Liu, W. Hu, F. Gao, Long-term battery voltage, power, and surface temperature prediction using a model-based extreme learning machine, Energies 11 (2018) 86.

[140] R. Guo, W. Shen, A data-model fusion method for online state of power estimation of lithium-ion batteries at high discharge rate in electric vehicles, Energy 254 (2022) 124270. https://doi.org/10.1016/j.energy.2022.124270.

[141] B. Li, C. Hu, Multifunctional Estimation and Analysis of Lithium-Ion Battery State Based on Data Model Fusion under Multiple Constraints, J. Electrochem. Soc. 169 (2022) 110548. https://doi.org/10.1149/1945-7111/aca2ee.

[142] R. Li, K. Li, P. Liu, X. Zhang, Research on Multi-Time Scale SOP Estimation of Lithium–Ion Battery Based on H∞ Filter, Batteries 9 (2023) 191.

[143] X. Hu, H. Jiang, F. Feng, C. Zou, A Novel Multi-scale Co-estimation Framework of State of Charge, State of Health, and State of Power for Lithium-Ion Batteries, DEStech Trans. Environ. Energy Earth Sci. (2019). https://doi.org/10.12783/dteees/iceee2018/27824.

[144] X. Hu, H. Jiang, F. Feng, B. Liu, An enhanced multi-state estimation hierarchy for advanced lithium-ion battery management, Appl. Energy 257 (2020) 114019. https://doi.org/10.1016/j.apenergy.2019.114019.

[145] X. Shu, G. Li, J. Shen, Z. Lei, Z. Chen, Y. Liu, An adaptive multi-state estimation algorithm for lithium-ion batteries incorporating temperature compensation, Energy 207 (2020) 118262. https://doi.org/10.1016/j.energy.2020.118262.

[146] T. Zhang, N. Guo, X. Sun, J. Fan, N. Yang, J. Song, Y. Zou, A Systematic Framework for State of Charge, State of Health and State of Power Co-Estimation of Lithium-Ion Battery in Electric Vehicles, Sustainability 13 (2021) 5166. https://doi.org/10.3390/su13095166.

[147] Y. Wang, R. Xu, C. Zhou, X. Kang, Z. Chen, Digital twin and cloud-side-end collaboration for intelligent battery management system, J. Manuf. Syst. 62 (2022) 124–134. https://doi.org/10.1016/j.jmsy.2021.11.006.

[148] X. Li, J. Xu, J. Hong, J. Tian, Y. Tian, State of energy estimation for a series-connected lithium-ion battery pack based on an adaptive weighted strategy, Energy 214 (2021) 118858. https://doi.org/10.1016/j.energy.2020.118858.

[149] B. Roşca, S. Wilkins, J. Jacob, E. Hoedemaekers, S. Van Den Hoek, Predictive model based battery constraints for electric motor control within EV Powertrains, in: 2014 IEEE Int. Electr. Veh. Conf. IEVC, IEEE, 2014: pp. 1–8.

[150] M. Ouyang, S. Gao, L. Lu, X. Feng, D. Ren, J. Li, Y. Zheng, P. Shen, Determination of the battery pack capacity considering the estimation error using a Capacity–Quantity diagram, Appl. Energy 177 (2016) 384–392. https://doi.org/10.1016/j.apenergy.2016.05.137.

[151] C. Zhang, C. Zhang, S.M. Sharkh, Estimation of Real-Time Peak Power Capability of a Traction Battery Pack Used in an HEV, in: 2010 Asia-Pac. Power Energy Eng. Conf., IEEE, Chengdu, China, 2010: pp. 1–6. https://doi.org/10.1109/APPEEC.2010.5448755.

[152] Z. Zhou, Y. Kang, Y. Shang, N. Cui, C. Zhang, B. Duan, Peak power prediction for series-connected LiNCM battery pack based on representative cells, J. Clean. Prod. 230 (2019) 1061–1073. https://doi.org/10.1016/j.jclepro.2019.05.144.

[153] B. Jiang, H. Dai, X. Wei, L. Zhu, Z. Sun, Online reliable peak charge/discharge power estimation of series-connected lithium-ion battery packs, Energies 10 (2017) 390.

[154] W. Han, F. Altaf, C. Zou, T. Wik, State of Power Prediction for Battery Systems with Parallel-Connected Units, IEEE Trans. Transp. Electrification (2021) 1–1. https://doi.org/10.1109/TTE.2021.3101242.

[155] L. Wang, Y. Cheng, J. Zou, Battery available power prediction of hybrid electric vehicle based on improved Dynamic Matrix Control algorithms, J. Power Sources 261 (2014) 337–347. https://doi.org/10.1016/j.jpowsour.2014.03.091.

[156] M.J. Esfandyari, M.R. Hairi Yazdi, V. Esfahanian, M. Masih-Tehrani, H. Nehzati, O. Shekoofa, A hybrid model predictive and fuzzy logic based control method for state of power estimation of series-connected Lithium-ion batteries in HEVs, J. Energy Storage 24 (2019) 100758. https://doi.org/10.1016/j.est.2019.100758.

[157] K. Qi, W. Zhang, W. Zhou, J. Cheng, Integrated battery power capability prediction and driving torque regulation for electric vehicles: A reduced order

MPC approach, Appl. Energy 317 (2022) 119179. https://doi.org/10.1016/j.apenergy.2022.119179.

[158] W. Waag, C. Fleischer, D.U. Sauer, On-line estimation of lithium-ion battery impedance parameters using a novel varied-parameters approach, J. Power Sources 237 (2013) 260–269. https://doi.org/10.1016/j.jpowsour.2013.03.034.

[159] Z. Chen, J. Lu, Y. Yang, R. Xiong, Online estimation of state of power for lithium-ion battery considering the battery aging, in: 2017 Chin. Autom. Congr. CAC, IEEE, Jinan, 2017: pp. 3112–3116. https://doi.org/10.1109/CAC.2017.8243310.

[160] J. Lu, Z. Chen, Y. Yang, M. L.V., Online Estimation of State of Power for Lithium-Ion Batteries in Electric Vehicles Using Genetic Algorithm, IEEE Access 6 (2018) 20868–20880. https://doi.org/10.1109/ACCESS.2018.2824559.

[161] X. Hu, R. Xiong, Model-Based Dynamic Power Assessment of Lithium-Ion Batteries Considering Different Operating Conditions, IEEE Trans. Ind. Inform. 10 (2014) 12.

[162] F. Zheng, Y. Xing, J. Jiang, B. Sun, J. Kim, M. Pecht, Influence of different open circuit voltage tests on state of charge online estimation for lithium-ion batteries, Appl. Energy 183 (2016) 513–525. https://doi.org/10.1016/j.apenergy.2016.09.010.

[163] M. Petzl, M.A. Danzer, Advancements in OCV Measurement and Analysis for Lithium-Ion Batteries, IEEE Trans. Energy Convers. 28 (2013) 675–681. https://doi.org/10.1109/TEC.2013.2259490.

[164] M.A. Roscher, D.U. Sauer, Dynamic electric behavior and open-circuit-voltage modeling of LiFePO4-based lithium ion secondary batteries, J. Power Sources 196 (2011) 331–336. https://doi.org/10.1016/j.jpowsour.2010.06.098.

[165] G. Liu, M. Ouyang, L. Lu, J. Li, X. Han, Analysis of the heat generation of lithium-ion battery during charging and discharging considering different influencing factors, J. Therm. Anal. Calorim. 116 (2014) 1001–1010. https://doi.org/10.1007/s10973-013-3599-9.

[166] F. Yang, D. Wang, Y. Zhao, K.-L. Tsui, S.J. Bae, A study of the relationship between coulombic efficiency and capacity degradation of commercial lithium-ion batteries, Energy 145 (2018) 486–495. https://doi.org/10.1016/j.energy.2017.12.144.

[167] F. Yang, X. Song, G. Dong, K.-L. Tsui, A coulombic efficiency-based model for prognostics and health estimation of lithium-ion batteries, Energy 171 (2019) 1173–1182. https://doi.org/10.1016/j.energy.2019.01.083.

[168] A.J. Smith, J.C. Burns, J.R. Dahn, A High Precision Study of the Coulombic Efficiency of Li-Ion Batteries, Electrochem. Solid-State Lett. 13 (2010) A177. https://doi.org/10.1149/1.3487637.

[169] G. Bhatti, H. Mohan, R. Raja Singh, Towards the future of smart electric vehicles: Digital twin technology, Renew. Sustain. Energy Rev. 141 (2021) 110801. https://doi.org/10.1016/j.rser.2021.110801.

[170] C. Hendricks, N. Williard, S. Mathew, M. Pecht, A failure modes, mechanisms, and effects analysis (FMMEA) of lithium-ion batteries, J. Power Sources 297 (2015) 113–120.

[171] X. Zhang, Y. Gao, B. Guo, C. Zhu, X. Zhou, L. Wang, J. Cao, A novel quantitative electrochemical aging model considering side reactions for lithium-ion batteries, Electrochimica Acta 343 (2020) 136070.

[172] Y. Hua, A. Cordoba-Arenas, N. Warner, G. Rizzoni, A multi time-scale state-of-charge and state-of-health estimation framework using nonlinear predictive filter

[173] Y. Zou, X. Hu, H. Ma, S.E. Li, Combined State of Charge and State of Health estimation over lithium-ion battery cell cycle lifespan for electric vehicles, J. Power Sources 273 (2015) 793–803. https://doi.org/10.1016/j.jpowsour.2014.09.146.

for lithium-ion battery pack with passive balance control, J. Power Sources 280 (2015) 293–312. https://doi.org/10.1016/j.jpowsour.2015.01.112.

[174] R. Guo, Y. Xu, C. Hu, W. Shen, Self-Adaptive Neural Network-Based Fractional-Order Nonlinear Observer Design for State of Charge Estimation of Lithium-Ion Batteries, IEEEASME Trans. Mechatron. (2023).

[175] B. Pattavathi, V. Surendran, S. Palani, M.M. Shaijumon, Artificial neural network-enabled approaches toward mass balancing and cell optimization of lithium dual ion batteries, J. Energy Storage 68 (2023) 107878.

[176] L. Yao, J. Zheng, Y. Xiao, C. Zhang, L. Zhang, X. Gong, G. Cui, An intelligent fault diagnosis method for lithium-ion battery pack based on empirical mode decomposition and convolutional neural network, J. Energy Storage 72 (2023) 108181.

[177] V. Klass, M. Behm, G. Lindbergh, A support vector machine-based state-of-health estimation method for lithium-ion batteries under electric vehicle operation, J. Power Sources 270 (2014) 262–272. https://doi.org/10.1016/j.jpowsour.2014.07.116.

[178] C. Hu, G. Jain, C. Schmidt, C. Strief, M. Sullivan, Online estimation of lithium-ion battery capacity using sparse Bayesian learning, J. Power Sources 289 (2015) 105–113.

[179] P. Singh, R. Vinjamuri, X. Wang, D. Reisner, Design and implementation of a fuzzy logic-based state-of-charge meter for Li-ion batteries used in portable defibrillators, J. Power Sources 162 (2006) 829–836.

[180] Y. Li, K. Li, X. Liu, Y. Wang, L. Zhang, Lithium-ion battery capacity estimation—A pruned convolutional neural network approach assisted with transfer learning, Appl. Energy 285 (2021) 116410.

[181] Y. Zhang, R. Xiong, H. He, M.G. Pecht, Long short-term memory recurrent neural network for remaining useful life prediction of lithium-ion batteries, IEEE Trans. Veh. Technol. 67 (2018) 5695–5705.

[182] X. Gu, K. See, P. Li, K. Shan, Y. Wang, L. Zhao, K.C. Lim, N. Zhang, A novel state-of-health estimation for the lithium-ion battery using a convolutional neural network and transformer model, Energy 262 (2023) 125501.

[183] J. Tian, R. Xiong, J. Lu, C. Chen, W. Shen, Battery state-of-charge estimation amid dynamic usage with physics-informed deep learning, Energy Storage Mater. 50 (2022) 718–729. https://doi.org/10.1016/j.ensm.2022.06.007.

[184] F. Naaz, A. Herle, J. Channegowda, A. Raj, M. Lakshminarayanan, A generative adversarial network-based synthetic data augmentation technique for battery condition evaluation, Int. J. Energy Res. 45 (2021) 19120–19135.

[185] R. Jiao, K. Peng, J. Dong, Remaining useful life prediction of lithium-ion batteries based on conditional variational autoencoders-particle filter, IEEE Trans. Instrum. Meas. 69 (2020) 8831–8843.

[186] H. Chaoui, C.C. Ibe-Ekeocha, State of Charge and State of Health Estimation for Lithium Batteries Using Recurrent Neural Networks, IEEE Trans. Veh. Technol. 66 (2017) 8773–8783. https://doi.org/10.1109/TVT.2017.2715333.

[187] T. Sun, S. Wang, S. Jiang, B. Xu, X. Han, X. Lai, Y. Zheng, A cloud-edge collaborative strategy for capacity prognostic of lithium-ion batteries based on dynamic weight allocation and machine learning, Energy 239 (2022) 122185.

[188] H. Liao, Z. Jia, Z. Zhou, Y. Wang, H. Zhang, S. Mumtaz, Cloud-Edge-End Collaboration in Air–Ground Integrated Power IoT: A Semidistributed Learning Approach, IEEE Trans. Ind. Inform. 18 (2022) 8047–8057.

[189] S. Ci, N. Lin, D. Wu, Reconfigurable battery techniques and systems: A survey, IEEE Access 4 (2016) 1175–1189.

[190] F. Feng, X. Hu, L. Hu, F. Hu, Y. Li, L. Zhang, Propagation mechanisms and diagnosis of parameter inconsistency within Li-Ion battery packs, Renew. Sustain. Energy Rev. 112 (2019) 102–113. https://doi.org/10.1016/j.rser.2019.05.042.

[191] L. Zhang, X. Hu, Z. Wang, F. Sun, J. Deng, D.G. Dorrell, Multiobjective optimal sizing of hybrid energy storage system for electric vehicles, IEEE Trans. Veh. Technol. 67 (2017) 1027–1035.

[192] X. Tang, K. Liu, Q. Liu, Q. Peng, F. Gao, Comprehensive study and improvement of experimental methods for obtaining referenced battery state-of-power, J. Power Sources 512 (2021) 230462. https://doi.org/10.1016/j.jpowsour.2021.230462.

[193] J.-N. Shen, Q.-K. Wang, Z. Ma, Y. He, Nonlinear optimization strategy for state of power estimation of lithium-ion batteries: A systematical uncertainty analysis of key impact parameters, IEEE Trans. Ind. Inform. (2021) 1–1. https://doi.org/10.1109/TII.2021.3111539.